\documentstyle [12pt]{article}
\def\fnote#1#2{\begingroup\def\thefootnote{#1}\footnote{#2}\addtocounter
{footnote}{-1}\endgroup}
\font\twentybf=cmbx10 at 20pt
\font\fourteenbf=cmbx7 at 14pt
\def\nn{\nonumber}
\def\beq{\begin{equation}}
\def\eeq{\end{equation}}
\def\bea{\begin{eqnarray}}
\def\eea{\end{eqnarray}}
\def\lleq#1{\label{#1}\eeq}
\def\llea#1{\label{#1}\eea}
\def\bigboldcp{{\twentybf\relax I\kern-.18em P}\kern-3pt
 \lower5pt\hbox{\fourteenbf 4}}
\def\slantcp{\relax{\sl I\kern-.18em P}\kern-1.5pt${}_4$}
\def\plot#1#2{\vskip\parskip
 \vbox{\hrule width\hsize
 \hbox{\kern-0.2pt\vrule height#1
 \vbox{\hfill}\kern-0.6pt
 \vrule}\hrule width\hsize}
 \setbox0=\hbox{#2} \dimen0=\wd0 \divide\dimen0 by 2
 \setbox0=\hbox{\kern-\dimen0 #2}
 \dimen3=#1}
\def\hmark{\kern-0.2pt\lower10pt\hbox{\vrule height 5pt}}
\def\leftscalemark{\vbox{\hrule width5pt}}
\def\rightscalemark{\kern-5pt\vbox{\hrule width5pt}}

\def\Place#1#2#3{
 \count10=#1 \advance\count10 by 960
 \dimen1=\hsize \divide\dimen1 by 1920 \multiply\dimen1 by \count10
 \dimen2=\dimen3 \divide\dimen2 by 550 \multiply\dimen2 by #2
 \vbox to 0pt{\kern-\parskip\kern-18.2truept\kern-\dimen2
 \hbox{\kern\dimen1#3}\vss}\nointerlineskip}
\def\datum#1#2{\Place{#1}{#2}{\copy0}}

\def\inbar{\vrule height1.5ex width.4pt depth0pt}
\def\IB{\relax{\rm I\kern-.18em B}}
\def\IC{\relax\,\hbox{$\inbar\kern-.3em{\rm C}$}}
\def\ID{\relax{\rm I\kern-.18em D}}
\def\IE{\relax{\rm I\kern-.18em E}}
\def\IF{\relax{\rm I\kern-.18em F}}
\def\IG{\relax\,\hbox{$\inbar\kern-.3em{\rm G}$}}
\def\IH{\relax{\rm I\kern-.18em H}}
\def\II{\relax{\rm I\kern-.18em I}}
\def\IK{\relax{\rm I\kern-.18em K}}
\def\IL{\relax{\rm I\kern-.18em L}}
\def\IM{\relax{\rm I\kern-.18em M}}
\def\IN{\relax{\rm I\kern-.18em N}}
\def\IO{\relax\,\hbox{$\inbar\kern-.3em{\rm O}$}}
\def\IP{\relax{\rm I\kern-.18em P}}
\def\IQ{\relax\,\hbox{$\inbar\kern-.3em{\rm Q}$}}
\def\IR{\relax{\rm I\kern-.18em R}}
\def\ZZ{\relax{\sf Z\kern-.4em Z}}
\def\fnote#1#2{\begingroup\def\thefootnote{#1}\footnote{#2}\addtocounter
 {footnote}{-1}\endgroup}
\def\tabroom{\hbox to0pt{\phantom{\Huge A}\hss}}
\def\notin{\ \hbox{{$\in$}\kern-.51em\hbox{/}}}
\def\a{\alpha} \def\b{\beta}  
 \def\l{\lambda} \def\om{\omega}
\def\Om{\Omega}  \def\th{\theta}
\def\cA{{\cal A}}   \def\cD{{\cal D}}
 \def\cG{{\cal G}} \def\cH{{\cal H}} \def\cI{{\cal I}}
   \def\cM{{\cal M}}
 \def\cO{{\cal O}} \def\cP{{\cal P}} 
\def\cR{{\cal R}} 
\def\phiti{\tilde {\Phi}}
\def\bn{\bar n} \def\bt{\bar t} \def\bz{\bar z}
\def\bPhi{\bar \Phi} \def\bth{\bar \theta}
\def\cym{Calabi--Yau manifold}
\def\cys{Calabi--Yau manifolds}
\def\cy{Calabi--Yau}
\def\del{\partial}
\def\etc{{\it etc\/}}
\def\0{\over } \def\1{\vec }
\let\qqd=\qquad 

\begin{document}
\hfill {
{HD--THEP--92--17}}
\vskip 1truein
\centerline {\LARGE THE CONSTRUCTION OF MIRROR SYMMETRY
 \fnote{\dagger}{Dedicated to the memory of B.Warr and T.I.A.Hu--Man.}
 \fnote{\star}{Based in part on talks presented at the
 Workshop on Mirror Symmetry, MSRI,
 Berkeley, May 1991 and the Workshop on
 Geometry and Quantum Field Theory, Johns Hopkins
 University, Baltimore, March 1992.
 }
 }
\vskip .4truein
\centerline {\sc Rolf Schimmrigk
 \fnote{\diamond}{E--mail: q25@dhdurz1.bitnet, netah@cernvm.bitnet}}

\vskip .2truein
\centerline {\it Institut f\"ur Theoretische Physik,
 Universit\"at Heidelberg}
\centerline {\it Philosophenweg 16,~ 6900 Heidelberg,~FRG}
\vskip .05truein
\centerline {and}
\vskip .05truein
\centerline {\it Theory Division, CERN}
\centerline {\it CH--1211 Geneva 23,~Switzerland}

\vskip 1.2truein

\centerline{\bf ABSTRACT}

\vskip .1truein

\noindent
The construction of mirror symmetry in the heterotic string is reviewed
in the context of Calabi--Yau and Landau--Ginzburg compactifications. This
framework has the virtue of providing a large subspace of the configuration
space of the heterotic string, probing its structure far beyond the present
reaches of solvable models.
The construction proceeds in two stages: First all singularities/catastrophes
which lead to ground states of the heterotic string are found. It is then
shown that not all ground states described in this way are independent but
that certain classes of these Landau--Ginzburg/Calabi--Yau string vacua can
be related to other, simpler, theories via a process involving fractional
transformations of the order parameters as well as orbifolding.
This construction has far reaching consequences. Firstly it allows for
a systematic identification of mirror pairs that appear abundantly in this
class of string vacua, thereby showing that the emerging mirror symmetry
is not accidental. This is important because models with mirror flipped
spectra are a priori independent theories, described by distinct
Calabi--Yau/Landau--Ginzburg models. It also shows that mirror symmetry
is not restricted to the space of string vacua described by theories based
on Fermat potentials (corresponding to minimal tensor models). Furthermore
it shows the need for a better set of coordinates of the configuration space
or else the structure of this space will remain obscure.
While the space of Landau--Ginzburg vacua is {\it not} completely mirror
symmetric, results described in the last part suggest that the space of
Landau--Ginburg {\it orbifolds} possesses this symmetry.

\renewcommand\thepage{}

\vfill
\eject

\parskip .1truein
\pagenumbering{arabic}
\parindent=20pt
\baselineskip=20pt

\section{Introduction}

\noindent
Even though the general structure of the configuration space of the
heterotic
string remains to a large extent terra incognita, some of its important
properties have been uncovered. Perhaps the most interesting of these
is the
recently discovered mirror symmetry of the space of (2,2)--supersymmetric
vacua \cite{cls}\cite{gp} .
Ideally questions about the space of ground states should be analyzed
starting from first principles, given an appropriate parametrization of
this manifold. Not too much progress however has
been made along this avenue. Instead one proceeds somewhat indirectly.
The symmetry principles of string theory are used to formulate a set of
consistency conditions which
are solved explicitly. Unfortunately this
introduces some uncertainty as to whether the part of the space of vacua
that has been uncovered via these constructions represents a typical slice
of the whole space. Properties that are generic in specific
constructions
may not at all be features of the total space one is interested in
but instead could merely be artefacts of the techniques employed.

An example of such an artefact is furnished by the class of
heterotic string vacua described by
{\bf c}omplete {\bf i}ntersection {\bf C}alabi--{\bf Y}au
manifolds embedded in products of projective spaces (CICYs). In this class
the number of generations and antigenerations of the models are
parametrized
 by the only two
independent Hodge numbers $(h^{(1,1)},h^{(2,1)})$ that exist on such
manifolds and the number of light
generations of these theories is measured by the Euler number
$\chi=2(h^{(1,1)}-h^{(2,1)})$.
The results for the latter turn out to lie in the range
$-200 \leq \chi \leq 0$ \cite{cdls}.
In Fig.1 the Euler number of all CICY vacua is plotted versus the
sum of the two independent Hodge numbers \cite{ghl}.

\vskip .1truein

\plot{2.5truein}{\scriptsize{$\bullet$}}
\nobreak
\Place{-960}{50}{\leftscalemark~~50}
\Place{-960}{100}{\leftscalemark~~100}
\Place{-960}{150}{\leftscalemark~~150}
\Place{-960}{200}{\leftscalemark~~200}
\Place{-960}{250}{\leftscalemark~~250}
\Place{-960}{300}{\leftscalemark~~300}
\Place{-960}{350}{\leftscalemark~~350}
\Place{-960}{400}{\leftscalemark~~400}
\Place{-960}{450}{\leftscalemark~~450}
\Place{-960}{500}{\leftscalemark~~500}
\Place{960}{50}{\rightscalemark\vphantom{0}}
\Place{960}{100}{\rightscalemark\vphantom{0}}
\Place{960}{150}{\rightscalemark\vphantom{0}}
\Place{960}{200}{\rightscalemark\vphantom{0}}
\Place{960}{250}{\rightscalemark\vphantom{0}}
\Place{960}{300}{\rightscalemark\vphantom{0}}
\Place{960}{350}{\rightscalemark\vphantom{0}}
\Place{960}{400}{\rightscalemark\vphantom{0}}
\Place{960}{450}{\rightscalemark\vphantom{0}}
\Place{960}{500}{\rightscalemark\vphantom{0}}
\Place{-960}{0}{\hmark\lower18pt\hbox{-960}}
\Place{-720}{0}{\hmark\lower18pt\hbox{-720}}
\Place{-480}{0}{\hmark\lower18pt\hbox{-480}}
\Place{-240}{0}{\hmark\lower18pt\hbox{-240}}
\Place{0}{0}{\hmark\lower18pt\hbox{0}}
\Place{240}{0}{\hmark\lower18pt\hbox{240}}
\Place{480}{0}{\hmark\lower18pt\hbox{480}}
\Place{720}{0}{\hmark\lower18pt\hbox{720}}
\Place{960}{0}{\hmark\lower18pt\hbox{960}}
\Place{-720}{550}{\hmark}
\Place{-480}{550}{\hmark}
\Place{-240}{550}{\hmark}
\Place{0}{550}{\hmark}
\Place{240}{550}{\hmark}
\Place{480}{550}{\hmark}
\Place{720}{550}{\hmark}
\Place{960}{550}{\hmark}
\nobreak
\datum{ 0}{ 30}
\datum{ 0}{ 38}
\datum{ -4}{ 30}
\datum{ -8}{ 30}
\datum{ -8}{ 32}
\datum{ -8}{ 36}
\datum{ -12}{ 30}
\datum{ -12}{ 32}
\datum{ -12}{ 36}
\datum{ -14}{ 31}
\datum{ -16}{ 30}
\datum{ -16}{ 32}
\datum{ -16}{ 34}
\datum{ -16}{ 36}
\datum{ -18}{ 31}
\datum{ -18}{ 33}
\datum{ -18}{ 37}
\datum{ -20}{ 30}
\datum{ -20}{ 32}
\datum{ -20}{ 34}
\datum{ -20}{ 36}
\datum{ -22}{ 31}
\datum{ -22}{ 33}
\datum{ -24}{ 30}
\datum{ -24}{ 32}
\datum{ -24}{ 34}
\datum{ -24}{ 36}
\datum{ -24}{ 38}
\datum{ -26}{ 31}
\datum{ -26}{ 33}
\datum{ -28}{ 30}
\datum{ -28}{ 32}
\datum{ -28}{ 34}
\datum{ -28}{ 36}
\datum{ -30}{ 31}
\datum{ -30}{ 33}
\datum{ -30}{ 35}
\datum{ -30}{ 37}
\datum{ -30}{ 39}
\datum{ -32}{ 30}
\datum{ -32}{ 32}
\datum{ -32}{ 34}
\datum{ -32}{ 36}
\datum{ -32}{ 38}
\datum{ -32}{ 40}
\datum{ -34}{ 31}
\datum{ -34}{ 33}
\datum{ -34}{ 35}
\datum{ -34}{ 37}
\datum{ -34}{ 39}
\datum{ -36}{ 30}
\datum{ -36}{ 32}
\datum{ -36}{ 34}
\datum{ -36}{ 36}
\datum{ -36}{ 37}
\datum{ -36}{ 38}
\datum{ -36}{ 40}
\datum{ -38}{ 31}
\datum{ -38}{ 33}
\datum{ -38}{ 35}
\datum{ -38}{ 37}
\datum{ -38}{ 39}
\datum{ -40}{ 30}
\datum{ -40}{ 32}
\datum{ -40}{ 34}
\datum{ -40}{ 36}
\datum{ -40}{ 37}
\datum{ -40}{ 38}
\datum{ -40}{ 40}
\datum{ -40}{ 42}
\datum{ -42}{ 31}
\datum{ -42}{ 33}
\datum{ -42}{ 35}
\datum{ -42}{ 37}
\datum{ -42}{ 39}
\datum{ -42}{ 41}
\datum{ -44}{ 30}
\datum{ -44}{ 32}
\datum{ -44}{ 34}
\datum{ -44}{ 36}
\datum{ -44}{ 37}
\datum{ -44}{ 38}
\datum{ -44}{ 40}
\datum{ -44}{ 42}
\datum{ -46}{ 33}
\datum{ -46}{ 35}
\datum{ -46}{ 37}
\datum{ -46}{ 39}
\datum{ -46}{ 41}
\datum{ -48}{ 30}
\datum{ -48}{ 32}
\datum{ -48}{ 34}
\datum{ -48}{ 36}
\datum{ -48}{ 37}
\datum{ -48}{ 38}
\datum{ -48}{ 40}
\datum{ -48}{ 42}
\datum{ -48}{ 44}
\datum{ -50}{ 33}
\datum{ -50}{ 35}
\datum{ -50}{ 37}
\datum{ -50}{ 39}
\datum{ -50}{ 41}
\datum{ -50}{ 45}
\datum{ -52}{ 34}
\datum{ -52}{ 36}
\datum{ -52}{ 37}
\datum{ -52}{ 38}
\datum{ -52}{ 40}
\datum{ -52}{ 41}
\datum{ -52}{ 42}
\datum{ -52}{ 43}
\datum{ -52}{ 44}
\datum{ -54}{ 35}
\datum{ -54}{ 37}
\datum{ -54}{ 39}
\datum{ -54}{ 41}
\datum{ -54}{ 43}
\datum{ -54}{ 45}
\datum{ -56}{ 34}
\datum{ -56}{ 36}
\datum{ -56}{ 37}
\datum{ -56}{ 38}
\datum{ -56}{ 40}
\datum{ -56}{ 41}
\datum{ -56}{ 42}
\datum{ -56}{ 43}
\datum{ -56}{ 44}
\datum{ -56}{ 46}
\datum{ -58}{ 37}
\datum{ -58}{ 39}
\datum{ -58}{ 41}
\datum{ -58}{ 43}
\datum{ -58}{ 45}
\datum{ -60}{ 36}
\datum{ -60}{ 38}
\datum{ -60}{ 40}
\datum{ -60}{ 41}
\datum{ -60}{ 42}
\datum{ -60}{ 43}
\datum{ -60}{ 44}
\datum{ -60}{ 46}
\datum{ -60}{ 48}
\datum{ -62}{ 39}
\datum{ -62}{ 41}
\datum{ -62}{ 43}
\datum{ -62}{ 45}
\datum{ -64}{ 38}
\datum{ -64}{ 40}
\datum{ -64}{ 41}
\datum{ -64}{ 42}
\datum{ -64}{ 43}
\datum{ -64}{ 44}
\datum{ -64}{ 46}
\datum{ -64}{ 48}
\datum{ -66}{ 39}
\datum{ -66}{ 41}
\datum{ -66}{ 43}
\datum{ -66}{ 45}
\datum{ -66}{ 47}
\datum{ -68}{ 40}
\datum{ -68}{ 42}
\datum{ -68}{ 43}
\datum{ -68}{ 44}
\datum{ -68}{ 45}
\datum{ -68}{ 46}
\datum{ -68}{ 47}
\datum{ -68}{ 48}
\datum{ -68}{ 49}
\datum{ -70}{ 41}
\datum{ -70}{ 43}
\datum{ -70}{ 45}
\datum{ -70}{ 46}
\datum{ -70}{ 47}
\datum{ -70}{ 51}
\datum{ -72}{ 42}
\datum{ -72}{ 44}
\datum{ -72}{ 45}
\datum{ -72}{ 46}
\datum{ -72}{ 47}
\datum{ -72}{ 48}
\datum{ -72}{ 49}
\datum{ -72}{ 50}
\datum{ -72}{ 51}
\datum{ -74}{ 43}
\datum{ -74}{ 45}
\datum{ -74}{ 46}
\datum{ -74}{ 47}
\datum{ -74}{ 48}
\datum{ -74}{ 49}
\datum{ -76}{ 44}
\datum{ -76}{ 46}
\datum{ -76}{ 47}
\datum{ -76}{ 48}
\datum{ -76}{ 49}
\datum{ -76}{ 50}
\datum{ -78}{ 47}
\datum{ -78}{ 48}
\datum{ -78}{ 49}
\datum{ -78}{ 51}
\datum{ -80}{ 46}
\datum{ -80}{ 47}
\datum{ -80}{ 48}
\datum{ -80}{ 49}
\datum{ -80}{ 50}
\datum{ -80}{ 52}
\datum{ -80}{ 54}
\datum{ -82}{ 47}
\datum{ -82}{ 48}
\datum{ -82}{ 49}
\datum{ -82}{ 51}
\datum{ -84}{ 48}
\datum{ -84}{ 49}
\datum{ -84}{ 50}
\datum{ -84}{ 52}
\datum{ -84}{ 54}
\datum{ -86}{ 48}
\datum{ -86}{ 49}
\datum{ -86}{ 51}
\datum{ -86}{ 53}
\datum{ -88}{ 48}
\datum{ -88}{ 50}
\datum{ -88}{ 52}
\datum{ -88}{ 54}
\datum{ -88}{ 55}
\datum{ -88}{ 56}
\datum{ -90}{ 48}
\datum{ -90}{ 49}
\datum{ -90}{ 51}
\datum{ -90}{ 53}
\datum{ -90}{ 55}
\datum{ -90}{ 57}
\datum{ -92}{ 52}
\datum{ -92}{ 54}
\datum{ -92}{ 55}
\datum{ -92}{ 56}
\datum{ -94}{ 53}
\datum{ -94}{ 55}
\datum{ -94}{ 56}
\datum{ -96}{ 52}
\datum{ -96}{ 54}
\datum{ -96}{ 56}
\datum{ -96}{ 57}
\datum{ -96}{ 58}
\datum{ -96}{ 60}
\datum{ -98}{ 45}
\datum{ -98}{ 55}
\datum{ -98}{ 56}
\datum{ -98}{ 57}
\datum{ -100}{ 54}
\datum{ -100}{ 56}
\datum{ -100}{ 57}
\datum{ -100}{ 60}
\datum{ -102}{ 57}
\datum{ -102}{ 59}
\datum{ -102}{ 61}
\datum{ -104}{ 56}
\datum{ -104}{ 57}
\datum{ -104}{ 58}
\datum{ -104}{ 59}
\datum{ -104}{ 60}
\datum{ -104}{ 62}
\datum{ -106}{ 57}
\datum{ -108}{ 58}
\datum{ -108}{ 59}
\datum{ -108}{ 60}
\datum{ -108}{ 62}
\datum{ -108}{ 64}
\datum{ -110}{ 57}
\datum{ -112}{ 54}
\datum{ -112}{ 60}
\datum{ -112}{ 62}
\datum{ -112}{ 63}
\datum{ -112}{ 64}
\datum{ -112}{ 66}
\datum{ -114}{ 61}
\datum{ -114}{ 63}
\datum{ -114}{ 65}
\datum{ -116}{ 64}
\datum{ -116}{ 66}
\datum{ -120}{ 64}
\datum{ -120}{ 65}
\datum{ -120}{ 66}
\datum{ -120}{ 68}
\datum{ -124}{ 66}
\datum{ -126}{ 66}
\datum{ -126}{ 69}
\datum{ -126}{ 71}
\datum{ -128}{ 63}
\datum{ -128}{ 66}
\datum{ -128}{ 67}
\datum{ -128}{ 68}
\datum{ -128}{ 72}
\datum{ -132}{ 70}
\datum{ -132}{ 71}
\datum{ -132}{ 72}
\datum{ -138}{ 70}
\datum{ -140}{ 73}
\datum{ -144}{ 74}
\datum{ -144}{ 78}
\datum{ -148}{ 78}
\datum{ -150}{ 79}
\datum{ -162}{ 85}
\datum{ -168}{ 88}
\datum{ -176}{ 90}
\datum{ -200}{ 102}

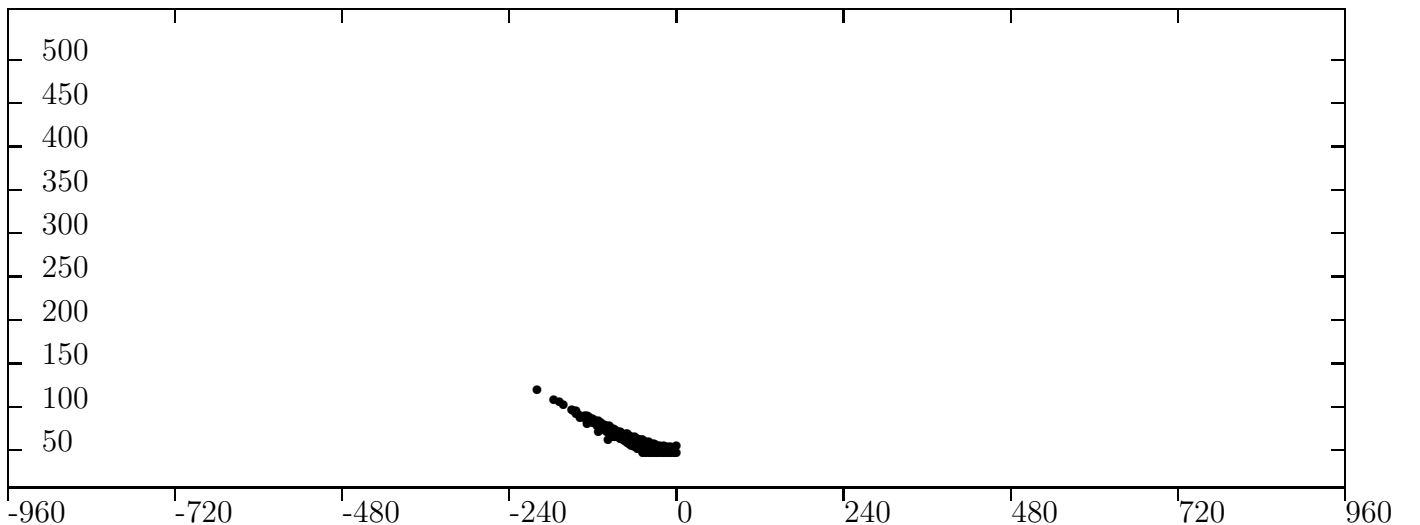
{\bf Fig. 1}~~{\it A plot of the Euler number versus
$(h^{(1,1)} +h^{(2,1)})$
for the spectra of the 7868 CICYs.}

At the time when the class of CICYs was constructed only very few
manifolds with
positive Euler number were known and hence one might naively have concluded
that the vast majority of groundstates of the heterotic string have negative
Euler numbers. This conclusion is
 reinforced by the complete construction \cite{ls3,fkss1} of the set
of heterotic vacua based on tensor products of minimal $N=2$ superconformal
field theories \cite{g}.
The resulting space is again rather asymmetric even though {\it some}
mirror pairs appear in this construction. Fig.2 contains again a plot
of the Euler numbers versus the sum of generations and antigenerations for
those theories.

\vskip .1truein

\plot{3.5truein}{\tiny{$\bullet$}}
\nobreak
\Place{-960}{50}{\leftscalemark~~50}
\Place{-960}{100}{\leftscalemark~~100}
\Place{-960}{150}{\leftscalemark~~150}
\Place{-960}{200}{\leftscalemark~~200}
\Place{-960}{250}{\leftscalemark~~250}
\Place{-960}{300}{\leftscalemark~~300}
\Place{-960}{350}{\leftscalemark~~350}
\Place{-960}{400}{\leftscalemark~~400}
\Place{-960}{450}{\leftscalemark~~450}
\Place{-960}{500}{\leftscalemark~~500}
\Place{960}{50}{\rightscalemark\vphantom{0}}
\Place{960}{100}{\rightscalemark\vphantom{0}}
\Place{960}{150}{\rightscalemark\vphantom{0}}
\Place{960}{200}{\rightscalemark\vphantom{0}}
\Place{960}{250}{\rightscalemark\vphantom{0}}
\Place{960}{300}{\rightscalemark\vphantom{0}}
\Place{960}{350}{\rightscalemark\vphantom{0}}
\Place{960}{400}{\rightscalemark\vphantom{0}}
\Place{960}{450}{\rightscalemark\vphantom{0}}
\Place{960}{500}{\rightscalemark\vphantom{0}}
\Place{-960}{0}{\hmark\lower18pt\hbox{-960}}
\Place{-720}{0}{\hmark\lower18pt\hbox{-720}}
\Place{-480}{0}{\hmark\lower18pt\hbox{-480}}
\Place{-240}{0}{\hmark\lower18pt\hbox{-240}}
\Place{0}{0}{\hmark\lower18pt\hbox{0}}
\Place{240}{0}{\hmark\lower18pt\hbox{240}}
\Place{480}{0}{\hmark\lower18pt\hbox{480}}
\Place{720}{0}{\hmark\lower18pt\hbox{720}}
\Place{960}{0}{\hmark\lower18pt\hbox{960}}
\Place{-720}{550}{\hmark}
\Place{-480}{550}{\hmark}
\Place{-240}{550}{\hmark}
\Place{0}{550}{\hmark}
\Place{240}{550}{\hmark}
\Place{480}{550}{\hmark}
\Place{720}{550}{\hmark}
\Place{960}{550}{\hmark}
\nobreak
\datum{-168}{ 84}
\datum{-104}{ 54}
\datum{-120}{ 62}
\datum{-144}{ 74}
\datum{-152}{ 78}
\datum{-168}{ 86}
\datum{-200}{ 102}
\datum{-204}{ 104}
\datum{-288}{ 146}
\datum{-296}{ 150}
\datum{-108}{ 58}
\datum{-112}{ 60}
\datum{-120}{ 64}
\datum{-144}{ 76}
\datum{-168}{ 88}
\datum{-208}{ 108}
\datum{-216}{ 112}
\datum{-240}{ 124}
\datum{-252}{ 130}
\datum{-540}{ 274}
\datum{ -96}{ 54}
\datum{-108}{ 60}
\datum{-132}{ 72}
\datum{-144}{ 78}
\datum{-192}{ 102}
\datum{-204}{ 108}
\datum{-324}{ 168}
\datum{-480}{ 246}
\datum{ -72}{ 44}
\datum{-144}{ 80}
\datum{-180}{ 98}
\datum{-216}{ 116}
\datum{-288}{ 152}
\datum{-408}{ 212}
\datum{ -72}{ 46}
\datum{ -96}{ 58}
\datum{-120}{ 70}
\datum{-144}{ 82}
\datum{-168}{ 94}
\datum{-192}{ 106}
\datum{-312}{ 166}
\datum{-360}{ 190}
\datum{-492}{ 256}
\datum{-108}{ 66}
\datum{-120}{ 72}
\datum{-168}{ 96}
\datum{-216}{ 120}
\datum{-228}{ 126}
\datum{-348}{ 186}
\datum{ -96}{ 62}
\datum{-108}{ 68}
\datum{-112}{ 70}
\datum{-144}{ 86}
\datum{-168}{ 98}
\datum{-192}{ 110}
\datum{-240}{ 134}
\datum{-272}{ 150}
\datum{-288}{ 158}
\datum{-528}{ 278}
\datum{ -54}{ 43}
\datum{ -72}{ 52}
\datum{-120}{ 76}
\datum{-144}{ 88}
\datum{-156}{ 94}
\datum{-192}{ 112}
\datum{-216}{ 124}
\datum{-264}{ 148}
\datum{-312}{ 172}
\datum{ -48}{ 42}
\datum{ -60}{ 48}
\datum{ -96}{ 66}
\datum{-120}{ 78}
\datum{-144}{ 90}
\datum{-204}{ 120}
\datum{-216}{ 126}
\datum{-240}{ 138}
\datum{-288}{ 162}
\datum{-624}{ 330}
\datum{ -72}{ 56}
\datum{-120}{ 80}
\datum{-192}{ 116}
\datum{-336}{ 188}
\datum{-408}{ 224}
\datum{ -48}{ 46}
\datum{ -80}{ 62}
\datum{ -84}{ 64}
\datum{ -96}{ 70}
\datum{-144}{ 94}
\datum{-160}{ 102}
\datum{-192}{ 118}
\datum{-240}{ 142}
\datum{-288}{ 166}
\datum{-312}{ 178}
\datum{-432}{ 238}
\datum{-960}{ 502}
\datum{ -72}{ 60}
\datum{-168}{ 108}
\datum{-228}{ 138}
\datum{ -48}{ 50}
\datum{ -72}{ 62}
\datum{ -96}{ 74}
\datum{-120}{ 86}
\datum{-144}{ 98}
\datum{-192}{ 122}
\datum{-216}{ 134}
\datum{-288}{ 170}
\datum{-336}{ 194}
\datum{-432}{ 242}
\datum{ -72}{ 64}
\datum{-168}{ 112}
\datum{-456}{ 256}
\datum{ -48}{ 54}
\datum{ -96}{ 78}
\datum{-264}{ 162}
\datum{-336}{ 198}
\datum{ -72}{ 68}
\datum{-120}{ 92}
\datum{-168}{ 116}
\datum{-192}{ 128}
\datum{ -24}{ 46}
\datum{ -48}{ 58}
\datum{ -96}{ 82}
\datum{-120}{ 94}
\datum{-168}{ 118}
\datum{-276}{ 172}
\datum{-312}{ 190}
\datum{-720}{ 394}
\datum{ -48}{ 60}
\datum{-168}{ 120}
\datum{-216}{ 144}
\datum{-408}{ 240}
\datum{ -16}{ 46}
\datum{ -48}{ 62}
\datum{ -96}{ 86}
\datum{-144}{ 110}
\datum{ -24}{ 52}
\datum{ -72}{ 76}
\datum{-108}{ 94}
\datum{-168}{ 124}
\datum{-312}{ 196}
\datum{ 0}{ 42}
\datum{-192}{ 138}
\datum{ -72}{ 80}
\datum{ -96}{ 92}
\datum{-120}{ 104}
\datum{-216}{ 152}
\datum{ 0}{ 46}
\datum{ -24}{ 58}
\datum{ -48}{ 70}
\datum{ -72}{ 82}
\datum{ -96}{ 94}
\datum{-240}{ 166}
\datum{-480}{ 286}
\datum{-624}{ 358}
\datum{ -24}{ 60}
\datum{-120}{ 108}
\datum{-360}{ 228}
\datum{ -24}{ 62}
\datum{-120}{ 110}
\datum{-144}{ 122}
\datum{-120}{ 112}
\datum{-144}{ 124}
\datum{-264}{ 184}
\datum{ 0}{ 54}
\datum{ -64}{ 86}
\datum{ -48}{ 80}
\datum{-264}{ 188}
\datum{ 0}{ 58}
\datum{ -24}{ 70}
\datum{ -48}{ 82}
\datum{-384}{ 250}
\datum{ -96}{ 108}
\datum{-132}{ 126}
\datum{ 16}{ 54}
\datum{ 0}{ 62}
\datum{ -48}{ 86}
\datum{-120}{ 122}
\datum{-144}{ 134}
\datum{ 24}{ 54}
\datum{ -72}{ 102}
\datum{-144}{ 138}
\datum{-216}{ 174}
\datum{ -24}{ 80}
\datum{ -48}{ 92}
\datum{-168}{ 152}
\datum{-312}{ 224}
\datum{ 0}{ 70}
\datum{ -48}{ 94}
\datum{-216}{ 180}
\datum{ -24}{ 86}
\datum{ -96}{ 122}
\datum{ -72}{ 112}
\datum{ 0}{ 78}
\datum{ -80}{ 118}
\datum{ -96}{ 126}
\datum{-120}{ 138}
\datum{ -72}{ 116}
\datum{ 0}{ 82}
\datum{ -24}{ 94}
\datum{ -72}{ 118}
\datum{-216}{ 192}
\datum{ 0}{ 86}
\datum{ -48}{ 110}
\datum{ -36}{ 106}
\datum{ -72}{ 124}
\datum{ 0}{ 90}
\datum{ -96}{ 138}
\datum{-120}{ 152}
\datum{ 0}{ 94}
\datum{-480}{ 334}
\datum{ 24}{ 84}
\datum{-264}{ 228}
\datum{ 0}{ 98}
\datum{ -48}{ 122}
\datum{-168}{ 184}
\datum{ 0}{ 106}
\datum{ -72}{ 142}
\datum{-240}{ 226}
\datum{ 0}{ 110}
\datum{ -48}{ 138}
\datum{ 0}{ 118}
\datum{ -48}{ 142}
\datum{-144}{ 190}
\datum{-120}{ 190}
\datum{-216}{ 240}
\datum{ -24}{ 148}
\datum{-120}{ 196}
\datum{ 0}{ 138}
\datum{ 0}{ 142}
\datum{-144}{ 214}
\datum{ -24}{ 160}
\datum{ 0}{ 158}
\datum{ 48}{ 138}
\datum{ 0}{ 166}
\datum{ -72}{ 212}
\datum{ 0}{ 178}
\datum{-336}{ 358}
\datum{ 0}{ 194}
\datum{ -96}{ 250}
\datum{ 0}{ 214}
\datum{ 0}{ 238}
\datum{ -96}{ 286}
\datum{ 144}{ 190}
\datum{-240}{ 394}
\datum{ 0}{ 286}
\datum{ 96}{ 286}
\datum{ 0}{ 502}
\datum{ 240}{ 394}
\begin{center}
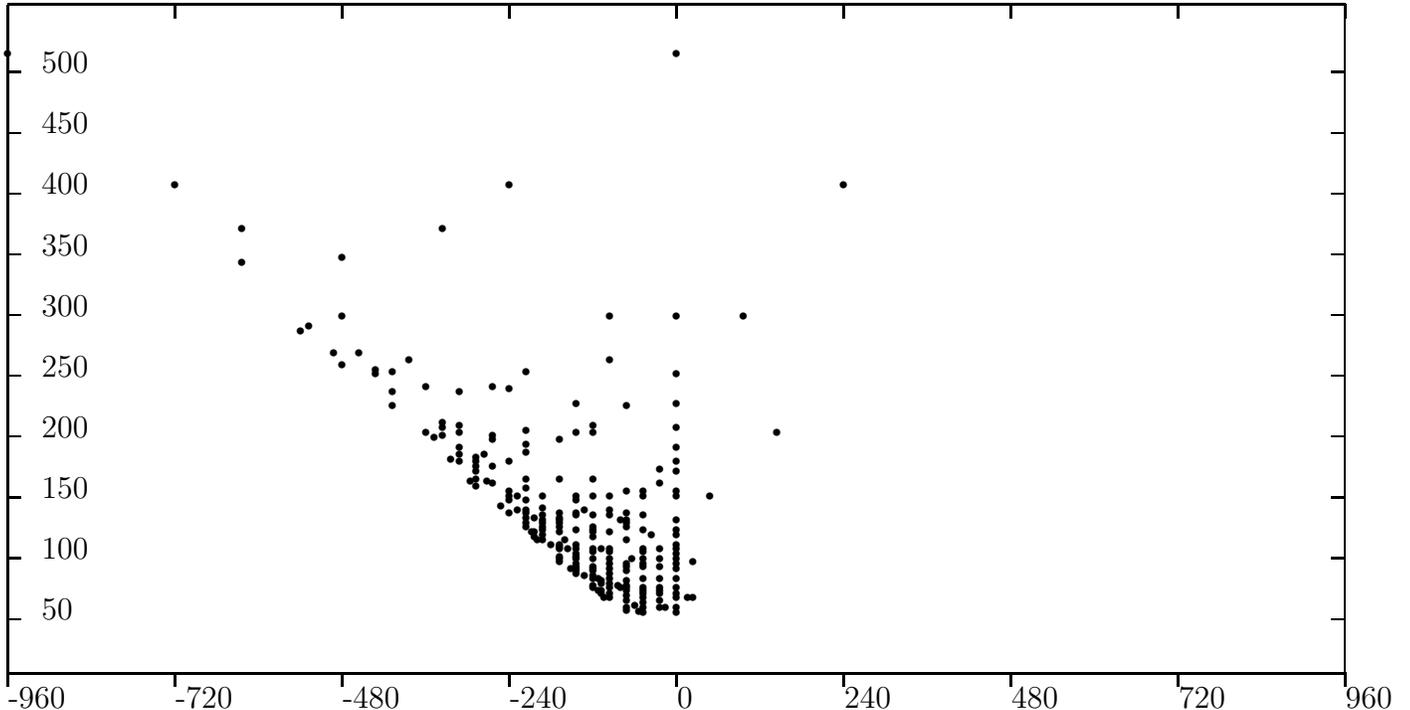
\parbox{6.4truein}{\noindent {\bf Fig. 2}~~{\it A plot of the Euler
number
versus $(h^{(1,1)} +h^{(2,1)})$ for all ADE tensor models.}}
\end{center}

As it turns out, however, that the idea of an asymmetric space of ground
states is not correct.
The purpose of this review is to first describe in Part I a class of
string vacua whose spectra are almost
symmetrically distributed in a range of positive and negative Euler
numbers,
secondly to show in Part II that this is not an accident and thirdly to
present evidence, in Part III, that mirror symmetry is a `robust'
property in that it is not an artefact of any one construction.

Sections 2 and 3 are devoted to the construction of a class of
Calabi--Yau
manifolds which may be realized by polynomials in weighted
$\IP_4's$. The result of this investigation \cite{cls} are some
6,500 examples
\fnote{1}{It was shown in \cite{ls4} and will be discussed in later
sections
 that not all these spaces are distinct.}.
This class of vacua is of considerable interest because it
interpolates between the previously studied class the CICYs \cite{cdls}
mentioned above,
which have negative Euler numbers and the orbifolds of tori which have
positive Euler number.

More recently the construction of all Calabi--Yau manifolds embedded in
weighted $\IP_4$ and, more generally, of all Landau--Ginzburg vacua with
an arbitrary number of fields was completed in \cite{ks}\cite{krsk}. The
total
class consists of some 10,000 Landau--Ginzburg configurations, the
resulting spectra of which have been plotted in Fig.3.

The most remarkable feature of this class of manifolds is
immediately apparent from Fig~3.
It is evident that the manifolds are very
evenly divided between positive and negative Euler numbers the
distribution
exhibitting an approximate but compelling symmetry under
$\chi\to -\chi$. This
resonates with the observation made with regard to conformal field
theories
that the distinction between particles and antiparticles is purely one of
convention and the suggestion that for every \cym\ with Euler number
$\chi$ there should be one with Euler number $-\chi$.

\begin{center}

\plot{5.6truein}{\tiny{$\bullet$}}
\nobreak
\Place{-960}{50}{\leftscalemark~~50}
\Place{-960}{100}{\leftscalemark~~100}
\Place{-960}{150}{\leftscalemark~~150}
\Place{-960}{200}{\leftscalemark~~200}
\Place{-960}{250}{\leftscalemark~~250}
\Place{-960}{300}{\leftscalemark~~300}
\Place{-960}{350}{\leftscalemark~~350}
\Place{-960}{400}{\leftscalemark~~400}
\Place{-960}{450}{\leftscalemark~~450}
\Place{-960}{500}{\leftscalemark~~500}
\Place{960}{50}{\rightscalemark\vphantom{0}}
\Place{960}{100}{\rightscalemark\vphantom{0}}
\Place{960}{150}{\rightscalemark\vphantom{0}}
\Place{960}{200}{\rightscalemark\vphantom{0}}
\Place{960}{250}{\rightscalemark\vphantom{0}}
\Place{960}{300}{\rightscalemark\vphantom{0}}
\Place{960}{350}{\rightscalemark\vphantom{0}}
\Place{960}{400}{\rightscalemark\vphantom{0}}
\Place{960}{450}{\rightscalemark\vphantom{0}}
\Place{960}{500}{\rightscalemark\vphantom{0}}
\Place{-960}{0}{\hmark\lower18pt\hbox{-960}}
\Place{-720}{0}{\hmark\lower18pt\hbox{-720}}
\Place{-480}{0}{\hmark\lower18pt\hbox{-480}}
\Place{-240}{0}{\hmark\lower18pt\hbox{-240}}
\Place{0}{0}{\hmark\lower18pt\hbox{0}}
\Place{240}{0}{\hmark\lower18pt\hbox{240}}
\Place{480}{0}{\hmark\lower18pt\hbox{480}}
\Place{720}{0}{\hmark\lower18pt\hbox{720}}
\Place{960}{0}{\hmark\lower18pt\hbox{960}}
\Place{-720}{550}{\hmark}
\Place{-480}{550}{\hmark}
\Place{-240}{550}{\hmark}
\Place{0}{550}{\hmark}
\Place{240}{550}{\hmark}
\Place{480}{550}{\hmark}
\Place{720}{550}{\hmark}
\Place{960}{550}{\hmark}
\nobreak
\begin{center}
\datum{ 0}{ 20}
\datum{ 0}{ 22}
\datum{ 0}{ 26}
\datum{ 0}{ 28}
\datum{ 0}{ 30}
\datum{ 0}{ 32}
\datum{ 0}{ 34}
\datum{ 0}{ 36}
\datum{ 0}{ 38}
\datum{ 0}{ 40}
\datum{ 0}{ 42}
\datum{ 0}{ 44}
\datum{ 0}{ 46}
\datum{ 0}{ 48}
\datum{ 0}{ 50}
\datum{ 0}{ 52}
\datum{ 0}{ 54}
\datum{ 0}{ 56}
\datum{ 0}{ 58}
\datum{ 0}{ 60}
\datum{ 0}{ 62}
\datum{ 0}{ 64}
\datum{ 0}{ 66}
\datum{ 0}{ 68}
\datum{ 0}{ 70}
\datum{ 0}{ 74}
\datum{ 0}{ 76}
\datum{ 0}{ 78}
\datum{ 0}{ 82}
\datum{ 0}{ 86}
\datum{ 0}{ 88}
\datum{ 0}{ 90}
\datum{ 0}{ 94}
\datum{ 0}{ 98}
\datum{ 0}{ 104}
\datum{ 0}{ 106}
\datum{ 0}{ 110}
\datum{ 0}{ 112}
\datum{ 0}{ 114}
\datum{ 0}{ 118}
\datum{ 0}{ 122}
\datum{ 0}{ 124}
\datum{ 0}{ 126}
\datum{ 0}{ 130}
\datum{ 0}{ 134}
\datum{ 0}{ 138}
\datum{ 0}{ 142}
\datum{ 0}{ 150}
\datum{ 0}{ 154}
\datum{ 0}{ 156}
\datum{ 0}{ 158}
\datum{ 0}{ 162}
\datum{ 0}{ 166}
\datum{ 0}{ 170}
\datum{ 0}{ 174}
\datum{ 0}{ 178}
\datum{ 0}{ 182}
\datum{ 0}{ 190}
\datum{ 0}{ 194}
\datum{ 0}{ 206}
\datum{ 0}{ 214}
\datum{ 0}{ 222}
\datum{ 0}{ 238}
\datum{ 0}{ 242}
\datum{ 0}{ 246}
\datum{ 0}{ 262}
\datum{ 0}{ 286}
\datum{ 0}{ 298}
\datum{ 0}{ 302}
\datum{ 0}{ 358}
\datum{ 0}{ 446}
\datum{ 0}{ 502}
\datum{ 2}{ 25}
\datum{ 4}{ 44}
\datum{ 4}{ 50}
\datum{ 4}{ 54}
\datum{ 4}{ 66}
\datum{ 4}{ 68}
\datum{ 4}{ 82}
\datum{ 4}{ 92}
\datum{ 4}{ 102}
\datum{ 6}{ 29}
\datum{ 6}{ 33}
\datum{ 6}{ 35}
\datum{ 6}{ 37}
\datum{ 6}{ 43}
\datum{ 6}{ 55}
\datum{ 6}{ 57}
\datum{ 6}{ 61}
\datum{ 6}{ 65}
\datum{ 6}{ 71}
\datum{ 6}{ 77}
\datum{ 6}{ 87}
\datum{ 6}{ 97}
\datum{ 6}{ 137}
\datum{ 6}{ 151}
\datum{ 8}{ 34}
\datum{ 8}{ 36}
\datum{ 8}{ 40}
\datum{ 8}{ 42}
\datum{ 8}{ 44}
\datum{ 8}{ 46}
\datum{ 8}{ 50}
\datum{ 8}{ 52}
\datum{ 8}{ 54}
\datum{ 8}{ 56}
\datum{ 8}{ 60}
\datum{ 8}{ 62}
\datum{ 8}{ 66}
\datum{ 8}{ 68}
\datum{ 8}{ 70}
\datum{ 8}{ 78}
\datum{ 8}{ 82}
\datum{ 8}{ 84}
\datum{ 8}{ 90}
\datum{ 8}{ 102}
\datum{ 8}{ 116}
\datum{ -4}{ 38}
\datum{ -4}{ 54}
\datum{ -4}{ 64}
\datum{ -4}{ 66}
\datum{ -4}{ 82}
\datum{ -4}{ 92}
\datum{ -4}{ 96}
\datum{ -4}{ 110}
\datum{ -4}{ 156}
\datum{ -6}{ 29}
\datum{ -6}{ 37}
\datum{ -6}{ 43}
\datum{ -6}{ 45}
\datum{ -6}{ 49}
\datum{ -6}{ 55}
\datum{ -6}{ 61}
\datum{ -6}{ 67}
\datum{ -6}{ 73}
\datum{ -6}{ 77}
\datum{ -6}{ 83}
\datum{ -6}{ 97}
\datum{ -6}{ 99}
\datum{ -6}{ 117}
\datum{ -8}{ 30}
\datum{ -8}{ 34}
\datum{ -8}{ 36}
\datum{ -8}{ 38}
\datum{ -8}{ 44}
\datum{ -8}{ 50}
\datum{ -8}{ 52}
\datum{ -8}{ 54}
\datum{ -8}{ 56}
\datum{ -8}{ 58}
\datum{ -8}{ 60}
\datum{ -8}{ 62}
\datum{ -8}{ 66}
\datum{ -8}{ 70}
\datum{ -8}{ 84}
\datum{ -8}{ 90}
\datum{ -8}{ 96}
\datum{ -8}{ 98}
\datum{ -8}{ 102}
\datum{ -8}{ 116}
\datum{ 10}{ 53}
\datum{ 10}{ 83}
\datum{ 10}{ 141}
\datum{ 12}{ 22}
\datum{ 12}{ 26}
\datum{ 12}{ 30}
\datum{ 12}{ 32}
\datum{ 12}{ 34}
\datum{ 12}{ 36}
\datum{ 12}{ 38}
\datum{ 12}{ 42}
\datum{ 12}{ 44}
\datum{ 12}{ 48}
\datum{ 12}{ 50}
\datum{ 12}{ 52}
\datum{ 12}{ 54}
\datum{ 12}{ 56}
\datum{ 12}{ 58}
\datum{ 12}{ 64}
\datum{ 12}{ 66}
\datum{ 12}{ 70}
\datum{ 12}{ 72}
\datum{ 12}{ 76}
\datum{ 12}{ 80}
\datum{ 12}{ 82}
\datum{ 12}{ 86}
\datum{ 12}{ 88}
\datum{ 12}{ 94}
\datum{ 12}{ 96}
\datum{ 12}{ 98}
\datum{ 12}{ 100}
\datum{ 12}{ 108}
\datum{ 12}{ 116}
\datum{ 12}{ 122}
\datum{ 12}{ 128}
\datum{ 12}{ 130}
\datum{ 12}{ 146}
\datum{ 12}{ 156}
\datum{ 14}{ 45}
\datum{ 14}{ 61}
\datum{ 16}{ 26}
\datum{ 16}{ 34}
\datum{ 16}{ 36}
\datum{ 16}{ 38}
\datum{ 16}{ 42}
\datum{ 16}{ 44}
\datum{ 16}{ 46}
\datum{ 16}{ 48}
\datum{ 16}{ 50}
\datum{ 16}{ 52}
\datum{ 16}{ 54}
\datum{ 16}{ 56}
\datum{ 16}{ 58}
\datum{ 16}{ 62}
\datum{ 16}{ 64}
\datum{ 16}{ 66}
\datum{ 16}{ 74}
\datum{ 16}{ 76}
\datum{ 16}{ 78}
\datum{ 16}{ 82}
\datum{ 16}{ 84}
\datum{ 16}{ 86}
\datum{ 16}{ 92}
\datum{ 16}{ 94}
\datum{ 16}{ 122}
\datum{ 18}{ 45}
\datum{ 18}{ 47}
\datum{ 18}{ 49}
\datum{ 18}{ 55}
\datum{ 18}{ 61}
\datum{ 18}{ 63}
\datum{ 18}{ 67}
\datum{ 18}{ 69}
\datum{ 18}{ 79}
\datum{ 18}{ 81}
\datum{ 18}{ 85}
\datum{ 18}{ 89}
\datum{ 18}{ 91}
\datum{ 18}{ 97}
\datum{ 18}{ 103}
\datum{ 18}{ 105}
\datum{ 20}{ 46}
\datum{ 20}{ 48}
\datum{ 20}{ 58}
\datum{ 20}{ 84}
\datum{ 20}{ 90}
\datum{ 20}{ 108}
\datum{ 20}{ 114}
\datum{ 20}{ 198}
\datum{ 20}{ 234}
\datum{ 24}{ 26}
\datum{ 24}{ 28}
\datum{ 24}{ 30}
\datum{ 24}{ 32}
\datum{ 24}{ 34}
\datum{ 24}{ 36}
\datum{ 24}{ 38}
\datum{ 24}{ 40}
\datum{ 24}{ 42}
\datum{ 24}{ 44}
\datum{ 24}{ 46}
\datum{ 24}{ 48}
\datum{ 24}{ 50}
\datum{ 24}{ 52}
\datum{ 24}{ 54}
\datum{ 24}{ 56}
\datum{ 24}{ 58}
\datum{ 24}{ 60}
\datum{ 24}{ 62}
\datum{ 24}{ 64}
\datum{ 24}{ 66}
\datum{ 24}{ 68}
\datum{ 24}{ 70}
\datum{ 24}{ 72}
\datum{ 24}{ 74}
\datum{ 24}{ 76}
\datum{ 24}{ 80}
\datum{ 24}{ 82}
\datum{ 24}{ 84}
\datum{ 24}{ 86}
\datum{ 24}{ 88}
\datum{ 24}{ 90}
\datum{ 24}{ 92}
\datum{ 24}{ 94}
\datum{ 24}{ 98}
\datum{ 24}{ 100}
\datum{ 24}{ 102}
\datum{ 24}{ 104}
\datum{ 24}{ 112}
\datum{ 24}{ 114}
\datum{ 24}{ 116}
\datum{ 24}{ 120}
\datum{ 24}{ 128}
\datum{ 24}{ 130}
\datum{ 24}{ 132}
\datum{ 24}{ 134}
\datum{ 24}{ 148}
\datum{ 24}{ 160}
\datum{ 24}{ 164}
\datum{ 24}{ 166}
\datum{ 24}{ 232}
\datum{ 28}{ 38}
\datum{ 28}{ 54}
\datum{ 28}{ 58}
\datum{ 28}{ 62}
\datum{ 28}{ 98}
\datum{ 28}{ 146}
\datum{ 30}{ 35}
\datum{ 30}{ 43}
\datum{ 30}{ 49}
\datum{ 30}{ 51}
\datum{ 30}{ 53}
\datum{ 30}{ 63}
\datum{ 30}{ 73}
\datum{ 30}{ 77}
\datum{ 30}{ 79}
\datum{ 30}{ 83}
\datum{ 30}{ 103}
\datum{ 30}{ 163}
\datum{ 32}{ 30}
\datum{ 32}{ 38}
\datum{ 32}{ 42}
\datum{ 32}{ 46}
\datum{ 32}{ 50}
\datum{ 32}{ 52}
\datum{ 32}{ 54}
\datum{ 32}{ 60}
\datum{ 32}{ 62}
\datum{ 32}{ 66}
\datum{ 32}{ 70}
\datum{ 32}{ 74}
\datum{ 32}{ 76}
\datum{ 32}{ 78}
\datum{ 32}{ 82}
\datum{ 32}{ 84}
\datum{ 32}{ 94}
\datum{ 32}{ 104}
\datum{ 32}{ 110}
\datum{ 32}{ 114}
\datum{ 32}{ 118}
\datum{ 32}{ 190}
\datum{ 34}{ 45}
\datum{ 34}{ 69}
\datum{ 34}{ 83}
\datum{ 36}{ 30}
\datum{ 36}{ 34}
\datum{ 36}{ 36}
\datum{ 36}{ 38}
\datum{ 36}{ 40}
\datum{ 36}{ 42}
\datum{ 36}{ 46}
\datum{ 36}{ 50}
\datum{ 36}{ 52}
\datum{ 36}{ 54}
\datum{ 36}{ 56}
\datum{ 36}{ 58}
\datum{ 36}{ 60}
\datum{ 36}{ 62}
\datum{ 36}{ 66}
\datum{ 36}{ 68}
\datum{ 36}{ 70}
\datum{ 36}{ 74}
\datum{ 36}{ 76}
\datum{ 36}{ 78}
\datum{ 36}{ 82}
\datum{ 36}{ 84}
\datum{ 36}{ 86}
\datum{ 36}{ 88}
\datum{ 36}{ 94}
\datum{ 36}{ 100}
\datum{ 36}{ 102}
\datum{ 36}{ 106}
\datum{ 36}{ 108}
\datum{ 36}{ 110}
\datum{ 36}{ 112}
\datum{ 36}{ 114}
\datum{ 36}{ 118}
\datum{ 36}{ 126}
\datum{ 36}{ 130}
\datum{ 36}{ 134}
\datum{ 36}{ 142}
\datum{ 36}{ 144}
\datum{ 36}{ 156}
\datum{ 36}{ 162}
\datum{ 36}{ 172}
\datum{ 36}{ 184}
\datum{ 36}{ 202}
\datum{ 36}{ 212}
\datum{ 36}{ 214}
\datum{ 36}{ 222}
\datum{ 36}{ 314}
\datum{ 40}{ 30}
\datum{ 40}{ 38}
\datum{ 40}{ 44}
\datum{ 40}{ 46}
\datum{ 40}{ 48}
\datum{ 40}{ 50}
\datum{ 40}{ 52}
\datum{ 40}{ 54}
\datum{ 40}{ 56}
\datum{ 40}{ 58}
\datum{ 40}{ 62}
\datum{ 40}{ 68}
\datum{ 40}{ 70}
\datum{ 40}{ 72}
\datum{ 40}{ 78}
\datum{ 40}{ 80}
\datum{ 40}{ 86}
\datum{ 40}{ 90}
\datum{ 40}{ 94}
\datum{ 40}{ 110}
\datum{ 40}{ 116}
\datum{ 40}{ 118}
\datum{ 40}{ 120}
\datum{ 40}{ 130}
\datum{ 40}{ 148}
\datum{ 42}{ 43}
\datum{ 42}{ 49}
\datum{ 42}{ 53}
\datum{ 42}{ 55}
\datum{ 42}{ 61}
\datum{ 42}{ 79}
\datum{ 42}{ 81}
\datum{ 42}{ 89}
\datum{ 42}{ 91}
\datum{ 42}{ 93}
\datum{ 42}{ 115}
\datum{ 44}{ 44}
\datum{ 44}{ 52}
\datum{ 44}{ 56}
\datum{ 44}{ 58}
\datum{ 44}{ 64}
\datum{ 44}{ 66}
\datum{ 44}{ 78}
\datum{ 44}{ 80}
\datum{ 44}{ 94}
\datum{ 44}{ 108}
\datum{ 48}{ 30}
\datum{ 48}{ 34}
\datum{ 48}{ 38}
\datum{ 48}{ 40}
\datum{ 48}{ 42}
\datum{ 48}{ 44}
\datum{ 48}{ 46}
\datum{ 48}{ 48}
\datum{ 48}{ 50}
\datum{ 48}{ 52}
\datum{ 48}{ 54}
\datum{ 48}{ 56}
\datum{ 48}{ 58}
\datum{ 48}{ 60}
\datum{ 48}{ 62}
\datum{ 48}{ 64}
\datum{ 48}{ 66}
\datum{ 48}{ 68}
\datum{ 48}{ 70}
\datum{ 48}{ 72}
\datum{ 48}{ 74}
\datum{ 48}{ 76}
\datum{ 48}{ 78}
\datum{ 48}{ 80}
\datum{ 48}{ 82}
\datum{ 48}{ 84}
\datum{ 48}{ 86}
\datum{ 48}{ 88}
\datum{ 48}{ 92}
\datum{ 48}{ 94}
\datum{ 48}{ 96}
\datum{ 48}{ 98}
\datum{ 48}{ 100}
\datum{ 48}{ 102}
\datum{ 48}{ 106}
\datum{ 48}{ 108}
\datum{ 48}{ 110}
\datum{ 48}{ 112}
\datum{ 48}{ 118}
\datum{ 48}{ 122}
\datum{ 48}{ 124}
\datum{ 48}{ 134}
\datum{ 48}{ 138}
\datum{ 48}{ 142}
\datum{ 48}{ 158}
\datum{ 48}{ 166}
\datum{ 48}{ 178}
\datum{ 48}{ 202}
\datum{ 48}{ 230}
\datum{ 48}{ 266}
\datum{ 50}{ 45}
\datum{ 50}{ 63}
\datum{ 50}{ 93}
\datum{ 52}{ 64}
\datum{ 52}{ 108}
\datum{ 52}{ 116}
\datum{ 54}{ 41}
\datum{ 54}{ 43}
\datum{ 54}{ 47}
\datum{ 54}{ 49}
\datum{ 54}{ 57}
\datum{ 54}{ 59}
\datum{ 54}{ 61}
\datum{ 54}{ 73}
\datum{ 54}{ 77}
\datum{ 54}{ 79}
\datum{ 54}{ 83}
\datum{ 54}{ 85}
\datum{ 54}{ 103}
\datum{ 54}{ 117}
\datum{ 54}{ 133}
\datum{ 54}{ 157}
\datum{ 56}{ 38}
\datum{ 56}{ 52}
\datum{ 56}{ 54}
\datum{ 56}{ 58}
\datum{ 56}{ 62}
\datum{ 56}{ 66}
\datum{ 56}{ 68}
\datum{ 56}{ 70}
\datum{ 56}{ 72}
\datum{ 56}{ 74}
\datum{ 56}{ 84}
\datum{ 56}{ 86}
\datum{ 56}{ 88}
\datum{ 56}{ 90}
\datum{ 56}{ 108}
\datum{ 56}{ 114}
\datum{ 56}{ 124}
\datum{ 56}{ 126}
\datum{ 56}{ 150}
\datum{ 56}{ 166}
\datum{ 56}{ 180}
\datum{ 56}{ 214}
\datum{ 60}{ 36}
\datum{ 60}{ 40}
\datum{ 60}{ 42}
\datum{ 60}{ 44}
\datum{ 60}{ 46}
\datum{ 60}{ 48}
\datum{ 60}{ 52}
\datum{ 60}{ 54}
\datum{ 60}{ 58}
\datum{ 60}{ 60}
\datum{ 60}{ 62}
\datum{ 60}{ 64}
\datum{ 60}{ 66}
\datum{ 60}{ 68}
\datum{ 60}{ 70}
\datum{ 60}{ 74}
\datum{ 60}{ 78}
\datum{ 60}{ 82}
\datum{ 60}{ 84}
\datum{ 60}{ 88}
\datum{ 60}{ 96}
\datum{ 60}{ 98}
\datum{ 60}{ 110}
\datum{ 60}{ 112}
\datum{ 60}{ 120}
\datum{ 60}{ 122}
\datum{ 60}{ 132}
\datum{ 60}{ 144}
\datum{ 60}{ 166}
\datum{ 60}{ 178}
\datum{ 60}{ 196}
\datum{ 60}{ 218}
\datum{ 60}{ 222}
\datum{ 60}{ 234}
\datum{ 60}{ 358}
\datum{ 60}{ 474}
\datum{ 64}{ 50}
\datum{ 64}{ 54}
\datum{ 64}{ 60}
\datum{ 64}{ 62}
\datum{ 64}{ 66}
\datum{ 64}{ 70}
\datum{ 64}{ 76}
\datum{ 64}{ 80}
\datum{ 64}{ 82}
\datum{ 64}{ 86}
\datum{ 64}{ 90}
\datum{ 64}{ 102}
\datum{ 64}{ 106}
\datum{ 64}{ 110}
\datum{ 64}{ 134}
\datum{ 64}{ 154}
\datum{ 64}{ 166}
\datum{ 66}{ 61}
\datum{ 66}{ 63}
\datum{ 66}{ 65}
\datum{ 66}{ 69}
\datum{ 66}{ 73}
\datum{ 66}{ 75}
\datum{ 66}{ 79}
\datum{ 66}{ 87}
\datum{ 66}{ 97}
\datum{ 66}{ 129}
\datum{ 66}{ 137}
\datum{ 66}{ 159}
\datum{ 68}{ 64}
\datum{ 68}{ 78}
\datum{ 68}{ 110}
\datum{ 68}{ 150}
\datum{ 70}{ 75}
\datum{ 70}{ 89}
\datum{ 70}{ 95}
\datum{ 70}{ 101}
\datum{ 70}{ 135}
\datum{ 72}{ 40}
\datum{ 72}{ 44}
\datum{ 72}{ 46}
\datum{ 72}{ 48}
\datum{ 72}{ 50}
\datum{ 72}{ 52}
\datum{ 72}{ 54}
\datum{ 72}{ 56}
\datum{ 72}{ 58}
\datum{ 72}{ 60}
\datum{ 72}{ 62}
\datum{ 72}{ 64}
\datum{ 72}{ 66}
\datum{ 72}{ 68}
\datum{ 72}{ 70}
\datum{ 72}{ 74}
\datum{ 72}{ 76}
\datum{ 72}{ 78}
\datum{ 72}{ 80}
\datum{ 72}{ 82}
\datum{ 72}{ 84}
\datum{ 72}{ 86}
\datum{ 72}{ 88}
\datum{ 72}{ 90}
\datum{ 72}{ 92}
\datum{ 72}{ 94}
\datum{ 72}{ 96}
\datum{ 72}{ 98}
\datum{ 72}{ 100}
\datum{ 72}{ 102}
\datum{ 72}{ 104}
\datum{ 72}{ 112}
\datum{ 72}{ 116}
\datum{ 72}{ 118}
\datum{ 72}{ 120}
\datum{ 72}{ 124}
\datum{ 72}{ 126}
\datum{ 72}{ 128}
\datum{ 72}{ 132}
\datum{ 72}{ 136}
\datum{ 72}{ 140}
\datum{ 72}{ 142}
\datum{ 72}{ 144}
\datum{ 72}{ 148}
\datum{ 72}{ 150}
\datum{ 72}{ 158}
\datum{ 72}{ 160}
\datum{ 72}{ 164}
\datum{ 72}{ 182}
\datum{ 72}{ 198}
\datum{ 72}{ 212}
\datum{ 74}{ 121}
\datum{ 76}{ 68}
\datum{ 76}{ 94}
\datum{ 76}{ 100}
\datum{ 76}{ 124}
\datum{ 78}{ 61}
\datum{ 78}{ 79}
\datum{ 78}{ 83}
\datum{ 78}{ 89}
\datum{ 78}{ 91}
\datum{ 78}{ 93}
\datum{ 78}{ 95}
\datum{ 78}{ 109}
\datum{ 78}{ 111}
\datum{ 78}{ 153}
\datum{ 78}{ 163}
\datum{ 80}{ 54}
\datum{ 80}{ 58}
\datum{ 80}{ 60}
\datum{ 80}{ 62}
\datum{ 80}{ 66}
\datum{ 80}{ 68}
\datum{ 80}{ 72}
\datum{ 80}{ 74}
\datum{ 80}{ 76}
\datum{ 80}{ 78}
\datum{ 80}{ 94}
\datum{ 80}{ 96}
\datum{ 80}{ 98}
\datum{ 80}{ 100}
\datum{ 80}{ 102}
\datum{ 80}{ 106}
\datum{ 80}{ 118}
\datum{ 84}{ 50}
\datum{ 84}{ 52}
\datum{ 84}{ 54}
\datum{ 84}{ 58}
\datum{ 84}{ 64}
\datum{ 84}{ 66}
\datum{ 84}{ 68}
\datum{ 84}{ 70}
\datum{ 84}{ 72}
\datum{ 84}{ 74}
\datum{ 84}{ 78}
\datum{ 84}{ 80}
\datum{ 84}{ 82}
\datum{ 84}{ 84}
\datum{ 84}{ 86}
\datum{ 84}{ 88}
\datum{ 84}{ 90}
\datum{ 84}{ 94}
\datum{ 84}{ 96}
\datum{ 84}{ 98}
\datum{ 84}{ 102}
\datum{ 84}{ 106}
\datum{ 84}{ 124}
\datum{ 84}{ 130}
\datum{ 84}{ 134}
\datum{ 84}{ 138}
\datum{ 84}{ 154}
\datum{ 84}{ 162}
\datum{ 84}{ 164}
\datum{ 84}{ 166}
\datum{ 84}{ 172}
\datum{ 84}{ 174}
\datum{ 84}{ 178}
\datum{ 84}{ 190}
\datum{ 84}{ 194}
\datum{ 84}{ 262}
\datum{ 84}{ 322}
\datum{ 86}{ 57}
\datum{ 86}{ 87}
\datum{ 86}{ 127}
\datum{ 88}{ 54}
\datum{ 88}{ 60}
\datum{ 88}{ 62}
\datum{ 88}{ 68}
\datum{ 88}{ 70}
\datum{ 88}{ 82}
\datum{ 88}{ 84}
\datum{ 88}{ 102}
\datum{ 88}{ 104}
\datum{ 88}{ 128}
\datum{ 88}{ 158}
\datum{ 90}{ 53}
\datum{ 90}{ 63}
\datum{ 90}{ 65}
\datum{ 90}{ 67}
\datum{ 90}{ 71}
\datum{ 90}{ 73}
\datum{ 90}{ 79}
\datum{ 90}{ 83}
\datum{ 90}{ 93}
\datum{ 90}{ 95}
\datum{ 90}{ 97}
\datum{ 90}{ 105}
\datum{ 90}{ 131}
\datum{ 90}{ 133}
\datum{ 90}{ 171}
\datum{ 92}{ 94}
\datum{ 92}{ 150}
\datum{ 96}{ 50}
\datum{ 96}{ 54}
\datum{ 96}{ 58}
\datum{ 96}{ 60}
\datum{ 96}{ 62}
\datum{ 96}{ 64}
\datum{ 96}{ 66}
\datum{ 96}{ 70}
\datum{ 96}{ 74}
\datum{ 96}{ 76}
\datum{ 96}{ 78}
\datum{ 96}{ 80}
\datum{ 96}{ 82}
\datum{ 96}{ 84}
\datum{ 96}{ 86}
\datum{ 96}{ 88}
\datum{ 96}{ 90}
\datum{ 96}{ 92}
\datum{ 96}{ 94}
\datum{ 96}{ 98}
\datum{ 96}{ 102}
\datum{ 96}{ 106}
\datum{ 96}{ 108}
\datum{ 96}{ 110}
\datum{ 96}{ 112}
\datum{ 96}{ 114}
\datum{ 96}{ 118}
\datum{ 96}{ 120}
\datum{ 96}{ 122}
\datum{ 96}{ 126}
\datum{ 96}{ 134}
\datum{ 96}{ 138}
\datum{ 96}{ 142}
\datum{ 96}{ 146}
\datum{ 96}{ 154}
\datum{ 96}{ 158}
\datum{ 96}{ 166}
\datum{ 96}{ 186}
\datum{ 96}{ 190}
\datum{ 96}{ 204}
\datum{ 96}{ 212}
\datum{ 96}{ 250}
\datum{ 96}{ 286}
\datum{ 96}{ 342}
\datum{ 96}{ 402}
\datum{ -10}{ 43}
\datum{ -10}{ 53}
\datum{ -10}{ 63}
\datum{ -10}{ 73}
\datum{ -12}{ 28}
\datum{ -12}{ 32}
\datum{ -12}{ 34}
\datum{ -12}{ 36}
\datum{ -12}{ 38}
\datum{ -12}{ 40}
\datum{ -12}{ 42}
\datum{ -12}{ 44}
\datum{ -12}{ 46}
\datum{ -12}{ 48}
\datum{ -12}{ 50}
\datum{ -12}{ 52}
\datum{ -12}{ 54}
\datum{ -12}{ 56}
\datum{ -12}{ 58}
\datum{ -12}{ 64}
\datum{ -12}{ 66}
\datum{ -12}{ 70}
\datum{ -12}{ 72}
\datum{ -12}{ 76}
\datum{ -12}{ 78}
\datum{ -12}{ 80}
\datum{ -12}{ 82}
\datum{ -12}{ 86}
\datum{ -12}{ 90}
\datum{ -12}{ 94}
\datum{ -12}{ 100}
\datum{ -12}{ 104}
\datum{ -12}{ 108}
\datum{ -12}{ 116}
\datum{ -12}{ 128}
\datum{ -12}{ 146}
\datum{ -12}{ 156}
\datum{ -14}{ 47}
\datum{ -14}{ 137}
\datum{ -16}{ 28}
\datum{ -16}{ 30}
\datum{ -16}{ 36}
\datum{ -16}{ 38}
\datum{ -16}{ 42}
\datum{ -16}{ 44}
\datum{ -16}{ 46}
\datum{ -16}{ 48}
\datum{ -16}{ 52}
\datum{ -16}{ 54}
\datum{ -16}{ 56}
\datum{ -16}{ 58}
\datum{ -16}{ 60}
\datum{ -16}{ 62}
\datum{ -16}{ 64}
\datum{ -16}{ 66}
\datum{ -16}{ 70}
\datum{ -16}{ 72}
\datum{ -16}{ 74}
\datum{ -16}{ 76}
\datum{ -16}{ 78}
\datum{ -16}{ 82}
\datum{ -16}{ 94}
\datum{ -16}{ 150}
\datum{ -18}{ 25}
\datum{ -18}{ 29}
\datum{ -18}{ 37}
\datum{ -18}{ 39}
\datum{ -18}{ 47}
\datum{ -18}{ 49}
\datum{ -18}{ 55}
\datum{ -18}{ 61}
\datum{ -18}{ 63}
\datum{ -18}{ 65}
\datum{ -18}{ 79}
\datum{ -18}{ 83}
\datum{ -18}{ 85}
\datum{ -18}{ 97}
\datum{ -18}{ 109}
\datum{ -18}{ 115}
\datum{ -18}{ 149}
\datum{ -20}{ 26}
\datum{ -20}{ 30}
\datum{ -20}{ 40}
\datum{ -20}{ 44}
\datum{ -20}{ 48}
\datum{ -20}{ 56}
\datum{ -20}{ 58}
\datum{ -20}{ 64}
\datum{ -20}{ 68}
\datum{ -20}{ 74}
\datum{ -20}{ 108}
\datum{ -20}{ 114}
\datum{ -20}{ 116}
\datum{ -20}{ 198}
\datum{ -22}{ 41}
\datum{ -24}{ 26}
\datum{ -24}{ 28}
\datum{ -24}{ 30}
\datum{ -24}{ 32}
\datum{ -24}{ 34}
\datum{ -24}{ 36}
\datum{ -24}{ 38}
\datum{ -24}{ 40}
\datum{ -24}{ 42}
\datum{ -24}{ 44}
\datum{ -24}{ 46}
\datum{ -24}{ 48}
\datum{ -24}{ 50}
\datum{ -24}{ 52}
\datum{ -24}{ 54}
\datum{ -24}{ 56}
\datum{ -24}{ 58}
\datum{ -24}{ 60}
\datum{ -24}{ 62}
\datum{ -24}{ 64}
\datum{ -24}{ 66}
\datum{ -24}{ 68}
\datum{ -24}{ 70}
\datum{ -24}{ 72}
\datum{ -24}{ 74}
\datum{ -24}{ 76}
\datum{ -24}{ 78}
\datum{ -24}{ 80}
\datum{ -24}{ 82}
\datum{ -24}{ 84}
\datum{ -24}{ 86}
\datum{ -24}{ 88}
\datum{ -24}{ 90}
\datum{ -24}{ 92}
\datum{ -24}{ 94}
\datum{ -24}{ 96}
\datum{ -24}{ 98}
\datum{ -24}{ 100}
\datum{ -24}{ 102}
\datum{ -24}{ 104}
\datum{ -24}{ 112}
\datum{ -24}{ 114}
\datum{ -24}{ 116}
\datum{ -24}{ 120}
\datum{ -24}{ 128}
\datum{ -24}{ 130}
\datum{ -24}{ 132}
\datum{ -24}{ 134}
\datum{ -24}{ 142}
\datum{ -24}{ 144}
\datum{ -24}{ 148}
\datum{ -24}{ 160}
\datum{ -24}{ 162}
\datum{ -24}{ 164}
\datum{ -24}{ 166}
\datum{ -24}{ 184}
\datum{ -24}{ 232}
\datum{ -26}{ 63}
\datum{ -26}{ 89}
\datum{ -28}{ 36}
\datum{ -28}{ 48}
\datum{ -28}{ 54}
\datum{ -28}{ 62}
\datum{ -28}{ 72}
\datum{ -28}{ 186}
\datum{ -30}{ 33}
\datum{ -30}{ 43}
\datum{ -30}{ 49}
\datum{ -30}{ 53}
\datum{ -30}{ 61}
\datum{ -30}{ 63}
\datum{ -30}{ 69}
\datum{ -30}{ 71}
\datum{ -30}{ 73}
\datum{ -30}{ 79}
\datum{ -30}{ 83}
\datum{ -30}{ 91}
\datum{ -30}{ 93}
\datum{ -30}{ 133}
\datum{ -30}{ 163}
\datum{ -32}{ 30}
\datum{ -32}{ 32}
\datum{ -32}{ 36}
\datum{ -32}{ 38}
\datum{ -32}{ 42}
\datum{ -32}{ 44}
\datum{ -32}{ 46}
\datum{ -32}{ 48}
\datum{ -32}{ 50}
\datum{ -32}{ 54}
\datum{ -32}{ 56}
\datum{ -32}{ 62}
\datum{ -32}{ 66}
\datum{ -32}{ 70}
\datum{ -32}{ 74}
\datum{ -32}{ 76}
\datum{ -32}{ 78}
\datum{ -32}{ 86}
\datum{ -32}{ 104}
\datum{ -32}{ 118}
\datum{ -32}{ 190}
\datum{ -36}{ 28}
\datum{ -36}{ 34}
\datum{ -36}{ 36}
\datum{ -36}{ 38}
\datum{ -36}{ 40}
\datum{ -36}{ 46}
\datum{ -36}{ 48}
\datum{ -36}{ 50}
\datum{ -36}{ 52}
\datum{ -36}{ 54}
\datum{ -36}{ 58}
\datum{ -36}{ 62}
\datum{ -36}{ 66}
\datum{ -36}{ 70}
\datum{ -36}{ 72}
\datum{ -36}{ 74}
\datum{ -36}{ 76}
\datum{ -36}{ 78}
\datum{ -36}{ 82}
\datum{ -36}{ 84}
\datum{ -36}{ 86}
\datum{ -36}{ 88}
\datum{ -36}{ 94}
\datum{ -36}{ 100}
\datum{ -36}{ 102}
\datum{ -36}{ 104}
\datum{ -36}{ 106}
\datum{ -36}{ 108}
\datum{ -36}{ 112}
\datum{ -36}{ 118}
\datum{ -36}{ 120}
\datum{ -36}{ 126}
\datum{ -36}{ 134}
\datum{ -36}{ 142}
\datum{ -36}{ 144}
\datum{ -36}{ 156}
\datum{ -36}{ 162}
\datum{ -36}{ 172}
\datum{ -36}{ 184}
\datum{ -36}{ 202}
\datum{ -36}{ 212}
\datum{ -36}{ 214}
\datum{ -36}{ 222}
\datum{ -36}{ 314}
\datum{ -38}{ 45}
\datum{ -40}{ 34}
\datum{ -40}{ 36}
\datum{ -40}{ 38}
\datum{ -40}{ 44}
\datum{ -40}{ 46}
\datum{ -40}{ 48}
\datum{ -40}{ 50}
\datum{ -40}{ 52}
\datum{ -40}{ 54}
\datum{ -40}{ 58}
\datum{ -40}{ 62}
\datum{ -40}{ 68}
\datum{ -40}{ 70}
\datum{ -40}{ 74}
\datum{ -40}{ 78}
\datum{ -40}{ 86}
\datum{ -40}{ 90}
\datum{ -40}{ 92}
\datum{ -40}{ 110}
\datum{ -40}{ 116}
\datum{ -40}{ 118}
\datum{ -40}{ 138}
\datum{ -40}{ 148}
\datum{ -40}{ 160}
\datum{ -42}{ 35}
\datum{ -42}{ 43}
\datum{ -42}{ 49}
\datum{ -42}{ 55}
\datum{ -42}{ 61}
\datum{ -42}{ 65}
\datum{ -42}{ 67}
\datum{ -42}{ 73}
\datum{ -42}{ 79}
\datum{ -42}{ 83}
\datum{ -42}{ 85}
\datum{ -42}{ 91}
\datum{ -42}{ 97}
\datum{ -42}{ 137}
\datum{ -44}{ 44}
\datum{ -44}{ 78}
\datum{ -44}{ 80}
\datum{ -44}{ 92}
\datum{ -44}{ 134}
\datum{ -46}{ 153}
\datum{ -48}{ 34}
\datum{ -48}{ 38}
\datum{ -48}{ 42}
\datum{ -48}{ 44}
\datum{ -48}{ 46}
\datum{ -48}{ 48}
\datum{ -48}{ 50}
\datum{ -48}{ 52}
\datum{ -48}{ 54}
\datum{ -48}{ 56}
\datum{ -48}{ 58}
\datum{ -48}{ 60}
\datum{ -48}{ 62}
\datum{ -48}{ 64}
\datum{ -48}{ 66}
\datum{ -48}{ 68}
\datum{ -48}{ 70}
\datum{ -48}{ 72}
\datum{ -48}{ 74}
\datum{ -48}{ 76}
\datum{ -48}{ 78}
\datum{ -48}{ 80}
\datum{ -48}{ 82}
\datum{ -48}{ 86}
\datum{ -48}{ 88}
\datum{ -48}{ 90}
\datum{ -48}{ 92}
\datum{ -48}{ 94}
\datum{ -48}{ 96}
\datum{ -48}{ 98}
\datum{ -48}{ 102}
\datum{ -48}{ 106}
\datum{ -48}{ 108}
\datum{ -48}{ 110}
\datum{ -48}{ 112}
\datum{ -48}{ 118}
\datum{ -48}{ 122}
\datum{ -48}{ 124}
\datum{ -48}{ 134}
\datum{ -48}{ 138}
\datum{ -48}{ 142}
\datum{ -48}{ 158}
\datum{ -48}{ 166}
\datum{ -48}{ 168}
\datum{ -48}{ 178}
\datum{ -48}{ 202}
\datum{ -48}{ 230}
\datum{ -48}{ 266}
\datum{ -50}{ 53}
\datum{ -50}{ 73}
\datum{ -52}{ 64}
\datum{ -54}{ 37}
\datum{ -54}{ 43}
\datum{ -54}{ 47}
\datum{ -54}{ 49}
\datum{ -54}{ 61}
\datum{ -54}{ 63}
\datum{ -54}{ 71}
\datum{ -54}{ 77}
\datum{ -54}{ 79}
\datum{ -54}{ 85}
\datum{ -54}{ 87}
\datum{ -54}{ 103}
\datum{ -54}{ 133}
\datum{ -54}{ 147}
\datum{ -56}{ 38}
\datum{ -56}{ 46}
\datum{ -56}{ 50}
\datum{ -56}{ 54}
\datum{ -56}{ 62}
\datum{ -56}{ 66}
\datum{ -56}{ 68}
\datum{ -56}{ 72}
\datum{ -56}{ 74}
\datum{ -56}{ 82}
\datum{ -56}{ 86}
\datum{ -56}{ 88}
\datum{ -56}{ 96}
\datum{ -56}{ 114}
\datum{ -56}{ 116}
\datum{ -56}{ 126}
\datum{ -56}{ 134}
\datum{ -56}{ 150}
\datum{ -56}{ 166}
\datum{ -56}{ 180}
\datum{ -56}{ 214}
\datum{ -58}{ 43}
\datum{ -58}{ 49}
\datum{ -58}{ 129}
\datum{ -60}{ 42}
\datum{ -60}{ 48}
\datum{ -60}{ 52}
\datum{ -60}{ 54}
\datum{ -60}{ 56}
\datum{ -60}{ 58}
\datum{ -60}{ 60}
\datum{ -60}{ 62}
\datum{ -60}{ 64}
\datum{ -60}{ 66}
\datum{ -60}{ 68}
\datum{ -60}{ 70}
\datum{ -60}{ 74}
\datum{ -60}{ 78}
\datum{ -60}{ 82}
\datum{ -60}{ 84}
\datum{ -60}{ 88}
\datum{ -60}{ 90}
\datum{ -60}{ 94}
\datum{ -60}{ 96}
\datum{ -60}{ 98}
\datum{ -60}{ 106}
\datum{ -60}{ 110}
\datum{ -60}{ 112}
\datum{ -60}{ 122}
\datum{ -60}{ 124}
\datum{ -60}{ 126}
\datum{ -60}{ 132}
\datum{ -60}{ 144}
\datum{ -60}{ 166}
\datum{ -60}{ 178}
\datum{ -60}{ 196}
\datum{ -60}{ 218}
\datum{ -60}{ 222}
\datum{ -60}{ 234}
\datum{ -60}{ 358}
\datum{ -60}{ 474}
\datum{ -62}{ 89}
\datum{ -62}{ 105}
\datum{ -62}{ 147}
\datum{ -64}{ 42}
\datum{ -64}{ 46}
\datum{ -64}{ 48}
\datum{ -64}{ 52}
\datum{ -64}{ 54}
\datum{ -64}{ 60}
\datum{ -64}{ 62}
\datum{ -64}{ 66}
\datum{ -64}{ 72}
\datum{ -64}{ 76}
\datum{ -64}{ 78}
\datum{ -64}{ 80}
\datum{ -64}{ 82}
\datum{ -64}{ 86}
\datum{ -64}{ 90}
\datum{ -64}{ 102}
\datum{ -64}{ 106}
\datum{ -64}{ 110}
\datum{ -64}{ 118}
\datum{ -64}{ 124}
\datum{ -64}{ 134}
\datum{ -64}{ 166}
\datum{ -66}{ 65}
\datum{ -66}{ 75}
\datum{ -66}{ 79}
\datum{ -66}{ 87}
\datum{ -66}{ 89}
\datum{ -66}{ 97}
\datum{ -68}{ 42}
\datum{ -68}{ 54}
\datum{ -68}{ 58}
\datum{ -68}{ 86}
\datum{ -68}{ 126}
\datum{ -70}{ 51}
\datum{ -70}{ 53}
\datum{ -70}{ 73}
\datum{ -72}{ 40}
\datum{ -72}{ 44}
\datum{ -72}{ 46}
\datum{ -72}{ 48}
\datum{ -72}{ 50}
\datum{ -72}{ 52}
\datum{ -72}{ 54}
\datum{ -72}{ 56}
\datum{ -72}{ 58}
\datum{ -72}{ 60}
\datum{ -72}{ 62}
\datum{ -72}{ 64}
\datum{ -72}{ 66}
\datum{ -72}{ 68}
\datum{ -72}{ 70}
\datum{ -72}{ 74}
\datum{ -72}{ 76}
\datum{ -72}{ 78}
\datum{ -72}{ 80}
\datum{ -72}{ 82}
\datum{ -72}{ 84}
\datum{ -72}{ 86}
\datum{ -72}{ 88}
\datum{ -72}{ 92}
\datum{ -72}{ 94}
\datum{ -72}{ 96}
\datum{ -72}{ 98}
\datum{ -72}{ 100}
\datum{ -72}{ 102}
\datum{ -72}{ 104}
\datum{ -72}{ 112}
\datum{ -72}{ 116}
\datum{ -72}{ 118}
\datum{ -72}{ 120}
\datum{ -72}{ 124}
\datum{ -72}{ 126}
\datum{ -72}{ 128}
\datum{ -72}{ 132}
\datum{ -72}{ 136}
\datum{ -72}{ 140}
\datum{ -72}{ 142}
\datum{ -72}{ 144}
\datum{ -72}{ 148}
\datum{ -72}{ 150}
\datum{ -72}{ 158}
\datum{ -72}{ 160}
\datum{ -72}{ 164}
\datum{ -72}{ 174}
\datum{ -72}{ 182}
\datum{ -72}{ 198}
\datum{ -72}{ 212}
\datum{ -74}{ 177}
\datum{ -76}{ 74}
\datum{ -76}{ 94}
\datum{ -76}{ 112}
\datum{ -76}{ 124}
\datum{ -76}{ 142}
\datum{ -78}{ 61}
\datum{ -78}{ 91}
\datum{ -78}{ 109}
\datum{ -78}{ 175}
\datum{ -80}{ 46}
\datum{ -80}{ 50}
\datum{ -80}{ 54}
\datum{ -80}{ 58}
\datum{ -80}{ 60}
\datum{ -80}{ 62}
\datum{ -80}{ 66}
\datum{ -80}{ 68}
\datum{ -80}{ 70}
\datum{ -80}{ 72}
\datum{ -80}{ 74}
\datum{ -80}{ 78}
\datum{ -80}{ 94}
\datum{ -80}{ 96}
\datum{ -80}{ 98}
\datum{ -80}{ 102}
\datum{ -80}{ 106}
\datum{ -80}{ 118}
\datum{ -80}{ 122}
\datum{ -80}{ 138}
\datum{ -80}{ 146}
\datum{ -84}{ 50}
\datum{ -84}{ 52}
\datum{ -84}{ 54}
\datum{ -84}{ 56}
\datum{ -84}{ 58}
\datum{ -84}{ 60}
\datum{ -84}{ 64}
\datum{ -84}{ 66}
\datum{ -84}{ 68}
\datum{ -84}{ 70}
\datum{ -84}{ 72}
\datum{ -84}{ 78}
\datum{ -84}{ 82}
\datum{ -84}{ 84}
\datum{ -84}{ 86}
\datum{ -84}{ 88}
\datum{ -84}{ 90}
\datum{ -84}{ 94}
\datum{ -84}{ 96}
\datum{ -84}{ 98}
\datum{ -84}{ 100}
\datum{ -84}{ 102}
\datum{ -84}{ 114}
\datum{ -84}{ 124}
\datum{ -84}{ 130}
\datum{ -84}{ 132}
\datum{ -84}{ 134}
\datum{ -84}{ 138}
\datum{ -84}{ 154}
\datum{ -84}{ 164}
\datum{ -84}{ 166}
\datum{ -84}{ 172}
\datum{ -84}{ 174}
\datum{ -84}{ 178}
\datum{ -84}{ 190}
\datum{ -84}{ 194}
\datum{ -84}{ 262}
\datum{ -84}{ 322}
\datum{ -86}{ 57}
\datum{ -86}{ 77}
\datum{ -86}{ 137}
\datum{ -88}{ 52}
\datum{ -88}{ 54}
\datum{ -88}{ 60}
\datum{ -88}{ 62}
\datum{ -88}{ 66}
\datum{ -88}{ 68}
\datum{ -88}{ 70}
\datum{ -88}{ 74}
\datum{ -88}{ 78}
\datum{ -88}{ 82}
\datum{ -88}{ 84}
\datum{ -88}{ 102}
\datum{ -88}{ 108}
\datum{ -88}{ 132}
\datum{ -88}{ 158}
\datum{ -90}{ 61}
\datum{ -90}{ 63}
\datum{ -90}{ 71}
\datum{ -90}{ 73}
\datum{ -90}{ 79}
\datum{ -90}{ 93}
\datum{ -90}{ 99}
\datum{ -90}{ 109}
\datum{ -90}{ 133}
\datum{ -90}{ 141}
\datum{ -92}{ 56}
\datum{ -92}{ 58}
\datum{ -92}{ 64}
\datum{ -92}{ 86}
\datum{ -92}{ 94}
\datum{ -92}{ 98}
\datum{ -92}{ 150}
\datum{ -94}{ 87}
\datum{ -96}{ 54}
\datum{ -96}{ 56}
\datum{ -96}{ 58}
\datum{ -96}{ 60}
\datum{ -96}{ 62}
\datum{ -96}{ 64}
\datum{ -96}{ 66}
\datum{ -96}{ 70}
\datum{ -96}{ 74}
\datum{ -96}{ 76}
\datum{ -96}{ 78}
\datum{ -96}{ 80}
\datum{ -96}{ 82}
\datum{ -96}{ 84}
\datum{ -96}{ 86}
\datum{ -96}{ 88}
\datum{ -96}{ 90}
\datum{ -96}{ 92}
\datum{ -96}{ 94}
\datum{ -96}{ 96}
\datum{ -96}{ 98}
\datum{ -96}{ 102}
\datum{ -96}{ 106}
\datum{ -96}{ 108}
\datum{ -96}{ 110}
\datum{ -96}{ 112}
\datum{ -96}{ 114}
\datum{ -96}{ 118}
\datum{ -96}{ 120}
\datum{ -96}{ 122}
\datum{ -96}{ 126}
\datum{ -96}{ 134}
\datum{ -96}{ 138}
\datum{ -96}{ 142}
\datum{ -96}{ 146}
\datum{ -96}{ 154}
\datum{ -96}{ 158}
\datum{ -96}{ 166}
\datum{ -96}{ 186}
\datum{ -96}{ 190}
\datum{ -96}{ 204}
\datum{ -96}{ 212}
\datum{ -96}{ 250}
\datum{ -96}{ 286}
\datum{ -96}{ 342}
\datum{ -96}{ 402}
\datum{ -98}{ 87}
\datum{ 100}{ 62}
\datum{ 100}{ 70}
\datum{ 100}{ 80}
\datum{ 100}{ 88}
\datum{ 100}{ 96}
\datum{ 100}{ 98}
\datum{ 100}{ 118}
\datum{ 100}{ 168}
\datum{ 102}{ 61}
\datum{ 102}{ 67}
\datum{ 102}{ 93}
\datum{ 102}{ 97}
\datum{ 102}{ 99}
\datum{ 104}{ 58}
\datum{ 104}{ 62}
\datum{ 104}{ 68}
\datum{ 104}{ 74}
\datum{ 104}{ 76}
\datum{ 104}{ 78}
\datum{ 104}{ 80}
\datum{ 104}{ 82}
\datum{ 104}{ 84}
\datum{ 104}{ 86}
\datum{ 104}{ 90}
\datum{ 104}{ 100}
\datum{ 104}{ 102}
\datum{ 104}{ 106}
\datum{ 104}{ 108}
\datum{ 104}{ 132}
\datum{ 104}{ 154}
\datum{ 106}{ 69}
\datum{ 106}{ 129}
\datum{ 108}{ 58}
\datum{ 108}{ 66}
\datum{ 108}{ 74}
\datum{ 108}{ 76}
\datum{ 108}{ 78}
\datum{ 108}{ 80}
\datum{ 108}{ 84}
\datum{ 108}{ 86}
\datum{ 108}{ 88}
\datum{ 108}{ 90}
\datum{ 108}{ 94}
\datum{ 108}{ 96}
\datum{ 108}{ 100}
\datum{ 108}{ 106}
\datum{ 108}{ 108}
\datum{ 108}{ 112}
\datum{ 108}{ 118}
\datum{ 108}{ 120}
\datum{ 108}{ 122}
\datum{ 108}{ 130}
\datum{ 108}{ 150}
\datum{ 108}{ 166}
\datum{ 108}{ 178}
\datum{ 108}{ 190}
\datum{ 108}{ 212}
\datum{ 110}{ 73}
\datum{ 112}{ 62}
\datum{ 112}{ 66}
\datum{ 112}{ 70}
\datum{ 112}{ 76}
\datum{ 112}{ 78}
\datum{ 112}{ 80}
\datum{ 112}{ 86}
\datum{ 112}{ 88}
\datum{ 112}{ 90}
\datum{ 112}{ 92}
\datum{ 112}{ 94}
\datum{ 112}{ 96}
\datum{ 112}{ 108}
\datum{ 112}{ 110}
\datum{ 112}{ 118}
\datum{ 112}{ 124}
\datum{ 112}{ 126}
\datum{ 112}{ 138}
\datum{ 112}{ 150}
\datum{ 114}{ 79}
\datum{ 114}{ 85}
\datum{ 114}{ 91}
\datum{ 114}{ 97}
\datum{ 114}{ 103}
\datum{ 114}{ 113}
\datum{ 114}{ 115}
\datum{ 114}{ 127}
\datum{ 114}{ 147}
\datum{ 114}{ 169}
\datum{ 116}{ 72}
\datum{ 116}{ 108}
\datum{ 116}{ 124}
\datum{ 120}{ 62}
\datum{ 120}{ 64}
\datum{ 120}{ 66}
\datum{ 120}{ 68}
\datum{ 120}{ 70}
\datum{ 120}{ 72}
\datum{ 120}{ 74}
\datum{ 120}{ 76}
\datum{ 120}{ 78}
\datum{ 120}{ 80}
\datum{ 120}{ 82}
\datum{ 120}{ 86}
\datum{ 120}{ 88}
\datum{ 120}{ 90}
\datum{ 120}{ 92}
\datum{ 120}{ 94}
\datum{ 120}{ 96}
\datum{ 120}{ 98}
\datum{ 120}{ 100}
\datum{ 120}{ 102}
\datum{ 120}{ 104}
\datum{ 120}{ 106}
\datum{ 120}{ 108}
\datum{ 120}{ 110}
\datum{ 120}{ 112}
\datum{ 120}{ 114}
\datum{ 120}{ 116}
\datum{ 120}{ 118}
\datum{ 120}{ 122}
\datum{ 120}{ 124}
\datum{ 120}{ 128}
\datum{ 120}{ 130}
\datum{ 120}{ 132}
\datum{ 120}{ 134}
\datum{ 120}{ 136}
\datum{ 120}{ 138}
\datum{ 120}{ 140}
\datum{ 120}{ 142}
\datum{ 120}{ 146}
\datum{ 120}{ 148}
\datum{ 120}{ 150}
\datum{ 120}{ 152}
\datum{ 120}{ 154}
\datum{ 120}{ 156}
\datum{ 120}{ 158}
\datum{ 120}{ 168}
\datum{ 120}{ 178}
\datum{ 120}{ 190}
\datum{ 120}{ 196}
\datum{ 120}{ 212}
\datum{ 120}{ 248}
\datum{ 120}{ 276}
\datum{ 120}{ 278}
\datum{ 124}{ 94}
\datum{ 124}{ 106}
\datum{ 124}{ 162}
\datum{ 126}{ 69}
\datum{ 126}{ 77}
\datum{ 126}{ 79}
\datum{ 126}{ 85}
\datum{ 126}{ 87}
\datum{ 126}{ 89}
\datum{ 126}{ 101}
\datum{ 126}{ 109}
\datum{ 126}{ 115}
\datum{ 126}{ 117}
\datum{ 126}{ 133}
\datum{ 126}{ 143}
\datum{ 128}{ 74}
\datum{ 128}{ 78}
\datum{ 128}{ 82}
\datum{ 128}{ 86}
\datum{ 128}{ 90}
\datum{ 128}{ 92}
\datum{ 128}{ 94}
\datum{ 128}{ 102}
\datum{ 128}{ 108}
\datum{ 128}{ 110}
\datum{ 128}{ 118}
\datum{ 128}{ 124}
\datum{ 128}{ 126}
\datum{ 128}{ 142}
\datum{ 130}{ 109}
\datum{ 132}{ 72}
\datum{ 132}{ 76}
\datum{ 132}{ 78}
\datum{ 132}{ 84}
\datum{ 132}{ 86}
\datum{ 132}{ 90}
\datum{ 132}{ 92}
\datum{ 132}{ 94}
\datum{ 132}{ 96}
\datum{ 132}{ 100}
\datum{ 132}{ 102}
\datum{ 132}{ 104}
\datum{ 132}{ 114}
\datum{ 132}{ 128}
\datum{ 132}{ 150}
\datum{ 132}{ 172}
\datum{ 132}{ 178}
\datum{ 132}{ 238}
\datum{ 132}{ 302}
\datum{ 136}{ 78}
\datum{ 136}{ 92}
\datum{ 136}{ 102}
\datum{ 136}{ 104}
\datum{ 136}{ 112}
\datum{ 136}{ 126}
\datum{ 136}{ 136}
\datum{ 136}{ 138}
\datum{ 136}{ 182}
\datum{ 138}{ 79}
\datum{ 138}{ 85}
\datum{ 138}{ 97}
\datum{ 138}{ 103}
\datum{ 138}{ 151}
\datum{ 140}{ 84}
\datum{ 140}{ 88}
\datum{ 140}{ 94}
\datum{ 140}{ 96}
\datum{ 140}{ 98}
\datum{ 140}{ 114}
\datum{ 140}{ 130}
\datum{ 140}{ 138}
\datum{ 140}{ 148}
\datum{ 140}{ 158}
\datum{ 142}{ 129}
\datum{ 144}{ 74}
\datum{ 144}{ 76}
\datum{ 144}{ 78}
\datum{ 144}{ 80}
\datum{ 144}{ 82}
\datum{ 144}{ 84}
\datum{ 144}{ 86}
\datum{ 144}{ 88}
\datum{ 144}{ 90}
\datum{ 144}{ 92}
\datum{ 144}{ 94}
\datum{ 144}{ 96}
\datum{ 144}{ 98}
\datum{ 144}{ 100}
\datum{ 144}{ 102}
\datum{ 144}{ 104}
\datum{ 144}{ 106}
\datum{ 144}{ 110}
\datum{ 144}{ 112}
\datum{ 144}{ 114}
\datum{ 144}{ 118}
\datum{ 144}{ 122}
\datum{ 144}{ 124}
\datum{ 144}{ 126}
\datum{ 144}{ 130}
\datum{ 144}{ 134}
\datum{ 144}{ 138}
\datum{ 144}{ 140}
\datum{ 144}{ 148}
\datum{ 144}{ 150}
\datum{ 144}{ 152}
\datum{ 144}{ 154}
\datum{ 144}{ 158}
\datum{ 144}{ 164}
\datum{ 144}{ 172}
\datum{ 144}{ 174}
\datum{ 144}{ 182}
\datum{ 144}{ 188}
\datum{ 144}{ 190}
\datum{ 144}{ 206}
\datum{ 144}{ 214}
\datum{ 146}{ 193}
\datum{ 148}{ 92}
\datum{ 148}{ 106}
\datum{ 150}{ 93}
\datum{ 150}{ 97}
\datum{ 150}{ 101}
\datum{ 150}{ 123}
\datum{ 150}{ 153}
\datum{ 150}{ 181}
\datum{ 152}{ 78}
\datum{ 152}{ 84}
\datum{ 152}{ 86}
\datum{ 152}{ 96}
\datum{ 152}{ 102}
\datum{ 152}{ 108}
\datum{ 152}{ 110}
\datum{ 152}{ 134}
\datum{ 154}{ 89}
\datum{ 154}{ 97}
\datum{ 156}{ 86}
\datum{ 156}{ 88}
\datum{ 156}{ 92}
\datum{ 156}{ 94}
\datum{ 156}{ 102}
\datum{ 156}{ 106}
\datum{ 156}{ 108}
\datum{ 156}{ 110}
\datum{ 156}{ 112}
\datum{ 156}{ 114}
\datum{ 156}{ 120}
\datum{ 156}{ 122}
\datum{ 156}{ 124}
\datum{ 156}{ 126}
\datum{ 156}{ 132}
\datum{ 156}{ 148}
\datum{ 156}{ 154}
\datum{ 156}{ 178}
\datum{ 156}{ 210}
\datum{ 156}{ 232}
\datum{ 156}{ 234}
\datum{ 156}{ 430}
\datum{ 160}{ 86}
\datum{ 160}{ 90}
\datum{ 160}{ 94}
\datum{ 160}{ 98}
\datum{ 160}{ 102}
\datum{ 160}{ 108}
\datum{ 160}{ 110}
\datum{ 160}{ 114}
\datum{ 160}{ 124}
\datum{ 160}{ 126}
\datum{ 160}{ 150}
\datum{ 160}{ 156}
\datum{ 160}{ 170}
\datum{ 160}{ 178}
\datum{ 162}{ 97}
\datum{ 162}{ 115}
\datum{ 162}{ 131}
\datum{ 162}{ 133}
\datum{ 164}{ 94}
\datum{ 168}{ 84}
\datum{ 168}{ 86}
\datum{ 168}{ 88}
\datum{ 168}{ 90}
\datum{ 168}{ 94}
\datum{ 168}{ 96}
\datum{ 168}{ 98}
\datum{ 168}{ 100}
\datum{ 168}{ 102}
\datum{ 168}{ 106}
\datum{ 168}{ 108}
\datum{ 168}{ 110}
\datum{ 168}{ 112}
\datum{ 168}{ 114}
\datum{ 168}{ 116}
\datum{ 168}{ 118}
\datum{ 168}{ 120}
\datum{ 168}{ 122}
\datum{ 168}{ 124}
\datum{ 168}{ 126}
\datum{ 168}{ 128}
\datum{ 168}{ 134}
\datum{ 168}{ 138}
\datum{ 168}{ 140}
\datum{ 168}{ 142}
\datum{ 168}{ 144}
\datum{ 168}{ 148}
\datum{ 168}{ 152}
\datum{ 168}{ 164}
\datum{ 168}{ 168}
\datum{ 168}{ 178}
\datum{ 168}{ 184}
\datum{ 168}{ 256}
\datum{ 170}{ 93}
\datum{ 170}{ 117}
\datum{ 172}{ 114}
\datum{ 172}{ 122}
\datum{ 174}{ 97}
\datum{ 174}{ 143}
\datum{ 174}{ 157}
\datum{ 174}{ 167}
\datum{ 176}{ 92}
\datum{ 176}{ 98}
\datum{ 176}{ 100}
\datum{ 176}{ 102}
\datum{ 176}{ 106}
\datum{ 176}{ 108}
\datum{ 176}{ 114}
\datum{ 176}{ 126}
\datum{ 176}{ 132}
\datum{ 176}{ 174}
\datum{ 180}{ 90}
\datum{ 180}{ 98}
\datum{ 180}{ 100}
\datum{ 180}{ 104}
\datum{ 180}{ 106}
\datum{ 180}{ 108}
\datum{ 180}{ 110}
\datum{ 180}{ 112}
\datum{ 180}{ 114}
\datum{ 180}{ 118}
\datum{ 180}{ 120}
\datum{ 180}{ 124}
\datum{ 180}{ 126}
\datum{ 180}{ 134}
\datum{ 180}{ 136}
\datum{ 180}{ 138}
\datum{ 180}{ 142}
\datum{ 180}{ 144}
\datum{ 180}{ 148}
\datum{ 180}{ 154}
\datum{ 180}{ 158}
\datum{ 180}{ 168}
\datum{ 180}{ 174}
\datum{ 180}{ 184}
\datum{ 180}{ 226}
\datum{ 180}{ 228}
\datum{ 180}{ 286}
\datum{ 180}{ 366}
\datum{ 184}{ 102}
\datum{ 184}{ 106}
\datum{ 184}{ 110}
\datum{ 184}{ 116}
\datum{ 184}{ 122}
\datum{ 184}{ 134}
\datum{ 184}{ 136}
\datum{ 184}{ 148}
\datum{ 184}{ 180}
\datum{ 186}{ 97}
\datum{ 186}{ 115}
\datum{ 186}{ 123}
\datum{ 186}{ 167}
\datum{ 192}{ 102}
\datum{ 192}{ 106}
\datum{ 192}{ 108}
\datum{ 192}{ 110}
\datum{ 192}{ 112}
\datum{ 192}{ 114}
\datum{ 192}{ 116}
\datum{ 192}{ 118}
\datum{ 192}{ 122}
\datum{ 192}{ 124}
\datum{ 192}{ 126}
\datum{ 192}{ 128}
\datum{ 192}{ 130}
\datum{ 192}{ 134}
\datum{ 192}{ 138}
\datum{ 192}{ 142}
\datum{ 192}{ 150}
\datum{ 192}{ 154}
\datum{ 192}{ 162}
\datum{ 192}{ 166}
\datum{ 192}{ 170}
\datum{ 192}{ 178}
\datum{ 192}{ 184}
\datum{ 192}{ 190}
\datum{ 192}{ 198}
\datum{ 192}{ 206}
\datum{ 192}{ 218}
\datum{ 192}{ 226}
\datum{ 192}{ 250}
\datum{ 196}{ 106}
\datum{ 196}{ 150}
\datum{ 196}{ 172}
\datum{ 198}{ 125}
\datum{ 198}{ 131}
\datum{ 200}{ 102}
\datum{ 200}{ 106}
\datum{ 200}{ 108}
\datum{ 200}{ 116}
\datum{ 200}{ 118}
\datum{ 200}{ 128}
\datum{ 200}{ 134}
\datum{ 200}{ 136}
\datum{ 200}{ 148}
\datum{ 200}{ 150}
\datum{ 200}{ 156}
\datum{ 200}{ 168}
\datum{ 204}{ 104}
\datum{ 204}{ 108}
\datum{ 204}{ 114}
\datum{ 204}{ 118}
\datum{ 204}{ 120}
\datum{ 204}{ 124}
\datum{ 204}{ 126}
\datum{ 204}{ 130}
\datum{ 204}{ 134}
\datum{ 204}{ 142}
\datum{ 204}{ 164}
\datum{ 204}{ 168}
\datum{ 204}{ 190}
\datum{ 208}{ 108}
\datum{ 208}{ 118}
\datum{ 208}{ 126}
\datum{ 208}{ 138}
\datum{ 210}{ 113}
\datum{ 210}{ 117}
\datum{ 210}{ 133}
\datum{ 210}{ 143}
\datum{ 210}{ 145}
\datum{ 210}{ 151}
\datum{ 210}{ 171}
\datum{ 210}{ 221}
\datum{ 212}{ 150}
\datum{ 212}{ 158}
\datum{ 216}{ 112}
\datum{ 216}{ 116}
\datum{ 216}{ 120}
\datum{ 216}{ 124}
\datum{ 216}{ 126}
\datum{ 216}{ 128}
\datum{ 216}{ 132}
\datum{ 216}{ 134}
\datum{ 216}{ 136}
\datum{ 216}{ 138}
\datum{ 216}{ 140}
\datum{ 216}{ 142}
\datum{ 216}{ 144}
\datum{ 216}{ 150}
\datum{ 216}{ 152}
\datum{ 216}{ 154}
\datum{ 216}{ 158}
\datum{ 216}{ 164}
\datum{ 216}{ 168}
\datum{ 216}{ 174}
\datum{ 216}{ 180}
\datum{ 216}{ 182}
\datum{ 216}{ 188}
\datum{ 216}{ 192}
\datum{ 216}{ 212}
\datum{ 216}{ 240}
\datum{ 216}{ 244}
\datum{ 216}{ 274}
\datum{ 216}{ 292}
\datum{ 220}{ 126}
\datum{ 220}{ 158}
\datum{ 220}{ 218}
\datum{ 222}{ 133}
\datum{ 222}{ 135}
\datum{ 224}{ 122}
\datum{ 224}{ 124}
\datum{ 224}{ 126}
\datum{ 224}{ 134}
\datum{ 224}{ 138}
\datum{ 224}{ 142}
\datum{ 224}{ 178}
\datum{ 228}{ 126}
\datum{ 228}{ 130}
\datum{ 228}{ 132}
\datum{ 228}{ 138}
\datum{ 228}{ 140}
\datum{ 228}{ 142}
\datum{ 228}{ 146}
\datum{ 228}{ 170}
\datum{ 228}{ 176}
\datum{ 228}{ 178}
\datum{ 228}{ 196}
\datum{ 228}{ 202}
\datum{ 230}{ 129}
\datum{ 232}{ 126}
\datum{ 232}{ 134}
\datum{ 232}{ 158}
\datum{ 234}{ 131}
\datum{ 234}{ 133}
\datum{ 234}{ 141}
\datum{ 234}{ 217}
\datum{ 236}{ 172}
\datum{ 240}{ 124}
\datum{ 240}{ 126}
\datum{ 240}{ 130}
\datum{ 240}{ 134}
\datum{ 240}{ 138}
\datum{ 240}{ 140}
\datum{ 240}{ 142}
\datum{ 240}{ 144}
\datum{ 240}{ 146}
\datum{ 240}{ 150}
\datum{ 240}{ 158}
\datum{ 240}{ 162}
\datum{ 240}{ 166}
\datum{ 240}{ 174}
\datum{ 240}{ 178}
\datum{ 240}{ 188}
\datum{ 240}{ 206}
\datum{ 240}{ 218}
\datum{ 240}{ 226}
\datum{ 240}{ 232}
\datum{ 240}{ 268}
\datum{ 240}{ 394}
\datum{ 242}{ 139}
\datum{ 242}{ 189}
\datum{ 246}{ 167}
\datum{ 246}{ 185}
\datum{ 246}{ 237}
\datum{ 248}{ 134}
\datum{ 248}{ 174}
\datum{ 252}{ 130}
\datum{ 252}{ 138}
\datum{ 252}{ 142}
\datum{ 252}{ 158}
\datum{ 252}{ 162}
\datum{ 252}{ 166}
\datum{ 252}{ 174}
\datum{ 252}{ 178}
\datum{ 252}{ 202}
\datum{ 252}{ 206}
\datum{ 252}{ 278}
\datum{ 256}{ 134}
\datum{ 256}{ 156}
\datum{ 256}{ 158}
\datum{ 256}{ 166}
\datum{ 256}{ 174}
\datum{ 256}{ 234}
\datum{ 258}{ 151}
\datum{ 258}{ 163}
\datum{ 260}{ 134}
\datum{ 260}{ 150}
\datum{ 264}{ 144}
\datum{ 264}{ 148}
\datum{ 264}{ 150}
\datum{ 264}{ 154}
\datum{ 264}{ 158}
\datum{ 264}{ 160}
\datum{ 264}{ 162}
\datum{ 264}{ 164}
\datum{ 264}{ 172}
\datum{ 264}{ 178}
\datum{ 264}{ 184}
\datum{ 264}{ 188}
\datum{ 264}{ 214}
\datum{ 264}{ 228}
\datum{ 264}{ 262}
\datum{ 266}{ 169}
\datum{ 270}{ 223}
\datum{ 272}{ 148}
\datum{ 272}{ 150}
\datum{ 272}{ 166}
\datum{ 272}{ 178}
\datum{ 276}{ 150}
\datum{ 276}{ 154}
\datum{ 276}{ 162}
\datum{ 276}{ 172}
\datum{ 276}{ 174}
\datum{ 276}{ 186}
\datum{ 276}{ 192}
\datum{ 276}{ 212}
\datum{ 276}{ 214}
\datum{ 276}{ 222}
\datum{ 276}{ 234}
\datum{ 276}{ 262}
\datum{ 276}{ 330}
\datum{ 280}{ 148}
\datum{ 280}{ 150}
\datum{ 280}{ 166}
\datum{ 286}{ 177}
\datum{ 288}{ 146}
\datum{ 288}{ 152}
\datum{ 288}{ 158}
\datum{ 288}{ 162}
\datum{ 288}{ 166}
\datum{ 288}{ 170}
\datum{ 288}{ 182}
\datum{ 288}{ 186}
\datum{ 288}{ 188}
\datum{ 288}{ 190}
\datum{ 288}{ 206}
\datum{ 288}{ 214}
\datum{ 288}{ 222}
\datum{ 288}{ 226}
\datum{ 288}{ 270}
\datum{ 288}{ 374}
\datum{ 292}{ 158}
\datum{ 294}{ 159}
\datum{ 294}{ 187}
\datum{ 296}{ 150}
\datum{ 296}{ 166}
\datum{ 296}{ 174}
\datum{ 300}{ 168}
\datum{ 300}{ 178}
\datum{ 300}{ 180}
\datum{ 300}{ 194}
\datum{ 300}{ 198}
\datum{ 300}{ 218}
\datum{ 300}{ 226}
\datum{ 300}{ 230}
\datum{ 304}{ 176}
\datum{ 306}{ 169}
\datum{ 306}{ 217}
\datum{ 312}{ 166}
\datum{ 312}{ 172}
\datum{ 312}{ 174}
\datum{ 312}{ 178}
\datum{ 312}{ 180}
\datum{ 312}{ 190}
\datum{ 312}{ 192}
\datum{ 312}{ 196}
\datum{ 312}{ 224}
\datum{ 316}{ 262}
\datum{ 318}{ 197}
\datum{ 320}{ 170}
\datum{ 320}{ 174}
\datum{ 320}{ 190}
\datum{ 320}{ 198}
\datum{ 320}{ 206}
\datum{ 320}{ 222}
\datum{ 322}{ 193}
\datum{ 324}{ 168}
\datum{ 324}{ 182}
\datum{ 324}{ 184}
\datum{ 324}{ 222}
\datum{ 324}{ 230}
\datum{ 324}{ 232}
\datum{ 324}{ 262}
\datum{ 330}{ 181}
\datum{ 330}{ 221}
\datum{ 330}{ 261}
\datum{ 336}{ 178}
\datum{ 336}{ 188}
\datum{ 336}{ 194}
\datum{ 336}{ 198}
\datum{ 336}{ 202}
\datum{ 336}{ 206}
\datum{ 336}{ 222}
\datum{ 336}{ 230}
\datum{ 336}{ 312}
\datum{ 336}{ 358}
\datum{ 340}{ 198}
\datum{ 342}{ 233}
\datum{ 348}{ 186}
\datum{ 348}{ 198}
\datum{ 348}{ 226}
\datum{ 348}{ 238}
\datum{ 352}{ 190}
\datum{ 354}{ 259}
\datum{ 356}{ 202}
\datum{ 356}{ 220}
\datum{ 360}{ 190}
\datum{ 360}{ 192}
\datum{ 360}{ 194}
\datum{ 360}{ 206}
\datum{ 360}{ 212}
\datum{ 360}{ 228}
\datum{ 360}{ 258}
\datum{ 360}{ 306}
\datum{ 364}{ 194}
\datum{ 368}{ 204}
\datum{ 372}{ 194}
\datum{ 372}{ 202}
\datum{ 372}{ 226}
\datum{ 372}{ 258}
\datum{ 372}{ 262}
\datum{ 372}{ 346}
\datum{ 376}{ 214}
\datum{ 380}{ 198}
\datum{ 384}{ 204}
\datum{ 384}{ 218}
\datum{ 384}{ 222}
\datum{ 384}{ 232}
\datum{ 384}{ 234}
\datum{ 384}{ 242}
\datum{ 384}{ 250}
\datum{ 384}{ 262}
\datum{ 396}{ 214}
\datum{ 396}{ 222}
\datum{ 396}{ 262}
\datum{ 396}{ 340}
\datum{ 408}{ 212}
\datum{ 408}{ 224}
\datum{ 408}{ 232}
\datum{ 408}{ 240}
\datum{ 408}{ 268}
\datum{ 420}{ 218}
\datum{ 420}{ 230}
\datum{ 420}{ 248}
\datum{ 420}{ 250}
\datum{ 420}{ 306}
\datum{ 420}{ 334}
\datum{ 426}{ 265}
\datum{ 432}{ 238}
\datum{ 432}{ 242}
\datum{ 432}{ 266}
\datum{ 432}{ 274}
\datum{ 432}{ 334}
\datum{ 444}{ 234}
\datum{ 450}{ 303}
\datum{ 456}{ 234}
\datum{ 456}{ 248}
\datum{ 456}{ 256}
\datum{ 456}{ 262}
\datum{ 456}{ 264}
\datum{ 456}{ 272}
\datum{ 456}{ 302}
\datum{ 468}{ 286}
\datum{ 468}{ 306}
\datum{ 476}{ 270}
\datum{ 480}{ 246}
\datum{ 480}{ 262}
\datum{ 480}{ 278}
\datum{ 480}{ 286}
\datum{ 480}{ 334}
\datum{ 492}{ 256}
\datum{ 504}{ 276}
\datum{ 504}{ 312}
\datum{ 510}{ 331}
\datum{ 512}{ 286}
\datum{ 516}{ 302}
\datum{ 516}{ 330}
\datum{ 528}{ 278}
\datum{ 528}{ 286}
\datum{ 528}{ 318}
\datum{ 528}{ 334}
\datum{ 540}{ 274}
\datum{ 540}{ 298}
\datum{ 540}{ 334}
\datum{ 552}{ 306}
\datum{ 564}{ 322}
\datum{ 564}{ 330}
\datum{ 564}{ 340}
\datum{ 576}{ 302}
\datum{ 576}{ 314}
\datum{ 588}{ 346}
\datum{ 612}{ 330}
\datum{ 612}{ 346}
\datum{ 624}{ 330}
\datum{ 624}{ 358}
\datum{ 636}{ 342}
\datum{ 648}{ 358}
\datum{ 660}{ 366}
\datum{ 672}{ 374}
\datum{ 720}{ 394}
\datum{ 732}{ 386}
\datum{ 744}{ 402}
\datum{ 804}{ 430}
\datum{ 840}{ 446}
\datum{ 900}{ 474}
\datum{ 960}{ 502}
\datum{ -100}{ 62}
\datum{ -100}{ 66}
\datum{ -100}{ 70}
\datum{ -100}{ 76}
\datum{ -100}{ 80}
\datum{ -100}{ 98}
\datum{ -100}{ 118}
\datum{ -100}{ 120}
\datum{ -102}{ 61}
\datum{ -102}{ 67}
\datum{ -102}{ 73}
\datum{ -102}{ 81}
\datum{ -102}{ 83}
\datum{ -102}{ 93}
\datum{ -102}{ 113}
\datum{ -104}{ 54}
\datum{ -104}{ 58}
\datum{ -104}{ 62}
\datum{ -104}{ 68}
\datum{ -104}{ 74}
\datum{ -104}{ 78}
\datum{ -104}{ 80}
\datum{ -104}{ 84}
\datum{ -104}{ 86}
\datum{ -104}{ 92}
\datum{ -104}{ 100}
\datum{ -104}{ 104}
\datum{ -104}{ 106}
\datum{ -104}{ 126}
\datum{ -104}{ 140}
\datum{ -104}{ 154}
\datum{ -106}{ 117}
\datum{ -106}{ 133}
\datum{ -108}{ 58}
\datum{ -108}{ 60}
\datum{ -108}{ 66}
\datum{ -108}{ 68}
\datum{ -108}{ 70}
\datum{ -108}{ 72}
\datum{ -108}{ 74}
\datum{ -108}{ 76}
\datum{ -108}{ 80}
\datum{ -108}{ 82}
\datum{ -108}{ 84}
\datum{ -108}{ 86}
\datum{ -108}{ 88}
\datum{ -108}{ 90}
\datum{ -108}{ 94}
\datum{ -108}{ 100}
\datum{ -108}{ 106}
\datum{ -108}{ 108}
\datum{ -108}{ 112}
\datum{ -108}{ 120}
\datum{ -108}{ 122}
\datum{ -108}{ 130}
\datum{ -108}{ 150}
\datum{ -108}{ 166}
\datum{ -108}{ 178}
\datum{ -108}{ 186}
\datum{ -108}{ 190}
\datum{ -108}{ 212}
\datum{ -110}{ 83}
\datum{ -110}{ 119}
\datum{ -110}{ 141}
\datum{ -110}{ 193}
\datum{ -112}{ 60}
\datum{ -112}{ 62}
\datum{ -112}{ 66}
\datum{ -112}{ 70}
\datum{ -112}{ 74}
\datum{ -112}{ 76}
\datum{ -112}{ 78}
\datum{ -112}{ 88}
\datum{ -112}{ 90}
\datum{ -112}{ 92}
\datum{ -112}{ 94}
\datum{ -112}{ 96}
\datum{ -112}{ 102}
\datum{ -112}{ 108}
\datum{ -112}{ 110}
\datum{ -112}{ 118}
\datum{ -112}{ 124}
\datum{ -112}{ 126}
\datum{ -112}{ 160}
\datum{ -114}{ 67}
\datum{ -114}{ 77}
\datum{ -114}{ 85}
\datum{ -114}{ 97}
\datum{ -114}{ 127}
\datum{ -114}{ 145}
\datum{ -114}{ 147}
\datum{ -114}{ 187}
\datum{ -114}{ 197}
\datum{ -116}{ 72}
\datum{ -116}{ 124}
\datum{ -120}{ 62}
\datum{ -120}{ 64}
\datum{ -120}{ 66}
\datum{ -120}{ 68}
\datum{ -120}{ 70}
\datum{ -120}{ 72}
\datum{ -120}{ 76}
\datum{ -120}{ 78}
\datum{ -120}{ 80}
\datum{ -120}{ 82}
\datum{ -120}{ 86}
\datum{ -120}{ 88}
\datum{ -120}{ 90}
\datum{ -120}{ 92}
\datum{ -120}{ 94}
\datum{ -120}{ 96}
\datum{ -120}{ 98}
\datum{ -120}{ 100}
\datum{ -120}{ 102}
\datum{ -120}{ 104}
\datum{ -120}{ 108}
\datum{ -120}{ 110}
\datum{ -120}{ 112}
\datum{ -120}{ 116}
\datum{ -120}{ 118}
\datum{ -120}{ 122}
\datum{ -120}{ 124}
\datum{ -120}{ 128}
\datum{ -120}{ 130}
\datum{ -120}{ 132}
\datum{ -120}{ 134}
\datum{ -120}{ 136}
\datum{ -120}{ 138}
\datum{ -120}{ 142}
\datum{ -120}{ 146}
\datum{ -120}{ 148}
\datum{ -120}{ 150}
\datum{ -120}{ 152}
\datum{ -120}{ 154}
\datum{ -120}{ 156}
\datum{ -120}{ 158}
\datum{ -120}{ 168}
\datum{ -120}{ 178}
\datum{ -120}{ 190}
\datum{ -120}{ 196}
\datum{ -120}{ 212}
\datum{ -120}{ 248}
\datum{ -120}{ 276}
\datum{ -120}{ 278}
\datum{ -124}{ 74}
\datum{ -124}{ 84}
\datum{ -124}{ 106}
\datum{ -124}{ 136}
\datum{ -124}{ 150}
\datum{ -126}{ 69}
\datum{ -126}{ 79}
\datum{ -126}{ 87}
\datum{ -126}{ 89}
\datum{ -126}{ 91}
\datum{ -126}{ 99}
\datum{ -126}{ 115}
\datum{ -126}{ 143}
\datum{ -126}{ 179}
\datum{ -128}{ 78}
\datum{ -128}{ 82}
\datum{ -128}{ 86}
\datum{ -128}{ 90}
\datum{ -128}{ 92}
\datum{ -128}{ 94}
\datum{ -128}{ 98}
\datum{ -128}{ 102}
\datum{ -128}{ 108}
\datum{ -128}{ 110}
\datum{ -128}{ 114}
\datum{ -128}{ 142}
\datum{ -130}{ 93}
\datum{ -132}{ 72}
\datum{ -132}{ 74}
\datum{ -132}{ 76}
\datum{ -132}{ 78}
\datum{ -132}{ 80}
\datum{ -132}{ 86}
\datum{ -132}{ 94}
\datum{ -132}{ 96}
\datum{ -132}{ 100}
\datum{ -132}{ 102}
\datum{ -132}{ 104}
\datum{ -132}{ 114}
\datum{ -132}{ 126}
\datum{ -132}{ 148}
\datum{ -132}{ 178}
\datum{ -132}{ 202}
\datum{ -132}{ 238}
\datum{ -132}{ 302}
\datum{ -134}{ 87}
\datum{ -136}{ 72}
\datum{ -136}{ 78}
\datum{ -136}{ 82}
\datum{ -136}{ 86}
\datum{ -136}{ 94}
\datum{ -136}{ 102}
\datum{ -136}{ 108}
\datum{ -136}{ 112}
\datum{ -136}{ 118}
\datum{ -136}{ 126}
\datum{ -136}{ 136}
\datum{ -138}{ 85}
\datum{ -138}{ 97}
\datum{ -138}{ 105}
\datum{ -138}{ 125}
\datum{ -138}{ 151}
\datum{ -140}{ 88}
\datum{ -140}{ 96}
\datum{ -140}{ 98}
\datum{ -140}{ 110}
\datum{ -140}{ 158}
\datum{ -142}{ 127}
\datum{ -144}{ 74}
\datum{ -144}{ 76}
\datum{ -144}{ 78}
\datum{ -144}{ 80}
\datum{ -144}{ 82}
\datum{ -144}{ 84}
\datum{ -144}{ 86}
\datum{ -144}{ 88}
\datum{ -144}{ 90}
\datum{ -144}{ 92}
\datum{ -144}{ 94}
\datum{ -144}{ 96}
\datum{ -144}{ 98}
\datum{ -144}{ 100}
\datum{ -144}{ 102}
\datum{ -144}{ 104}
\datum{ -144}{ 106}
\datum{ -144}{ 110}
\datum{ -144}{ 112}
\datum{ -144}{ 114}
\datum{ -144}{ 118}
\datum{ -144}{ 122}
\datum{ -144}{ 124}
\datum{ -144}{ 126}
\datum{ -144}{ 130}
\datum{ -144}{ 134}
\datum{ -144}{ 138}
\datum{ -144}{ 140}
\datum{ -144}{ 148}
\datum{ -144}{ 150}
\datum{ -144}{ 154}
\datum{ -144}{ 158}
\datum{ -144}{ 164}
\datum{ -144}{ 172}
\datum{ -144}{ 174}
\datum{ -144}{ 182}
\datum{ -144}{ 188}
\datum{ -144}{ 190}
\datum{ -144}{ 206}
\datum{ -144}{ 214}
\datum{ -148}{ 92}
\datum{ -148}{ 96}
\datum{ -148}{ 158}
\datum{ -150}{ 93}
\datum{ -150}{ 97}
\datum{ -150}{ 103}
\datum{ -150}{ 109}
\datum{ -150}{ 133}
\datum{ -150}{ 151}
\datum{ -152}{ 78}
\datum{ -152}{ 84}
\datum{ -152}{ 86}
\datum{ -152}{ 96}
\datum{ -152}{ 108}
\datum{ -152}{ 110}
\datum{ -152}{ 130}
\datum{ -152}{ 134}
\datum{ -152}{ 150}
\datum{ -156}{ 86}
\datum{ -156}{ 88}
\datum{ -156}{ 92}
\datum{ -156}{ 94}
\datum{ -156}{ 102}
\datum{ -156}{ 106}
\datum{ -156}{ 108}
\datum{ -156}{ 110}
\datum{ -156}{ 112}
\datum{ -156}{ 114}
\datum{ -156}{ 118}
\datum{ -156}{ 120}
\datum{ -156}{ 122}
\datum{ -156}{ 124}
\datum{ -156}{ 132}
\datum{ -156}{ 150}
\datum{ -156}{ 210}
\datum{ -156}{ 232}
\datum{ -156}{ 234}
\datum{ -156}{ 430}
\datum{ -160}{ 90}
\datum{ -160}{ 96}
\datum{ -160}{ 98}
\datum{ -160}{ 102}
\datum{ -160}{ 108}
\datum{ -160}{ 110}
\datum{ -160}{ 124}
\datum{ -160}{ 126}
\datum{ -160}{ 150}
\datum{ -160}{ 156}
\datum{ -160}{ 178}
\datum{ -160}{ 198}
\datum{ -160}{ 314}
\datum{ -162}{ 121}
\datum{ -162}{ 131}
\datum{ -162}{ 133}
\datum{ -162}{ 137}
\datum{ -162}{ 185}
\datum{ -164}{ 94}
\datum{ -166}{ 127}
\datum{ -168}{ 84}
\datum{ -168}{ 86}
\datum{ -168}{ 88}
\datum{ -168}{ 90}
\datum{ -168}{ 94}
\datum{ -168}{ 96}
\datum{ -168}{ 98}
\datum{ -168}{ 100}
\datum{ -168}{ 102}
\datum{ -168}{ 106}
\datum{ -168}{ 108}
\datum{ -168}{ 110}
\datum{ -168}{ 112}
\datum{ -168}{ 114}
\datum{ -168}{ 116}
\datum{ -168}{ 118}
\datum{ -168}{ 120}
\datum{ -168}{ 122}
\datum{ -168}{ 124}
\datum{ -168}{ 128}
\datum{ -168}{ 134}
\datum{ -168}{ 138}
\datum{ -168}{ 140}
\datum{ -168}{ 142}
\datum{ -168}{ 144}
\datum{ -168}{ 148}
\datum{ -168}{ 152}
\datum{ -168}{ 168}
\datum{ -168}{ 174}
\datum{ -168}{ 178}
\datum{ -168}{ 184}
\datum{ -168}{ 256}
\datum{ -172}{ 106}
\datum{ -172}{ 144}
\datum{ -174}{ 97}
\datum{ -174}{ 167}
\datum{ -176}{ 92}
\datum{ -176}{ 102}
\datum{ -176}{ 106}
\datum{ -176}{ 108}
\datum{ -176}{ 114}
\datum{ -176}{ 126}
\datum{ -176}{ 150}
\datum{ -176}{ 174}
\datum{ -180}{ 90}
\datum{ -180}{ 98}
\datum{ -180}{ 108}
\datum{ -180}{ 112}
\datum{ -180}{ 118}
\datum{ -180}{ 120}
\datum{ -180}{ 124}
\datum{ -180}{ 126}
\datum{ -180}{ 136}
\datum{ -180}{ 138}
\datum{ -180}{ 142}
\datum{ -180}{ 148}
\datum{ -180}{ 154}
\datum{ -180}{ 158}
\datum{ -180}{ 168}
\datum{ -180}{ 174}
\datum{ -180}{ 184}
\datum{ -180}{ 194}
\datum{ -180}{ 226}
\datum{ -180}{ 228}
\datum{ -180}{ 286}
\datum{ -180}{ 366}
\datum{ -184}{ 102}
\datum{ -184}{ 110}
\datum{ -184}{ 112}
\datum{ -184}{ 120}
\datum{ -184}{ 134}
\datum{ -184}{ 148}
\datum{ -184}{ 168}
\datum{ -184}{ 174}
\datum{ -186}{ 97}
\datum{ -186}{ 117}
\datum{ -186}{ 127}
\datum{ -186}{ 151}
\datum{ -190}{ 129}
\datum{ -192}{ 102}
\datum{ -192}{ 106}
\datum{ -192}{ 108}
\datum{ -192}{ 110}
\datum{ -192}{ 112}
\datum{ -192}{ 114}
\datum{ -192}{ 116}
\datum{ -192}{ 118}
\datum{ -192}{ 122}
\datum{ -192}{ 124}
\datum{ -192}{ 126}
\datum{ -192}{ 128}
\datum{ -192}{ 130}
\datum{ -192}{ 134}
\datum{ -192}{ 138}
\datum{ -192}{ 150}
\datum{ -192}{ 154}
\datum{ -192}{ 162}
\datum{ -192}{ 166}
\datum{ -192}{ 170}
\datum{ -192}{ 178}
\datum{ -192}{ 184}
\datum{ -192}{ 190}
\datum{ -192}{ 198}
\datum{ -192}{ 206}
\datum{ -192}{ 218}
\datum{ -192}{ 226}
\datum{ -192}{ 250}
\datum{ -194}{ 113}
\datum{ -194}{ 119}
\datum{ -196}{ 106}
\datum{ -196}{ 172}
\datum{ -198}{ 113}
\datum{ -198}{ 139}
\datum{ -200}{ 102}
\datum{ -200}{ 106}
\datum{ -200}{ 116}
\datum{ -200}{ 130}
\datum{ -200}{ 134}
\datum{ -200}{ 150}
\datum{ -200}{ 168}
\datum{ -202}{ 133}
\datum{ -204}{ 104}
\datum{ -204}{ 108}
\datum{ -204}{ 114}
\datum{ -204}{ 118}
\datum{ -204}{ 120}
\datum{ -204}{ 126}
\datum{ -204}{ 130}
\datum{ -204}{ 134}
\datum{ -204}{ 142}
\datum{ -204}{ 164}
\datum{ -204}{ 178}
\datum{ -204}{ 222}
\datum{ -208}{ 108}
\datum{ -208}{ 126}
\datum{ -208}{ 138}
\datum{ -208}{ 178}
\datum{ -210}{ 113}
\datum{ -210}{ 131}
\datum{ -210}{ 145}
\datum{ -210}{ 151}
\datum{ -210}{ 171}
\datum{ -210}{ 221}
\datum{ -210}{ 241}
\datum{ -216}{ 112}
\datum{ -216}{ 116}
\datum{ -216}{ 120}
\datum{ -216}{ 124}
\datum{ -216}{ 126}
\datum{ -216}{ 128}
\datum{ -216}{ 132}
\datum{ -216}{ 134}
\datum{ -216}{ 136}
\datum{ -216}{ 142}
\datum{ -216}{ 144}
\datum{ -216}{ 150}
\datum{ -216}{ 152}
\datum{ -216}{ 154}
\datum{ -216}{ 158}
\datum{ -216}{ 164}
\datum{ -216}{ 174}
\datum{ -216}{ 180}
\datum{ -216}{ 182}
\datum{ -216}{ 188}
\datum{ -216}{ 192}
\datum{ -216}{ 212}
\datum{ -216}{ 240}
\datum{ -216}{ 244}
\datum{ -216}{ 274}
\datum{ -216}{ 292}
\datum{ -220}{ 158}
\datum{ -222}{ 133}
\datum{ -224}{ 122}
\datum{ -224}{ 126}
\datum{ -224}{ 142}
\datum{ -224}{ 146}
\datum{ -224}{ 152}
\datum{ -224}{ 170}
\datum{ -228}{ 126}
\datum{ -228}{ 130}
\datum{ -228}{ 132}
\datum{ -228}{ 138}
\datum{ -228}{ 140}
\datum{ -228}{ 142}
\datum{ -228}{ 146}
\datum{ -228}{ 170}
\datum{ -228}{ 176}
\datum{ -228}{ 196}
\datum{ -228}{ 346}
\datum{ -232}{ 126}
\datum{ -232}{ 134}
\datum{ -232}{ 158}
\datum{ -234}{ 217}
\datum{ -236}{ 162}
\datum{ -236}{ 172}
\datum{ -236}{ 182}
\datum{ -240}{ 124}
\datum{ -240}{ 126}
\datum{ -240}{ 134}
\datum{ -240}{ 138}
\datum{ -240}{ 140}
\datum{ -240}{ 142}
\datum{ -240}{ 144}
\datum{ -240}{ 146}
\datum{ -240}{ 150}
\datum{ -240}{ 158}
\datum{ -240}{ 162}
\datum{ -240}{ 166}
\datum{ -240}{ 174}
\datum{ -240}{ 178}
\datum{ -240}{ 188}
\datum{ -240}{ 206}
\datum{ -240}{ 218}
\datum{ -240}{ 226}
\datum{ -240}{ 232}
\datum{ -240}{ 268}
\datum{ -240}{ 394}
\datum{ -244}{ 162}
\datum{ -246}{ 167}
\datum{ -246}{ 237}
\datum{ -252}{ 130}
\datum{ -252}{ 138}
\datum{ -252}{ 142}
\datum{ -252}{ 158}
\datum{ -252}{ 162}
\datum{ -252}{ 178}
\datum{ -252}{ 202}
\datum{ -252}{ 206}
\datum{ -252}{ 278}
\datum{ -256}{ 134}
\datum{ -256}{ 166}
\datum{ -256}{ 174}
\datum{ -258}{ 161}
\datum{ -260}{ 134}
\datum{ -260}{ 150}
\datum{ -264}{ 144}
\datum{ -264}{ 148}
\datum{ -264}{ 150}
\datum{ -264}{ 154}
\datum{ -264}{ 158}
\datum{ -264}{ 162}
\datum{ -264}{ 164}
\datum{ -264}{ 172}
\datum{ -264}{ 184}
\datum{ -264}{ 188}
\datum{ -264}{ 214}
\datum{ -264}{ 228}
\datum{ -264}{ 262}
\datum{ -272}{ 150}
\datum{ -272}{ 178}
\datum{ -276}{ 150}
\datum{ -276}{ 154}
\datum{ -276}{ 172}
\datum{ -276}{ 174}
\datum{ -276}{ 192}
\datum{ -276}{ 214}
\datum{ -276}{ 234}
\datum{ -276}{ 262}
\datum{ -276}{ 330}
\datum{ -280}{ 148}
\datum{ -280}{ 166}
\datum{ -284}{ 166}
\datum{ -286}{ 163}
\datum{ -288}{ 146}
\datum{ -288}{ 152}
\datum{ -288}{ 158}
\datum{ -288}{ 162}
\datum{ -288}{ 166}
\datum{ -288}{ 170}
\datum{ -288}{ 182}
\datum{ -288}{ 186}
\datum{ -288}{ 188}
\datum{ -288}{ 190}
\datum{ -288}{ 206}
\datum{ -288}{ 214}
\datum{ -288}{ 222}
\datum{ -288}{ 226}
\datum{ -288}{ 258}
\datum{ -288}{ 270}
\datum{ -288}{ 374}
\datum{ -292}{ 158}
\datum{ -294}{ 259}
\datum{ -296}{ 150}
\datum{ -296}{ 162}
\datum{ -296}{ 166}
\datum{ -300}{ 168}
\datum{ -300}{ 178}
\datum{ -300}{ 180}
\datum{ -300}{ 198}
\datum{ -300}{ 218}
\datum{ -300}{ 222}
\datum{ -300}{ 226}
\datum{ -300}{ 230}
\datum{ -304}{ 176}
\datum{ -306}{ 169}
\datum{ -306}{ 177}
\datum{ -306}{ 217}
\datum{ -312}{ 166}
\datum{ -312}{ 172}
\datum{ -312}{ 178}
\datum{ -312}{ 180}
\datum{ -312}{ 190}
\datum{ -312}{ 192}
\datum{ -312}{ 196}
\datum{ -312}{ 224}
\datum{ -312}{ 318}
\datum{ -320}{ 174}
\datum{ -324}{ 168}
\datum{ -324}{ 182}
\datum{ -324}{ 184}
\datum{ -324}{ 232}
\datum{ -324}{ 262}
\datum{ -330}{ 181}
\datum{ -330}{ 221}
\datum{ -330}{ 261}
\datum{ -336}{ 178}
\datum{ -336}{ 188}
\datum{ -336}{ 194}
\datum{ -336}{ 198}
\datum{ -336}{ 202}
\datum{ -336}{ 206}
\datum{ -336}{ 222}
\datum{ -336}{ 230}
\datum{ -336}{ 312}
\datum{ -336}{ 358}
\datum{ -340}{ 198}
\datum{ -342}{ 233}
\datum{ -344}{ 224}
\datum{ -348}{ 186}
\datum{ -348}{ 198}
\datum{ -348}{ 226}
\datum{ -348}{ 238}
\datum{ -348}{ 258}
\datum{ -356}{ 234}
\datum{ -360}{ 190}
\datum{ -360}{ 192}
\datum{ -360}{ 212}
\datum{ -360}{ 228}
\datum{ -364}{ 204}
\datum{ -368}{ 204}
\datum{ -372}{ 194}
\datum{ -372}{ 202}
\datum{ -372}{ 226}
\datum{ -372}{ 258}
\datum{ -372}{ 262}
\datum{ -372}{ 306}
\datum{ -372}{ 346}
\datum{ -376}{ 214}
\datum{ -384}{ 204}
\datum{ -384}{ 218}
\datum{ -384}{ 232}
\datum{ -384}{ 242}
\datum{ -384}{ 250}
\datum{ -390}{ 303}
\datum{ -396}{ 222}
\datum{ -396}{ 262}
\datum{ -396}{ 340}
\datum{ -408}{ 212}
\datum{ -408}{ 224}
\datum{ -408}{ 232}
\datum{ -408}{ 240}
\datum{ -408}{ 268}
\datum{ -416}{ 262}
\datum{ -420}{ 218}
\datum{ -420}{ 230}
\datum{ -420}{ 248}
\datum{ -420}{ 250}
\datum{ -420}{ 306}
\datum{ -420}{ 334}
\datum{ -426}{ 265}
\datum{ -432}{ 238}
\datum{ -432}{ 242}
\datum{ -432}{ 266}
\datum{ -432}{ 274}
\datum{ -432}{ 334}
\datum{ -444}{ 234}
\datum{ -444}{ 330}
\datum{ -450}{ 331}
\datum{ -456}{ 234}
\datum{ -456}{ 248}
\datum{ -456}{ 256}
\datum{ -456}{ 262}
\datum{ -456}{ 264}
\datum{ -456}{ 272}
\datum{ -468}{ 286}
\datum{ -480}{ 246}
\datum{ -480}{ 262}
\datum{ -480}{ 278}
\datum{ -480}{ 286}
\datum{ -480}{ 306}
\datum{ -480}{ 334}
\datum{ -492}{ 256}
\datum{ -504}{ 276}
\datum{ -504}{ 312}
\datum{ -516}{ 302}
\datum{ -528}{ 278}
\datum{ -528}{ 286}
\datum{ -528}{ 334}
\datum{ -540}{ 274}
\datum{ -540}{ 334}
\datum{ -552}{ 306}
\datum{ -564}{ 322}
\datum{ -564}{ 330}
\datum{ -564}{ 340}
\datum{ -576}{ 314}
\datum{ -588}{ 346}
\datum{ -612}{ 330}
\datum{ -624}{ 330}
\datum{ -624}{ 358}
\datum{ -636}{ 342}
\datum{ -660}{ 366}
\datum{ -672}{ 374}
\datum{ -720}{ 394}
\datum{ -732}{ 386}
\datum{ -744}{ 402}
\datum{ -804}{ 430}
\datum{ -900}{ 474}
\datum{ -960}{ 502}
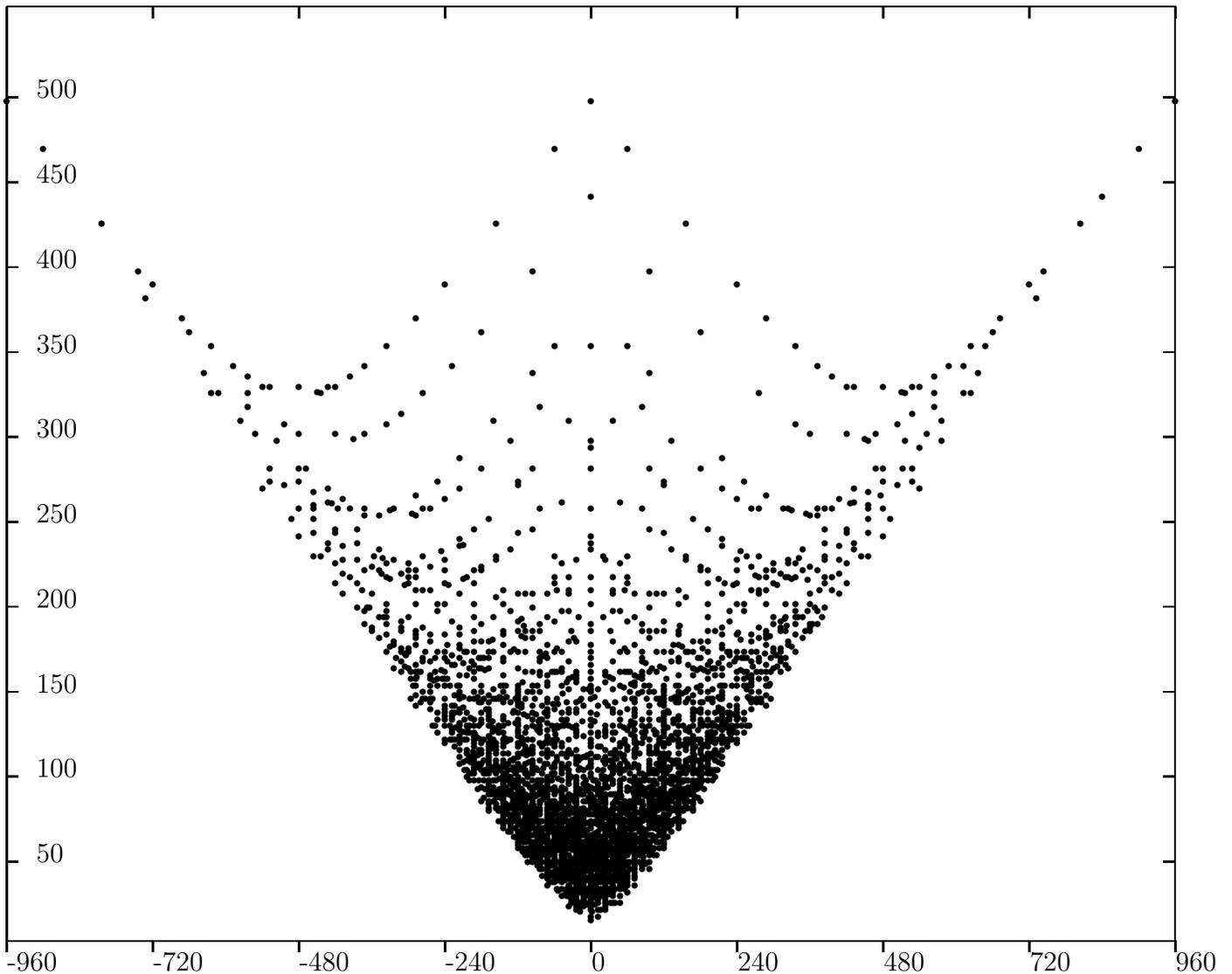
\parbox{6.4truein}{\noindent {\bf Fig. 3}~~{\it A plot of Euler numbers
against
 ${\bar n}_g+n_g$ for the 2997 spectra of all the LG potentials.
}}
\end{center}
\end{center}

The richness of the list
can be appreciated from the fact that the number of distinct spectra in
this class
is about an order of magnitude larger than in the previous constructions.
Mirror
symmetry emerges only when a sufficiently large subspace of the
configuration space
of the heterotic string is probed. Furthermore this class is also of
potential
phenomenological interest. Whereas it had previously proved very difficult
to find \cys\ with $\chi=\pm 6$ the construction of weighted CICYs leads to
some tens such manifolds. These spaces are not of immediate phenomenological
interest since they are all simply connected. Hence the best place to seek
interesting models may well be among manifolds with Euler number
$\pm 6k$, with
$k>1$, although it may also be possible to construct interesting
(2,0)--models
\cite{dg} from the manifolds of Euler number $\pm 6$ which have reduced
gauge groups
due to an enlarged background gauge group. This however is a question for
future work.

With benefit of hindsight it seems odd that no systematic construction of
such
manifolds was previously attempted. Hypersurfaces in weighted projective
spaces
were first discussed in the physics literature by Yau \cite{yau} and
Strominger and Witten\cite{hitnrun} who constructed a number of examples
with
large negative Euler number
$-200>\chi>-300$. Subsequently Kim, Koh and Yoon \cite{kky} constructed a
few further
examples with Euler numbers in a similar range. A complicating feature
of these
examples was the assumed need to avoid the singular sets of the embedding
$\IP_4$, coupled with the large values of $|\chi|$ achieved, this led to the
impression that the construction was contrived and the resulting
manifolds very
complicated. To those elucidating the connections between exactly soluble
conformal theories and Calabi--Yau manifolds \cite{m}\cite{gvw} however
it was
apparent that avoiding the
singular sets of the embedding space was unnecessary since the
singularities
on the resulting hypersurface can be resolved while maintaining the
condition
$c_1=0$. This greatly increases the number of manifolds that can be
constructed
in this way. In fact of the 10,000 odd examples that have been constructed
in refs. \cite{cls}\cite{ks} only a small subset consisting of spaces
described by
polynomials of Fermat
type do not intersect the singular sets. Resolution of singularities has the
effect of raising the Euler number and in this way the many Euler numbers
plotted in Fig.~3 are achieved including many values of moderate size that
may be of interest for model building.

Even though the high degree of symmetry of this space of groundstates
under the flip of the sign of the Euler numbers strongly suggests a
relation between dual pairs there is a priori no reason as far as the
different Landau--Ginzburg potentials are concerned why this should
be the case. Looking more closely at the Hodge pairs of \cite{cls} seems to
confuse the issue since generically one finds for a given dual pair of
Hodge numbers not two manifolds but {\it many} and it is unclear which
of the possible combinations (if any) are in fact related.

It is therefore of interest to ask whether a systematic procedure can be
established which relates pairs of vacua as mirror partners. This is indeed
the case. The construction, introduced in ref. \cite{ls4}, consists
of a combination of an orbifolding and a nonlinear transformation with
somewhat unusual properties in that it involves fractional powers. Even
though this procedure can be formulated in the manifold picture it is
most easily motivated in Landau--Ginzburg language. By applying this
construction to a certain class of mirror candidates it is possible to
establish a close connection between dual vacua. The description of this
construction comprises Part II of this review.

The starting point
is the well known equivalence of the affine D--invariant in the
N=2 superconformal minimal series to the $\ZZ_2$--orbifold of the diagonal
invariant at the same level. In section 4 this equivalence will be lifted to
the full vacuum described by a tensor product of $N=2$ minimal theories.
It is necessary to reformulate this equivalence in terms of the
Landau--Ginzburg (LG) potential and its projectivization as a
Calabi--Yau (CY) manifold since for most of our models
no exactly solvable theory is known.
The computations in this framework then either involve orbifoldizing
LG--potentials \cite{v} or modding out discrete groups of CY--manifolds and
resolving singularities \cite{ry}\cite{s3}.
Once this procedure has been formulated in the simple framework of the
tensor series of $N=2$ minimal models it will be clear how to proceed in
general. Sections 5 and 6 extend the discussion to more general potentials
and illustrate to what extent our LG--potentials can be viewed as orbifolds.
As already mentioned a general technique to find mirror pairs will emerge.
Another, more mathematical, aspect which lends itself to an analysis via
fractional transformation is the strange duality of Arnold.

Finally, in Part III, the emphasis is shifted from Landau--Ginzburg theories
to their orbifolds. Even though it is possible to map via the fractional
transformations of Part II many classes of orbifolds of the models described
in Part I to complete intersections it will become clear that not all
orbifolds can be understood this way, at least with the techniques
presently available. Hence
it is useful to consider orbifolds in their own right firstly as a
potential pool of new models and also in order to pursue the question
raised at the beginning of the introduction regarding the `robustness'
 the mirror
property against a change of technique
\fnote{2}{An orbifold analysis of minimal exactly solvable models has been
 performed in ref. \cite{gp}. The interesting result is that the
 set of all orbifolds emerging from a given diagonal theory is
 selfdual (see also ref. \cite{alr})}.
 In ref. \cite{kss} we have
constructed the complete set of Landau--Ginzburg potentials with an
arbitrary number of fields involving
only couplings between at most two fields (i.e. arbitrary combinations
of Fermat, 1--Tadpole and 1--Loop polynomials) and implemented some 40 odd
actions of phase symmetries. It turns out that for this class
of orbifolds the mirror property improves from about eighty percent for
the LG--potentials
to about 94\%. Figure 4 contains the plot of the resulting spectra.

\begin{center}
\plot{6truein}{\tiny{$\bullet$}}
\nobreak
\Place{-960}{50}{\leftscalemark~~50}
\Place{-960}{100}{\leftscalemark~~100}
\Place{-960}{150}{\leftscalemark~~150}
\Place{-960}{200}{\leftscalemark~~200}
\Place{-960}{250}{\leftscalemark~~250}
\Place{-960}{300}{\leftscalemark~~300}
\Place{-960}{350}{\leftscalemark~~350}
\Place{-960}{400}{\leftscalemark~~400}
\Place{-960}{450}{\leftscalemark~~450}
\Place{-960}{500}{\leftscalemark~~500}
\Place{960}{50}{\rightscalemark\vphantom{0}}
\Place{960}{100}{\rightscalemark\vphantom{0}}
\Place{960}{150}{\rightscalemark\vphantom{0}}
\Place{960}{200}{\rightscalemark\vphantom{0}}
\Place{960}{250}{\rightscalemark\vphantom{0}}
\Place{960}{300}{\rightscalemark\vphantom{0}}
\Place{960}{350}{\rightscalemark\vphantom{0}}
\Place{960}{400}{\rightscalemark\vphantom{0}}
\Place{960}{450}{\rightscalemark\vphantom{0}}
\Place{960}{500}{\rightscalemark\vphantom{0}}
\Place{-960}{0}{\hmark\lower18pt\hbox{-960}}
\Place{-720}{0}{\hmark\lower18pt\hbox{-720}}
\Place{-480}{0}{\hmark\lower18pt\hbox{-480}}
\Place{-240}{0}{\hmark\lower18pt\hbox{-240}}
\Place{0}{0}{\hmark\lower18pt\hbox{0}}
\Place{240}{0}{\hmark\lower18pt\hbox{240}}
\Place{480}{0}{\hmark\lower18pt\hbox{480}}
\Place{720}{0}{\hmark\lower18pt\hbox{720}}
\Place{960}{0}{\hmark\lower18pt\hbox{960}}
\Place{-720}{550}{\hmark}
\Place{-480}{550}{\hmark}
\Place{-240}{550}{\hmark}
\Place{0}{550}{\hmark}
\Place{240}{550}{\hmark}
\Place{480}{550}{\hmark}
\Place{720}{550}{\hmark}
\Place{960}{550}{\hmark}
\nobreak
\datum{ -72}{ 36}
\datum{ -108}{ 54}
\datum{ -168}{ 84}
\datum{ -180}{ 90}
\datum{ -40}{ 22}
\datum{ -72}{ 38}
\datum{ -96}{ 50}
\datum{ -104}{ 54}
\datum{ -108}{ 56}
\datum{ -120}{ 62}
\datum{ -144}{ 74}
\datum{ -152}{ 78}
\datum{ -168}{ 86}
\datum{ -200}{ 102}
\datum{ -204}{ 104}
\datum{ -288}{ 146}
\datum{ -296}{ 150}
\datum{ -72}{ 40}
\datum{ -108}{ 58}
\datum{ -112}{ 60}
\datum{ -120}{ 64}
\datum{ -144}{ 76}
\datum{ -168}{ 88}
\datum{ -176}{ 92}
\datum{ -186}{ 97}
\datum{ -208}{ 108}
\datum{ -216}{ 112}
\datum{ -240}{ 124}
\datum{ -252}{ 130}
\datum{ -260}{ 134}
\datum{ -540}{ 274}
\datum{ -72}{ 42}
\datum{ -80}{ 46}
\datum{ -84}{ 48}
\datum{ -96}{ 54}
\datum{ -104}{ 58}
\datum{ -108}{ 60}
\datum{ -112}{ 62}
\datum{ -120}{ 66}
\datum{ -126}{ 69}
\datum{ -132}{ 72}
\datum{ -144}{ 78}
\datum{ -160}{ 86}
\datum{ -168}{ 90}
\datum{ -192}{ 102}
\datum{ -200}{ 106}
\datum{ -204}{ 108}
\datum{ -240}{ 126}
\datum{ -256}{ 134}
\datum{ -324}{ 168}
\datum{ -456}{ 234}
\datum{ -480}{ 246}
\datum{ -72}{ 44}
\datum{ -84}{ 50}
\datum{ -96}{ 56}
\datum{ -120}{ 68}
\datum{ -132}{ 74}
\datum{ -144}{ 80}
\datum{ -180}{ 98}
\datum{ -196}{ 106}
\datum{ -216}{ 116}
\datum{ -280}{ 148}
\datum{ -288}{ 152}
\datum{ -372}{ 194}
\datum{ -408}{ 212}
\datum{ -420}{ 218}
\datum{ -48}{ 34}
\datum{ -56}{ 38}
\datum{ -64}{ 42}
\datum{ -72}{ 46}
\datum{ -88}{ 54}
\datum{ -96}{ 58}
\datum{ -102}{ 61}
\datum{ -104}{ 62}
\datum{ -120}{ 70}
\datum{ -128}{ 74}
\datum{ -132}{ 76}
\datum{ -136}{ 78}
\datum{ -144}{ 82}
\datum{ -156}{ 88}
\datum{ -160}{ 90}
\datum{ -168}{ 94}
\datum{ -184}{ 102}
\datum{ -192}{ 106}
\datum{ -232}{ 126}
\datum{ -312}{ 166}
\datum{ -336}{ 178}
\datum{ -360}{ 190}
\datum{ -492}{ 256}
\datum{ -24}{ 24}
\datum{ -48}{ 36}
\datum{ -60}{ 42}
\datum{ -72}{ 48}
\datum{ -84}{ 54}
\datum{ -96}{ 60}
\datum{ -100}{ 62}
\datum{ -108}{ 66}
\datum{ -120}{ 72}
\datum{ -132}{ 78}
\datum{ -164}{ 94}
\datum{ -168}{ 96}
\datum{ -192}{ 108}
\datum{ -216}{ 120}
\datum{ -228}{ 126}
\datum{ -252}{ 138}
\datum{ -264}{ 144}
\datum{ -276}{ 150}
\datum{ -292}{ 158}
\datum{ -348}{ 186}
\datum{ -360}{ 192}
\datum{ -384}{ 204}
\datum{ -444}{ 234}
\datum{ -42}{ 35}
\datum{ -48}{ 38}
\datum{ -54}{ 41}
\datum{ -60}{ 44}
\datum{ -72}{ 50}
\datum{ -80}{ 54}
\datum{ -96}{ 62}
\datum{ -108}{ 68}
\datum{ -112}{ 70}
\datum{ -120}{ 74}
\datum{ -128}{ 78}
\datum{ -132}{ 80}
\datum{ -144}{ 86}
\datum{ -168}{ 98}
\datum{ -176}{ 102}
\datum{ -192}{ 110}
\datum{ -240}{ 134}
\datum{ -272}{ 150}
\datum{ -288}{ 158}
\datum{ -528}{ 278}
\datum{ -36}{ 34}
\datum{ -48}{ 40}
\datum{ -54}{ 43}
\datum{ -60}{ 46}
\datum{ -64}{ 48}
\datum{ -72}{ 52}
\datum{ -76}{ 54}
\datum{ -84}{ 58}
\datum{ -88}{ 60}
\datum{ -100}{ 66}
\datum{ -102}{ 67}
\datum{ -104}{ 68}
\datum{ -108}{ 70}
\datum{ -120}{ 76}
\datum{ -126}{ 79}
\datum{ -138}{ 85}
\datum{ -144}{ 88}
\datum{ -156}{ 94}
\datum{ -168}{ 100}
\datum{ -180}{ 106}
\datum{ -192}{ 112}
\datum{ -204}{ 118}
\datum{ -216}{ 124}
\datum{ -228}{ 130}
\datum{ -252}{ 142}
\datum{ -264}{ 148}
\datum{ -276}{ 154}
\datum{ -312}{ 172}
\datum{ -372}{ 202}
\datum{ 0}{ 18}
\datum{ -24}{ 30}
\datum{ -36}{ 36}
\datum{ -48}{ 42}
\datum{ -60}{ 48}
\datum{ -64}{ 50}
\datum{ -72}{ 54}
\datum{ -80}{ 58}
\datum{ -90}{ 63}
\datum{ -96}{ 66}
\datum{ -108}{ 72}
\datum{ -120}{ 78}
\datum{ -136}{ 86}
\datum{ -140}{ 88}
\datum{ -144}{ 90}
\datum{ -148}{ 92}
\datum{ -150}{ 93}
\datum{ -160}{ 98}
\datum{ -168}{ 102}
\datum{ -176}{ 106}
\datum{ -180}{ 108}
\datum{ -184}{ 110}
\datum{ -192}{ 114}
\datum{ -204}{ 120}
\datum{ -216}{ 126}
\datum{ -228}{ 132}
\datum{ -232}{ 134}
\datum{ -240}{ 138}
\datum{ -288}{ 162}
\datum{ -296}{ 166}
\datum{ -300}{ 168}
\datum{ -624}{ 330}
\datum{ -12}{ 26}
\datum{ -24}{ 32}
\datum{ -30}{ 35}
\datum{ -32}{ 36}
\datum{ -48}{ 44}
\datum{ -54}{ 47}
\datum{ -60}{ 50}
\datum{ -72}{ 56}
\datum{ -80}{ 60}
\datum{ -84}{ 62}
\datum{ -96}{ 68}
\datum{ -108}{ 74}
\datum{ -112}{ 76}
\datum{ -120}{ 80}
\datum{ -144}{ 92}
\datum{ -184}{ 112}
\datum{ -192}{ 116}
\datum{ -216}{ 128}
\datum{ -260}{ 150}
\datum{ -336}{ 188}
\datum{ -368}{ 204}
\datum{ -408}{ 224}
\datum{ -420}{ 230}
\datum{ -456}{ 248}
\datum{ -732}{ 386}
\datum{ 0}{ 22}
\datum{ -12}{ 28}
\datum{ -16}{ 30}
\datum{ -24}{ 34}
\datum{ -32}{ 38}
\datum{ -36}{ 40}
\datum{ -42}{ 43}
\datum{ -44}{ 44}
\datum{ -48}{ 46}
\datum{ -54}{ 49}
\datum{ -56}{ 50}
\datum{ -60}{ 52}
\datum{ -64}{ 54}
\datum{ -72}{ 58}
\datum{ -78}{ 61}
\datum{ -80}{ 62}
\datum{ -84}{ 64}
\datum{ -90}{ 67}
\datum{ -96}{ 70}
\datum{ -104}{ 74}
\datum{ -108}{ 76}
\datum{ -112}{ 78}
\datum{ -116}{ 80}
\datum{ -120}{ 82}
\datum{ -128}{ 86}
\datum{ -144}{ 94}
\datum{ -160}{ 102}
\datum{ -168}{ 106}
\datum{ -180}{ 112}
\datum{ -192}{ 118}
\datum{ -208}{ 126}
\datum{ -240}{ 142}
\datum{ -288}{ 166}
\datum{ -312}{ 178}
\datum{ -324}{ 184}
\datum{ -432}{ 238}
\datum{ -480}{ 262}
\datum{ -528}{ 286}
\datum{ -960}{ 502}
\datum{ -24}{ 36}
\datum{ -36}{ 42}
\datum{ -40}{ 44}
\datum{ -48}{ 48}
\datum{ -72}{ 60}
\datum{ -84}{ 66}
\datum{ -88}{ 68}
\datum{ -96}{ 72}
\datum{ -108}{ 78}
\datum{ -120}{ 84}
\datum{ -126}{ 87}
\datum{ -144}{ 96}
\datum{ -168}{ 108}
\datum{ -204}{ 126}
\datum{ -216}{ 132}
\datum{ -228}{ 138}
\datum{ -240}{ 144}
\datum{ -252}{ 150}
\datum{ -304}{ 176}
\datum{ -348}{ 198}
\datum{ -396}{ 222}
\datum{ -612}{ 330}
\datum{ -900}{ 474}
\datum{ 0}{ 26}
\datum{ -12}{ 32}
\datum{ -24}{ 38}
\datum{ -32}{ 42}
\datum{ -36}{ 44}
\datum{ -48}{ 50}
\datum{ -72}{ 62}
\datum{ -88}{ 70}
\datum{ -96}{ 74}
\datum{ -108}{ 80}
\datum{ -120}{ 86}
\datum{ -132}{ 92}
\datum{ -144}{ 98}
\datum{ -152}{ 102}
\datum{ -168}{ 110}
\datum{ -176}{ 114}
\datum{ -192}{ 122}
\datum{ -216}{ 134}
\datum{ -228}{ 140}
\datum{ -264}{ 158}
\datum{ -288}{ 170}
\datum{ -336}{ 194}
\datum{ -376}{ 214}
\datum{ -384}{ 218}
\datum{ -432}{ 242}
\datum{ -12}{ 34}
\datum{ -24}{ 40}
\datum{ -32}{ 44}
\datum{ -36}{ 46}
\datum{ -42}{ 49}
\datum{ -44}{ 50}
\datum{ -48}{ 52}
\datum{ -60}{ 58}
\datum{ -72}{ 64}
\datum{ -80}{ 68}
\datum{ -96}{ 76}
\datum{ -104}{ 80}
\datum{ -108}{ 82}
\datum{ -120}{ 88}
\datum{ -128}{ 92}
\datum{ -138}{ 97}
\datum{ -156}{ 106}
\datum{ -160}{ 108}
\datum{ -168}{ 112}
\datum{ -180}{ 118}
\datum{ -192}{ 124}
\datum{ -204}{ 130}
\datum{ -228}{ 142}
\datum{ -340}{ 198}
\datum{ -408}{ 232}
\datum{ -456}{ 256}
\datum{ -804}{ 430}
\datum{ 12}{ 24}
\datum{ 0}{ 30}
\datum{ -12}{ 36}
\datum{ -16}{ 38}
\datum{ -24}{ 42}
\datum{ -40}{ 50}
\datum{ -48}{ 54}
\datum{ -60}{ 60}
\datum{ -64}{ 62}
\datum{ -72}{ 66}
\datum{ -80}{ 70}
\datum{ -96}{ 78}
\datum{ -100}{ 80}
\datum{ -102}{ 81}
\datum{ -108}{ 84}
\datum{ -112}{ 86}
\datum{ -120}{ 90}
\datum{ -144}{ 102}
\datum{ -156}{ 108}
\datum{ -160}{ 110}
\datum{ -168}{ 114}
\datum{ -180}{ 120}
\datum{ -224}{ 142}
\datum{ -240}{ 150}
\datum{ -264}{ 162}
\datum{ -300}{ 180}
\datum{ -336}{ 198}
\datum{ -552}{ 306}
\datum{ -744}{ 402}
\datum{ 0}{ 32}
\datum{ -12}{ 38}
\datum{ -18}{ 41}
\datum{ -24}{ 44}
\datum{ -36}{ 50}
\datum{ -48}{ 56}
\datum{ -60}{ 62}
\datum{ -66}{ 65}
\datum{ -72}{ 68}
\datum{ -80}{ 72}
\datum{ -96}{ 80}
\datum{ -102}{ 83}
\datum{ -108}{ 86}
\datum{ -120}{ 92}
\datum{ -144}{ 104}
\datum{ -152}{ 108}
\datum{ -156}{ 110}
\datum{ -168}{ 116}
\datum{ -192}{ 128}
\datum{ -204}{ 134}
\datum{ -228}{ 146}
\datum{ -264}{ 164}
\datum{ -360}{ 212}
\datum{ 12}{ 28}
\datum{ 0}{ 34}
\datum{ -8}{ 38}
\datum{ -12}{ 40}
\datum{ -16}{ 42}
\datum{ -24}{ 46}
\datum{ -32}{ 50}
\datum{ -36}{ 52}
\datum{ -40}{ 54}
\datum{ -42}{ 55}
\datum{ -48}{ 58}
\datum{ -56}{ 62}
\datum{ -64}{ 66}
\datum{ -72}{ 70}
\datum{ -80}{ 74}
\datum{ -96}{ 82}
\datum{ -104}{ 86}
\datum{ -108}{ 88}
\datum{ -120}{ 94}
\datum{ -132}{ 100}
\datum{ -144}{ 106}
\datum{ -152}{ 110}
\datum{ -156}{ 112}
\datum{ -168}{ 118}
\datum{ -200}{ 134}
\datum{ -216}{ 142}
\datum{ -276}{ 172}
\datum{ -312}{ 190}
\datum{ -336}{ 202}
\datum{ -720}{ 394}
\datum{ 24}{ 24}
\datum{ 0}{ 36}
\datum{ -6}{ 39}
\datum{ -12}{ 42}
\datum{ -16}{ 44}
\datum{ -24}{ 48}
\datum{ -36}{ 54}
\datum{ -48}{ 60}
\datum{ -60}{ 66}
\datum{ -64}{ 68}
\datum{ -66}{ 69}
\datum{ -80}{ 76}
\datum{ -84}{ 78}
\datum{ -108}{ 90}
\datum{ -112}{ 92}
\datum{ -120}{ 96}
\datum{ -132}{ 102}
\datum{ -144}{ 108}
\datum{ -156}{ 114}
\datum{ -168}{ 120}
\datum{ -216}{ 144}
\datum{ -252}{ 162}
\datum{ -276}{ 174}
\datum{ -312}{ 192}
\datum{ -408}{ 240}
\datum{ -456}{ 264}
\datum{ -660}{ 366}
\datum{ 12}{ 32}
\datum{ 0}{ 38}
\datum{ -12}{ 44}
\datum{ -16}{ 46}
\datum{ -20}{ 48}
\datum{ -24}{ 50}
\datum{ -32}{ 54}
\datum{ -36}{ 56}
\datum{ -48}{ 62}
\datum{ -52}{ 64}
\datum{ -60}{ 68}
\datum{ -72}{ 74}
\datum{ -96}{ 86}
\datum{ -120}{ 98}
\datum{ -128}{ 102}
\datum{ -132}{ 104}
\datum{ -144}{ 110}
\datum{ -168}{ 122}
\datum{ -192}{ 134}
\datum{ -240}{ 158}
\datum{ -256}{ 166}
\datum{ -288}{ 182}
\datum{ -336}{ 206}
\datum{ -672}{ 374}
\datum{ 12}{ 34}
\datum{ 0}{ 40}
\datum{ -6}{ 43}
\datum{ -12}{ 46}
\datum{ -24}{ 52}
\datum{ -36}{ 58}
\datum{ -44}{ 62}
\datum{ -48}{ 64}
\datum{ -56}{ 68}
\datum{ -60}{ 70}
\datum{ -72}{ 76}
\datum{ -78}{ 79}
\datum{ -84}{ 82}
\datum{ -96}{ 88}
\datum{ -108}{ 94}
\datum{ -112}{ 96}
\datum{ -120}{ 100}
\datum{ -132}{ 106}
\datum{ -144}{ 112}
\datum{ -160}{ 120}
\datum{ -168}{ 124}
\datum{ -180}{ 130}
\datum{ -264}{ 172}
\datum{ -312}{ 196}
\datum{ -372}{ 226}
\datum{ -564}{ 322}
\datum{ 40}{ 22}
\datum{ 24}{ 30}
\datum{ 12}{ 36}
\datum{ 8}{ 38}
\datum{ 6}{ 39}
\datum{ 0}{ 42}
\datum{ -12}{ 48}
\datum{ -16}{ 50}
\datum{ -24}{ 54}
\datum{ -30}{ 57}
\datum{ -40}{ 62}
\datum{ -48}{ 66}
\datum{ -72}{ 78}
\datum{ -84}{ 84}
\datum{ -96}{ 90}
\datum{ -108}{ 96}
\datum{ -120}{ 102}
\datum{ -156}{ 120}
\datum{ -184}{ 134}
\datum{ -192}{ 138}
\datum{ -216}{ 150}
\datum{ -232}{ 158}
\datum{ -240}{ 162}
\datum{ -272}{ 178}
\datum{ -288}{ 186}
\datum{ 24}{ 32}
\datum{ 12}{ 38}
\datum{ 0}{ 44}
\datum{ -12}{ 50}
\datum{ -16}{ 52}
\datum{ -24}{ 56}
\datum{ -48}{ 68}
\datum{ -54}{ 71}
\datum{ -60}{ 74}
\datum{ -64}{ 76}
\datum{ -72}{ 80}
\datum{ -96}{ 92}
\datum{ -120}{ 104}
\datum{ -128}{ 108}
\datum{ -168}{ 128}
\datum{ -216}{ 152}
\datum{ -236}{ 162}
\datum{ -288}{ 188}
\datum{ -456}{ 272}
\datum{ -516}{ 302}
\datum{ 24}{ 34}
\datum{ 16}{ 38}
\datum{ 12}{ 40}
\datum{ 6}{ 43}
\datum{ 0}{ 46}
\datum{ -12}{ 52}
\datum{ -16}{ 54}
\datum{ -18}{ 55}
\datum{ -24}{ 58}
\datum{ -30}{ 61}
\datum{ -48}{ 70}
\datum{ -56}{ 74}
\datum{ -60}{ 76}
\datum{ -64}{ 78}
\datum{ -72}{ 82}
\datum{ -84}{ 88}
\datum{ -96}{ 94}
\datum{ -120}{ 106}
\datum{ -128}{ 110}
\datum{ -144}{ 118}
\datum{ -156}{ 124}
\datum{ -192}{ 142}
\datum{ -240}{ 166}
\datum{ -256}{ 174}
\datum{ -288}{ 190}
\datum{ -480}{ 286}
\datum{ -624}{ 358}
\datum{ 24}{ 36}
\datum{ 12}{ 42}
\datum{ 0}{ 48}
\datum{ -4}{ 50}
\datum{ -8}{ 52}
\datum{ -12}{ 54}
\datum{ -16}{ 56}
\datum{ -18}{ 57}
\datum{ -24}{ 60}
\datum{ -30}{ 63}
\datum{ -32}{ 64}
\datum{ -40}{ 68}
\datum{ -48}{ 72}
\datum{ -60}{ 78}
\datum{ -64}{ 80}
\datum{ -72}{ 84}
\datum{ -84}{ 90}
\datum{ -90}{ 93}
\datum{ -120}{ 108}
\datum{ -128}{ 112}
\datum{ -132}{ 114}
\datum{ -156}{ 126}
\datum{ -180}{ 138}
\datum{ -300}{ 198}
\datum{ -360}{ 228}
\datum{ -564}{ 330}
\datum{ 30}{ 35}
\datum{ 24}{ 38}
\datum{ 18}{ 41}
\datum{ 16}{ 42}
\datum{ 12}{ 44}
\datum{ 0}{ 50}
\datum{ -12}{ 56}
\datum{ -24}{ 62}
\datum{ -36}{ 68}
\datum{ -40}{ 70}
\datum{ -48}{ 74}
\datum{ -72}{ 86}
\datum{ -96}{ 98}
\datum{ -120}{ 110}
\datum{ -144}{ 122}
\datum{ -168}{ 134}
\datum{ -200}{ 150}
\datum{ -216}{ 158}
\datum{ -384}{ 242}
\datum{ -432}{ 266}
\datum{ 24}{ 40}
\datum{ 16}{ 44}
\datum{ 12}{ 46}
\datum{ 4}{ 50}
\datum{ 0}{ 52}
\datum{ -4}{ 54}
\datum{ -12}{ 58}
\datum{ -24}{ 64}
\datum{ -36}{ 70}
\datum{ -48}{ 76}
\datum{ -60}{ 82}
\datum{ -72}{ 88}
\datum{ -84}{ 94}
\datum{ -112}{ 108}
\datum{ -120}{ 112}
\datum{ -144}{ 124}
\datum{ -252}{ 178}
\datum{ -264}{ 184}
\datum{ -468}{ 286}
\datum{ -588}{ 346}
\datum{ 36}{ 36}
\datum{ 24}{ 42}
\datum{ 16}{ 46}
\datum{ 12}{ 48}
\datum{ 0}{ 54}
\datum{ -20}{ 64}
\datum{ -24}{ 66}
\datum{ -32}{ 70}
\datum{ -48}{ 78}
\datum{ -60}{ 84}
\datum{ -64}{ 86}
\datum{ -66}{ 87}
\datum{ -96}{ 102}
\datum{ -108}{ 108}
\datum{ -112}{ 110}
\datum{ -152}{ 130}
\datum{ -156}{ 132}
\datum{ -192}{ 150}
\datum{ -236}{ 172}
\datum{ -240}{ 174}
\datum{ -276}{ 192}
\datum{ 24}{ 44}
\datum{ 12}{ 50}
\datum{ 8}{ 52}
\datum{ 4}{ 54}
\datum{ 0}{ 56}
\datum{ -8}{ 60}
\datum{ -12}{ 62}
\datum{ -16}{ 64}
\datum{ -24}{ 68}
\datum{ -32}{ 72}
\datum{ -36}{ 74}
\datum{ -48}{ 80}
\datum{ -72}{ 92}
\datum{ -76}{ 94}
\datum{ -80}{ 96}
\datum{ -84}{ 98}
\datum{ -120}{ 116}
\datum{ -168}{ 140}
\datum{ -216}{ 164}
\datum{ -228}{ 170}
\datum{ -264}{ 188}
\datum{ 48}{ 34}
\datum{ 24}{ 46}
\datum{ 20}{ 48}
\datum{ 16}{ 50}
\datum{ 12}{ 52}
\datum{ 4}{ 56}
\datum{ 0}{ 58}
\datum{ -8}{ 62}
\datum{ -12}{ 64}
\datum{ -16}{ 66}
\datum{ -24}{ 70}
\datum{ -36}{ 76}
\datum{ -40}{ 78}
\datum{ -42}{ 79}
\datum{ -44}{ 80}
\datum{ -48}{ 82}
\datum{ -54}{ 85}
\datum{ -56}{ 86}
\datum{ -60}{ 88}
\datum{ -64}{ 90}
\datum{ -72}{ 94}
\datum{ -88}{ 102}
\datum{ -96}{ 106}
\datum{ -108}{ 112}
\datum{ -120}{ 118}
\datum{ -128}{ 122}
\datum{ -136}{ 126}
\datum{ -168}{ 142}
\datum{ -192}{ 154}
\datum{ -240}{ 178}
\datum{ -384}{ 250}
\datum{ -432}{ 274}
\datum{ -564}{ 340}
\datum{ 36}{ 42}
\datum{ 32}{ 44}
\datum{ 24}{ 48}
\datum{ 16}{ 52}
\datum{ 12}{ 54}
\datum{ 0}{ 60}
\datum{ -12}{ 66}
\datum{ -20}{ 70}
\datum{ -24}{ 72}
\datum{ -32}{ 76}
\datum{ -36}{ 78}
\datum{ -56}{ 88}
\datum{ -60}{ 90}
\datum{ -72}{ 96}
\datum{ -84}{ 102}
\datum{ -96}{ 108}
\datum{ -108}{ 114}
\datum{ -132}{ 126}
\datum{ -168}{ 144}
\datum{ -504}{ 312}
\datum{ 48}{ 38}
\datum{ 36}{ 44}
\datum{ 24}{ 50}
\datum{ 16}{ 54}
\datum{ 12}{ 56}
\datum{ 0}{ 62}
\datum{ -6}{ 65}
\datum{ -8}{ 66}
\datum{ -18}{ 71}
\datum{ -24}{ 74}
\datum{ -32}{ 78}
\datum{ -48}{ 86}
\datum{ -72}{ 98}
\datum{ -80}{ 102}
\datum{ -96}{ 110}
\datum{ -120}{ 122}
\datum{ -144}{ 134}
\datum{ -228}{ 176}
\datum{ -288}{ 206}
\datum{ -336}{ 230}
\datum{ 48}{ 40}
\datum{ 42}{ 43}
\datum{ 24}{ 52}
\datum{ 18}{ 55}
\datum{ 16}{ 56}
\datum{ 12}{ 58}
\datum{ 8}{ 60}
\datum{ 0}{ 64}
\datum{ -6}{ 67}
\datum{ -8}{ 68}
\datum{ -12}{ 70}
\datum{ -16}{ 72}
\datum{ -24}{ 76}
\datum{ -48}{ 88}
\datum{ -64}{ 96}
\datum{ -66}{ 97}
\datum{ -72}{ 100}
\datum{ -96}{ 112}
\datum{ -120}{ 124}
\datum{ -168}{ 148}
\datum{ -348}{ 238}
\datum{ -396}{ 262}
\datum{ -408}{ 268}
\datum{ -540}{ 334}
\datum{ 48}{ 42}
\datum{ 44}{ 44}
\datum{ 32}{ 50}
\datum{ 24}{ 54}
\datum{ 18}{ 57}
\datum{ 8}{ 62}
\datum{ 0}{ 66}
\datum{ -8}{ 70}
\datum{ -12}{ 72}
\datum{ -16}{ 74}
\datum{ -24}{ 78}
\datum{ -36}{ 84}
\datum{ -40}{ 86}
\datum{ -48}{ 90}
\datum{ -72}{ 102}
\datum{ -96}{ 114}
\datum{ -108}{ 120}
\datum{ -116}{ 124}
\datum{ -144}{ 138}
\datum{ -216}{ 174}
\datum{ -480}{ 306}
\datum{ 48}{ 44}
\datum{ 36}{ 50}
\datum{ 24}{ 56}
\datum{ 12}{ 62}
\datum{ 6}{ 65}
\datum{ 0}{ 68}
\datum{ -24}{ 80}
\datum{ -30}{ 83}
\datum{ -48}{ 92}
\datum{ -60}{ 98}
\datum{ -100}{ 118}
\datum{ -120}{ 128}
\datum{ -136}{ 136}
\datum{ -168}{ 152}
\datum{ -180}{ 158}
\datum{ -240}{ 188}
\datum{ -312}{ 224}
\datum{ 48}{ 46}
\datum{ 42}{ 49}
\datum{ 40}{ 50}
\datum{ 36}{ 52}
\datum{ 32}{ 54}
\datum{ 24}{ 58}
\datum{ 12}{ 64}
\datum{ 8}{ 66}
\datum{ 6}{ 67}
\datum{ 0}{ 70}
\datum{ -12}{ 76}
\datum{ -16}{ 78}
\datum{ -24}{ 82}
\datum{ -36}{ 88}
\datum{ -40}{ 90}
\datum{ -48}{ 94}
\datum{ -96}{ 118}
\datum{ -120}{ 130}
\datum{ -160}{ 150}
\datum{ -192}{ 166}
\datum{ -288}{ 214}
\datum{ -324}{ 232}
\datum{ -528}{ 334}
\datum{ 60}{ 42}
\datum{ 48}{ 48}
\datum{ 36}{ 54}
\datum{ 30}{ 57}
\datum{ 24}{ 60}
\datum{ 16}{ 64}
\datum{ 12}{ 66}
\datum{ 8}{ 68}
\datum{ -12}{ 78}
\datum{ -24}{ 84}
\datum{ -48}{ 96}
\datum{ -72}{ 108}
\datum{ -96}{ 120}
\datum{ -120}{ 132}
\datum{ -216}{ 180}
\datum{ -372}{ 258}
\datum{ 60}{ 44}
\datum{ 48}{ 50}
\datum{ 40}{ 54}
\datum{ 24}{ 62}
\datum{ 16}{ 66}
\datum{ 8}{ 70}
\datum{ 0}{ 74}
\datum{ -24}{ 86}
\datum{ -48}{ 98}
\datum{ -64}{ 106}
\datum{ -96}{ 122}
\datum{ -120}{ 134}
\datum{ -192}{ 170}
\datum{ -196}{ 172}
\datum{ -216}{ 182}
\datum{ -276}{ 212}
\datum{ 60}{ 46}
\datum{ 54}{ 49}
\datum{ 48}{ 52}
\datum{ 36}{ 58}
\datum{ 30}{ 61}
\datum{ 24}{ 64}
\datum{ 12}{ 70}
\datum{ 0}{ 76}
\datum{ -12}{ 82}
\datum{ -24}{ 88}
\datum{ -36}{ 94}
\datum{ -48}{ 100}
\datum{ -72}{ 112}
\datum{ -120}{ 136}
\datum{ -144}{ 148}
\datum{ -156}{ 154}
\datum{ -160}{ 156}
\datum{ -252}{ 202}
\datum{ -300}{ 226}
\datum{ -372}{ 262}
\datum{ 72}{ 42}
\datum{ 60}{ 48}
\datum{ 48}{ 54}
\datum{ 36}{ 60}
\datum{ 30}{ 63}
\datum{ 24}{ 66}
\datum{ 12}{ 72}
\datum{ 0}{ 78}
\datum{ -16}{ 86}
\datum{ -24}{ 90}
\datum{ -48}{ 102}
\datum{ -80}{ 118}
\datum{ -96}{ 126}
\datum{ -120}{ 138}
\datum{ -128}{ 142}
\datum{ -144}{ 150}
\datum{ -180}{ 168}
\datum{ -288}{ 222}
\datum{ 72}{ 44}
\datum{ 60}{ 50}
\datum{ 32}{ 64}
\datum{ 20}{ 70}
\datum{ 18}{ 71}
\datum{ 16}{ 72}
\datum{ 0}{ 80}
\datum{ -12}{ 86}
\datum{ -32}{ 96}
\datum{ -48}{ 104}
\datum{ -72}{ 116}
\datum{ -252}{ 206}
\datum{ -300}{ 230}
\datum{ 72}{ 46}
\datum{ 64}{ 50}
\datum{ 60}{ 52}
\datum{ 48}{ 58}
\datum{ 40}{ 62}
\datum{ 24}{ 70}
\datum{ 16}{ 74}
\datum{ 12}{ 76}
\datum{ 0}{ 82}
\datum{ -24}{ 94}
\datum{ -48}{ 106}
\datum{ -60}{ 112}
\datum{ -72}{ 118}
\datum{ -120}{ 142}
\datum{ -192}{ 178}
\datum{ -228}{ 196}
\datum{ -288}{ 226}
\datum{ 72}{ 48}
\datum{ 48}{ 60}
\datum{ 44}{ 62}
\datum{ 12}{ 78}
\datum{ -12}{ 90}
\datum{ -36}{ 102}
\datum{ -48}{ 108}
\datum{ -72}{ 120}
\datum{ -180}{ 174}
\datum{ -216}{ 192}
\datum{ 80}{ 46}
\datum{ 72}{ 50}
\datum{ 64}{ 54}
\datum{ 48}{ 62}
\datum{ 36}{ 68}
\datum{ 32}{ 70}
\datum{ 16}{ 78}
\datum{ 0}{ 86}
\datum{ -16}{ 94}
\datum{ -24}{ 98}
\datum{ -48}{ 110}
\datum{ -56}{ 114}
\datum{ -76}{ 124}
\datum{ -96}{ 134}
\datum{ -176}{ 174}
\datum{ -240}{ 206}
\datum{ 72}{ 52}
\datum{ 60}{ 58}
\datum{ 48}{ 64}
\datum{ 40}{ 68}
\datum{ 36}{ 70}
\datum{ 32}{ 72}
\datum{ 24}{ 76}
\datum{ 12}{ 82}
\datum{ 0}{ 88}
\datum{ -12}{ 94}
\datum{ -24}{ 100}
\datum{ -32}{ 104}
\datum{ -36}{ 106}
\datum{ -72}{ 124}
\datum{ -84}{ 130}
\datum{ -120}{ 148}
\datum{ -140}{ 158}
\datum{ -192}{ 184}
\datum{ 84}{ 48}
\datum{ 72}{ 54}
\datum{ 60}{ 60}
\datum{ 56}{ 62}
\datum{ 52}{ 64}
\datum{ 48}{ 66}
\datum{ 24}{ 78}
\datum{ 0}{ 90}
\datum{ -24}{ 102}
\datum{ -36}{ 108}
\datum{ -40}{ 110}
\datum{ -52}{ 116}
\datum{ -96}{ 138}
\datum{ -168}{ 174}
\datum{ 72}{ 56}
\datum{ 60}{ 62}
\datum{ 48}{ 68}
\datum{ 36}{ 74}
\datum{ 32}{ 76}
\datum{ 24}{ 80}
\datum{ 12}{ 86}
\datum{ 0}{ 92}
\datum{ -72}{ 128}
\datum{ -120}{ 152}
\datum{ -144}{ 164}
\datum{ 80}{ 54}
\datum{ 72}{ 58}
\datum{ 64}{ 62}
\datum{ 52}{ 68}
\datum{ 48}{ 70}
\datum{ 36}{ 76}
\datum{ 32}{ 78}
\datum{ 24}{ 82}
\datum{ 16}{ 86}
\datum{ 0}{ 94}
\datum{ -12}{ 100}
\datum{ -36}{ 112}
\datum{ -48}{ 118}
\datum{ -60}{ 124}
\datum{ -96}{ 142}
\datum{ -168}{ 178}
\datum{ -180}{ 184}
\datum{ -192}{ 190}
\datum{ -480}{ 334}
\datum{ 84}{ 54}
\datum{ 72}{ 60}
\datum{ 56}{ 68}
\datum{ 48}{ 72}
\datum{ 36}{ 78}
\datum{ 24}{ 84}
\datum{ 12}{ 90}
\datum{ -12}{ 102}
\datum{ -40}{ 116}
\datum{ -72}{ 132}
\datum{ -84}{ 138}
\datum{ -96}{ 144}
\datum{ -120}{ 156}
\datum{ -264}{ 228}
\datum{ -276}{ 234}
\datum{ -420}{ 306}
\datum{ 88}{ 54}
\datum{ 80}{ 58}
\datum{ 72}{ 62}
\datum{ 64}{ 66}
\datum{ 60}{ 68}
\datum{ 54}{ 71}
\datum{ 48}{ 74}
\datum{ 40}{ 78}
\datum{ 30}{ 83}
\datum{ 24}{ 86}
\datum{ 0}{ 98}
\datum{ -20}{ 108}
\datum{ -40}{ 118}
\datum{ -48}{ 122}
\datum{ -56}{ 126}
\datum{ -120}{ 158}
\datum{ -160}{ 178}
\datum{ 84}{ 58}
\datum{ 80}{ 60}
\datum{ 78}{ 61}
\datum{ 72}{ 64}
\datum{ 64}{ 68}
\datum{ 48}{ 76}
\datum{ 42}{ 79}
\datum{ 24}{ 88}
\datum{ -24}{ 112}
\datum{ -48}{ 124}
\datum{ -144}{ 172}
\datum{ -168}{ 184}
\datum{ -324}{ 262}
\datum{ 96}{ 54}
\datum{ 80}{ 62}
\datum{ 72}{ 66}
\datum{ 66}{ 69}
\datum{ 56}{ 74}
\datum{ 48}{ 78}
\datum{ 44}{ 80}
\datum{ 36}{ 84}
\datum{ 24}{ 90}
\datum{ 16}{ 94}
\datum{ -12}{ 108}
\datum{ -24}{ 114}
\datum{ -32}{ 118}
\datum{ -144}{ 174}
\datum{ -192}{ 198}
\datum{ 96}{ 56}
\datum{ 84}{ 62}
\datum{ 72}{ 68}
\datum{ 60}{ 74}
\datum{ 48}{ 80}
\datum{ 0}{ 104}
\datum{ -24}{ 116}
\datum{ -72}{ 140}
\datum{ -216}{ 212}
\datum{ 96}{ 58}
\datum{ 84}{ 64}
\datum{ 72}{ 70}
\datum{ 60}{ 76}
\datum{ 48}{ 82}
\datum{ 40}{ 86}
\datum{ 36}{ 88}
\datum{ 24}{ 94}
\datum{ 12}{ 100}
\datum{ 0}{ 106}
\datum{ -72}{ 142}
\datum{ -96}{ 154}
\datum{ -240}{ 226}
\datum{ 108}{ 54}
\datum{ 96}{ 60}
\datum{ 84}{ 66}
\datum{ 80}{ 68}
\datum{ 64}{ 76}
\datum{ 60}{ 78}
\datum{ 32}{ 92}
\datum{ 12}{ 102}
\datum{ -16}{ 116}
\datum{ -24}{ 120}
\datum{ -36}{ 126}
\datum{ -60}{ 138}
\datum{ -72}{ 144}
\datum{ -120}{ 168}
\datum{ 96}{ 62}
\datum{ 80}{ 70}
\datum{ 72}{ 74}
\datum{ 64}{ 78}
\datum{ 48}{ 86}
\datum{ 40}{ 90}
\datum{ 24}{ 98}
\datum{ 0}{ 110}
\datum{ -12}{ 116}
\datum{ -40}{ 130}
\datum{ -48}{ 134}
\datum{ -96}{ 158}
\datum{ -144}{ 182}
\datum{ -192}{ 206}
\datum{ 102}{ 61}
\datum{ 90}{ 67}
\datum{ 88}{ 68}
\datum{ 80}{ 72}
\datum{ 72}{ 76}
\datum{ 64}{ 80}
\datum{ 60}{ 82}
\datum{ 54}{ 85}
\datum{ 48}{ 88}
\datum{ 36}{ 94}
\datum{ 32}{ 96}
\datum{ 24}{ 100}
\datum{ -36}{ 130}
\datum{ -72}{ 148}
\datum{ -108}{ 166}
\datum{ -132}{ 178}
\datum{ -240}{ 232}
\datum{ 96}{ 66}
\datum{ 88}{ 70}
\datum{ 80}{ 74}
\datum{ 72}{ 78}
\datum{ 60}{ 84}
\datum{ 56}{ 86}
\datum{ 48}{ 90}
\datum{ 24}{ 102}
\datum{ 12}{ 108}
\datum{ -48}{ 138}
\datum{ -60}{ 144}
\datum{ 80}{ 76}
\datum{ 72}{ 80}
\datum{ 48}{ 92}
\datum{ -36}{ 134}
\datum{ 102}{ 67}
\datum{ 96}{ 70}
\datum{ 78}{ 79}
\datum{ 72}{ 82}
\datum{ 64}{ 86}
\datum{ 60}{ 88}
\datum{ 48}{ 94}
\datum{ 20}{ 108}
\datum{ 0}{ 118}
\datum{ -24}{ 130}
\datum{ -48}{ 142}
\datum{ -96}{ 166}
\datum{ -144}{ 190}
\datum{ -432}{ 334}
\datum{ 108}{ 66}
\datum{ 104}{ 68}
\datum{ 96}{ 72}
\datum{ 84}{ 78}
\datum{ 72}{ 84}
\datum{ 66}{ 87}
\datum{ 60}{ 90}
\datum{ 48}{ 96}
\datum{ 36}{ 102}
\datum{ 32}{ 104}
\datum{ -24}{ 132}
\datum{ -372}{ 306}
\datum{ 120}{ 62}
\datum{ 108}{ 68}
\datum{ 96}{ 74}
\datum{ 72}{ 86}
\datum{ 64}{ 90}
\datum{ 48}{ 98}
\datum{ 12}{ 116}
\datum{ 0}{ 122}
\datum{ -12}{ 128}
\datum{ -24}{ 134}
\datum{ -48}{ 146}
\datum{ -72}{ 158}
\datum{ -84}{ 164}
\datum{ 120}{ 64}
\datum{ 108}{ 70}
\datum{ 96}{ 76}
\datum{ 92}{ 78}
\datum{ 84}{ 82}
\datum{ 72}{ 88}
\datum{ 48}{ 100}
\datum{ 36}{ 106}
\datum{ 24}{ 112}
\datum{ 16}{ 116}
\datum{ -12}{ 130}
\datum{ -84}{ 166}
\datum{ -108}{ 178}
\datum{ -420}{ 334}
\datum{ 120}{ 66}
\datum{ 112}{ 70}
\datum{ 108}{ 72}
\datum{ 104}{ 74}
\datum{ 96}{ 78}
\datum{ 84}{ 84}
\datum{ 48}{ 102}
\datum{ 36}{ 108}
\datum{ 24}{ 114}
\datum{ 0}{ 126}
\datum{ -288}{ 270}
\datum{ 120}{ 68}
\datum{ 108}{ 74}
\datum{ 96}{ 80}
\datum{ 72}{ 92}
\datum{ 60}{ 98}
\datum{ 48}{ 104}
\datum{ 24}{ 116}
\datum{ -72}{ 164}
\datum{ 120}{ 70}
\datum{ 108}{ 76}
\datum{ 100}{ 80}
\datum{ 96}{ 82}
\datum{ 84}{ 88}
\datum{ 72}{ 94}
\datum{ 66}{ 97}
\datum{ 48}{ 106}
\datum{ 40}{ 110}
\datum{ 36}{ 112}
\datum{ 0}{ 130}
\datum{ -24}{ 142}
\datum{ -84}{ 172}
\datum{ -120}{ 190}
\datum{ -192}{ 226}
\datum{ -264}{ 262}
\datum{ 126}{ 69}
\datum{ 120}{ 72}
\datum{ 112}{ 76}
\datum{ 108}{ 78}
\datum{ 104}{ 80}
\datum{ 102}{ 81}
\datum{ 84}{ 90}
\datum{ 76}{ 94}
\datum{ 72}{ 96}
\datum{ 48}{ 108}
\datum{ 24}{ 120}
\datum{ -84}{ 174}
\datum{ -156}{ 210}
\datum{ -216}{ 240}
\datum{ 120}{ 74}
\datum{ 112}{ 78}
\datum{ 108}{ 80}
\datum{ 102}{ 83}
\datum{ 96}{ 86}
\datum{ 72}{ 98}
\datum{ 48}{ 110}
\datum{ 32}{ 118}
\datum{ 12}{ 128}
\datum{ -48}{ 158}
\datum{ -64}{ 166}
\datum{ -144}{ 206}
\datum{ 120}{ 76}
\datum{ 96}{ 88}
\datum{ 84}{ 94}
\datum{ 80}{ 96}
\datum{ 72}{ 100}
\datum{ 40}{ 116}
\datum{ 12}{ 130}
\datum{ -24}{ 148}
\datum{ -60}{ 166}
\datum{ -108}{ 190}
\datum{ -120}{ 196}
\datum{ -132}{ 202}
\datum{ -180}{ 226}
\datum{ -216}{ 244}
\datum{ 132}{ 72}
\datum{ 128}{ 74}
\datum{ 120}{ 78}
\datum{ 116}{ 80}
\datum{ 108}{ 84}
\datum{ 104}{ 86}
\datum{ 96}{ 90}
\datum{ 90}{ 93}
\datum{ 72}{ 102}
\datum{ 64}{ 106}
\datum{ 40}{ 118}
\datum{ 0}{ 138}
\datum{ -36}{ 156}
\datum{ -56}{ 166}
\datum{ -96}{ 186}
\datum{ 132}{ 74}
\datum{ 120}{ 80}
\datum{ 108}{ 86}
\datum{ 96}{ 92}
\datum{ 84}{ 98}
\datum{ -12}{ 146}
\datum{ 132}{ 76}
\datum{ 126}{ 79}
\datum{ 120}{ 82}
\datum{ 112}{ 86}
\datum{ 108}{ 88}
\datum{ 96}{ 94}
\datum{ 80}{ 102}
\datum{ 60}{ 112}
\datum{ 56}{ 114}
\datum{ 52}{ 116}
\datum{ 48}{ 118}
\datum{ 24}{ 130}
\datum{ 0}{ 142}
\datum{ -48}{ 166}
\datum{ -96}{ 190}
\datum{ -144}{ 214}
\datum{ -396}{ 340}
\datum{ 132}{ 78}
\datum{ 120}{ 84}
\datum{ 108}{ 90}
\datum{ 72}{ 108}
\datum{ 36}{ 126}
\datum{ 24}{ 132}
\datum{ -336}{ 312}
\datum{ 144}{ 74}
\datum{ 136}{ 78}
\datum{ 132}{ 80}
\datum{ 120}{ 86}
\datum{ 96}{ 98}
\datum{ 48}{ 122}
\datum{ 24}{ 134}
\datum{ 144}{ 76}
\datum{ 120}{ 88}
\datum{ 112}{ 92}
\datum{ 108}{ 94}
\datum{ 48}{ 124}
\datum{ 36}{ 130}
\datum{ -24}{ 160}
\datum{ -60}{ 178}
\datum{ -84}{ 190}
\datum{ -240}{ 268}
\datum{ 144}{ 78}
\datum{ 128}{ 86}
\datum{ 126}{ 87}
\datum{ 120}{ 90}
\datum{ 108}{ 96}
\datum{ 96}{ 102}
\datum{ 40}{ 130}
\datum{ 0}{ 150}
\datum{ -12}{ 156}
\datum{ 120}{ 92}
\datum{ 112}{ 96}
\datum{ 72}{ 116}
\datum{ 36}{ 134}
\datum{ 12}{ 146}
\datum{ -120}{ 212}
\datum{ 144}{ 82}
\datum{ 138}{ 85}
\datum{ 136}{ 86}
\datum{ 120}{ 94}
\datum{ 96}{ 106}
\datum{ 72}{ 118}
\datum{ 56}{ 126}
\datum{ -36}{ 172}
\datum{ -156}{ 232}
\datum{ 128}{ 92}
\datum{ 120}{ 96}
\datum{ 96}{ 108}
\datum{ 72}{ 120}
\datum{ -96}{ 204}
\datum{ 144}{ 86}
\datum{ 140}{ 88}
\datum{ 132}{ 92}
\datum{ 120}{ 98}
\datum{ 96}{ 110}
\datum{ 80}{ 118}
\datum{ 48}{ 134}
\datum{ 0}{ 158}
\datum{ -108}{ 212}
\datum{ 144}{ 88}
\datum{ 120}{ 100}
\datum{ 96}{ 112}
\datum{ 72}{ 124}
\datum{ 24}{ 148}
\datum{ -372}{ 346}
\datum{ 144}{ 90}
\datum{ 120}{ 102}
\datum{ 96}{ 114}
\datum{ 76}{ 124}
\datum{ 72}{ 126}
\datum{ 48}{ 138}
\datum{ 12}{ 156}
\datum{ 0}{ 162}
\datum{ -312}{ 318}
\datum{ 144}{ 92}
\datum{ 120}{ 104}
\datum{ 112}{ 108}
\datum{ 72}{ 128}
\datum{ -96}{ 212}
\datum{ 160}{ 86}
\datum{ 156}{ 88}
\datum{ 148}{ 92}
\datum{ 144}{ 94}
\datum{ 138}{ 97}
\datum{ 132}{ 100}
\datum{ 128}{ 102}
\datum{ 120}{ 106}
\datum{ 112}{ 110}
\datum{ 108}{ 112}
\datum{ 96}{ 118}
\datum{ 48}{ 142}
\datum{ 0}{ 166}
\datum{ -60}{ 196}
\datum{ 168}{ 84}
\datum{ 150}{ 93}
\datum{ 144}{ 96}
\datum{ 132}{ 102}
\datum{ 120}{ 108}
\datum{ 108}{ 114}
\datum{ 72}{ 132}
\datum{ 60}{ 138}
\datum{ 160}{ 90}
\datum{ 144}{ 98}
\datum{ 132}{ 104}
\datum{ 120}{ 110}
\datum{ 96}{ 122}
\datum{ 48}{ 146}
\datum{ 168}{ 88}
\datum{ 156}{ 94}
\datum{ 132}{ 106}
\datum{ 128}{ 108}
\datum{ 120}{ 112}
\datum{ 84}{ 130}
\datum{ 24}{ 160}
\datum{ -24}{ 184}
\datum{ -132}{ 238}
\datum{ -168}{ 256}
\datum{ 144}{ 102}
\datum{ 128}{ 110}
\datum{ 108}{ 120}
\datum{ 96}{ 126}
\datum{ 60}{ 144}
\datum{ 36}{ 156}
\datum{ -32}{ 190}
\datum{ 144}{ 104}
\datum{ 128}{ 112}
\datum{ 120}{ 116}
\datum{ 72}{ 140}
\datum{ -72}{ 212}
\datum{ -84}{ 218}
\datum{ 168}{ 94}
\datum{ 160}{ 98}
\datum{ 152}{ 102}
\datum{ 144}{ 106}
\datum{ 120}{ 118}
\datum{ 72}{ 142}
\datum{ 0}{ 178}
\datum{ -48}{ 202}
\datum{ 180}{ 90}
\datum{ 168}{ 96}
\datum{ 144}{ 108}
\datum{ 84}{ 138}
\datum{ 72}{ 144}
\datum{ 168}{ 98}
\datum{ 160}{ 102}
\datum{ 144}{ 110}
\datum{ 120}{ 122}
\datum{ 116}{ 124}
\datum{ 96}{ 134}
\datum{ 48}{ 158}
\datum{ 0}{ 182}
\datum{ 168}{ 100}
\datum{ 156}{ 106}
\datum{ 152}{ 108}
\datum{ 144}{ 112}
\datum{ 120}{ 124}
\datum{ 72}{ 148}
\datum{ -36}{ 202}
\datum{ -216}{ 292}
\datum{ 156}{ 108}
\datum{ 152}{ 110}
\datum{ 128}{ 122}
\datum{ 96}{ 138}
\datum{ -56}{ 214}
\datum{ 180}{ 98}
\datum{ 160}{ 108}
\datum{ 156}{ 110}
\datum{ 120}{ 128}
\datum{ -60}{ 218}
\datum{ -120}{ 248}
\datum{ 176}{ 102}
\datum{ 168}{ 106}
\datum{ 160}{ 110}
\datum{ 156}{ 112}
\datum{ 144}{ 118}
\datum{ 120}{ 130}
\datum{ 96}{ 142}
\datum{ 48}{ 166}
\datum{ 36}{ 172}
\datum{ 0}{ 190}
\datum{ -336}{ 358}
\datum{ 168}{ 108}
\datum{ 156}{ 114}
\datum{ 132}{ 126}
\datum{ 120}{ 132}
\datum{ 96}{ 144}
\datum{ -276}{ 330}
\datum{ 184}{ 102}
\datum{ 176}{ 106}
\datum{ 168}{ 110}
\datum{ 144}{ 122}
\datum{ 136}{ 126}
\datum{ 120}{ 134}
\datum{ 72}{ 158}
\datum{ 56}{ 166}
\datum{ 0}{ 194}
\datum{ -36}{ 212}
\datum{ 180}{ 106}
\datum{ 168}{ 112}
\datum{ 144}{ 124}
\datum{ 120}{ 136}
\datum{ 60}{ 166}
\datum{ -36}{ 214}
\datum{ -180}{ 286}
\datum{ 192}{ 102}
\datum{ 168}{ 114}
\datum{ 156}{ 120}
\datum{ 120}{ 138}
\datum{ 64}{ 166}
\datum{ 168}{ 116}
\datum{ 160}{ 120}
\datum{ 72}{ 164}
\datum{ 200}{ 102}
\datum{ 192}{ 106}
\datum{ 184}{ 110}
\datum{ 180}{ 112}
\datum{ 168}{ 118}
\datum{ 156}{ 124}
\datum{ 96}{ 154}
\datum{ -96}{ 250}
\datum{ 196}{ 106}
\datum{ 184}{ 112}
\datum{ 168}{ 120}
\datum{ 156}{ 126}
\datum{ 136}{ 136}
\datum{ -36}{ 222}
\datum{ 204}{ 104}
\datum{ 200}{ 106}
\datum{ 192}{ 110}
\datum{ 168}{ 122}
\datum{ 152}{ 130}
\datum{ 144}{ 134}
\datum{ 128}{ 142}
\datum{ 96}{ 158}
\datum{ 84}{ 164}
\datum{ 32}{ 190}
\datum{ 0}{ 206}
\datum{ -48}{ 230}
\datum{ 192}{ 112}
\datum{ 180}{ 118}
\datum{ 168}{ 124}
\datum{ 120}{ 148}
\datum{ 84}{ 166}
\datum{ 60}{ 178}
\datum{ 204}{ 108}
\datum{ 192}{ 114}
\datum{ 180}{ 120}
\datum{ 156}{ 132}
\datum{ 144}{ 138}
\datum{ 208}{ 108}
\datum{ 192}{ 116}
\datum{ 168}{ 128}
\datum{ 120}{ 152}
\datum{ 192}{ 118}
\datum{ 96}{ 166}
\datum{ 84}{ 172}
\datum{ 0}{ 214}
\datum{ 120}{ 156}
\datum{ 84}{ 174}
\datum{ -120}{ 276}
\datum{ 192}{ 122}
\datum{ 168}{ 134}
\datum{ 120}{ 158}
\datum{ -120}{ 278}
\datum{ 192}{ 124}
\datum{ 144}{ 148}
\datum{ 108}{ 166}
\datum{ 36}{ 202}
\datum{ 204}{ 120}
\datum{ 144}{ 150}
\datum{ 216}{ 116}
\datum{ 192}{ 128}
\datum{ 168}{ 140}
\datum{ 184}{ 134}
\datum{ 168}{ 142}
\datum{ 60}{ 196}
\datum{ 48}{ 202}
\datum{ 216}{ 120}
\datum{ 204}{ 126}
\datum{ 180}{ 138}
\datum{ 168}{ 144}
\datum{ 140}{ 158}
\datum{ 120}{ 168}
\datum{ 192}{ 134}
\datum{ 160}{ 150}
\datum{ 36}{ 212}
\datum{ -288}{ 374}
\datum{ 216}{ 124}
\datum{ 204}{ 130}
\datum{ 168}{ 148}
\datum{ 156}{ 154}
\datum{ 108}{ 178}
\datum{ 84}{ 190}
\datum{ 36}{ 214}
\datum{ -228}{ 346}
\datum{ 216}{ 126}
\datum{ 200}{ 134}
\datum{ 192}{ 138}
\datum{ 96}{ 186}
\datum{ 216}{ 128}
\datum{ 204}{ 134}
\datum{ 168}{ 152}
\datum{ 160}{ 156}
\datum{ 144}{ 164}
\datum{ 192}{ 142}
\datum{ 96}{ 190}
\datum{ 0}{ 238}
\datum{ -96}{ 286}
\datum{ 228}{ 126}
\datum{ 216}{ 132}
\datum{ 36}{ 222}
\datum{ 232}{ 126}
\datum{ 216}{ 134}
\datum{ 56}{ 214}
\datum{ 0}{ 242}
\datum{ -48}{ 266}
\datum{ 240}{ 124}
\datum{ 228}{ 130}
\datum{ 144}{ 172}
\datum{ 132}{ 178}
\datum{ 108}{ 190}
\datum{ 240}{ 126}
\datum{ 228}{ 132}
\datum{ 192}{ 150}
\datum{ 144}{ 174}
\datum{ 0}{ 246}
\datum{ 180}{ 158}
\datum{ 72}{ 212}
\datum{ 60}{ 218}
\datum{ 232}{ 134}
\datum{ 216}{ 142}
\datum{ 200}{ 150}
\datum{ 192}{ 154}
\datum{ 120}{ 190}
\datum{ 228}{ 138}
\datum{ 216}{ 144}
\datum{ 96}{ 204}
\datum{ 240}{ 134}
\datum{ 228}{ 140}
\datum{ 224}{ 142}
\datum{ 144}{ 182}
\datum{ 48}{ 230}
\datum{ 252}{ 130}
\datum{ 120}{ 196}
\datum{ 240}{ 138}
\datum{ 216}{ 150}
\datum{ 180}{ 168}
\datum{ 168}{ 174}
\datum{ 228}{ 146}
\datum{ 216}{ 152}
\datum{ 96}{ 212}
\datum{ 84}{ 218}
\datum{ 240}{ 142}
\datum{ 192}{ 166}
\datum{ 176}{ 174}
\datum{ 168}{ 178}
\datum{ 144}{ 190}
\datum{ 0}{ 262}
\datum{ 260}{ 134}
\datum{ 240}{ 144}
\datum{ 180}{ 174}
\datum{ 216}{ 158}
\datum{ 192}{ 170}
\datum{ 108}{ 212}
\datum{ 252}{ 142}
\datum{ 168}{ 184}
\datum{ 132}{ 202}
\datum{ 240}{ 150}
\datum{ 196}{ 172}
\datum{ 216}{ 164}
\datum{ 120}{ 212}
\datum{ 232}{ 158}
\datum{ 192}{ 178}
\datum{ 180}{ 184}
\datum{ -240}{ 394}
\datum{ 264}{ 144}
\datum{ 252}{ 150}
\datum{ -180}{ 366}
\datum{ 240}{ 158}
\datum{ 144}{ 206}
\datum{ 264}{ 148}
\datum{ 260}{ 150}
\datum{ 236}{ 162}
\datum{ 192}{ 184}
\datum{ -84}{ 322}
\datum{ 240}{ 162}
\datum{ 216}{ 174}
\datum{ 228}{ 170}
\datum{ 272}{ 150}
\datum{ 240}{ 166}
\datum{ 192}{ 190}
\datum{ 144}{ 214}
\datum{ 0}{ 286}
\datum{ 280}{ 148}
\datum{ 276}{ 150}
\datum{ 252}{ 162}
\datum{ 216}{ 180}
\datum{ 156}{ 210}
\datum{ 288}{ 146}
\datum{ 264}{ 158}
\datum{ 228}{ 176}
\datum{ 216}{ 182}
\datum{ 48}{ 266}
\datum{ 276}{ 154}
\datum{ 204}{ 190}
\datum{ 264}{ 162}
\datum{ 256}{ 166}
\datum{ 240}{ 174}
\datum{ 192}{ 198}
\datum{ 288}{ 152}
\datum{ 264}{ 164}
\datum{ 296}{ 150}
\datum{ 240}{ 178}
\datum{ 96}{ 250}
\datum{ 216}{ 192}
\datum{ 288}{ 158}
\datum{ 256}{ 174}
\datum{ 192}{ 206}
\datum{ 0}{ 302}
\datum{ 264}{ 172}
\datum{ 252}{ 178}
\datum{ 132}{ 238}
\datum{ 288}{ 162}
\datum{ 240}{ 188}
\datum{ 120}{ 248}
\datum{ 288}{ 166}
\datum{ 276}{ 172}
\datum{ 228}{ 196}
\datum{ 156}{ 232}
\datum{ 276}{ 174}
\datum{ 296}{ 166}
\datum{ 288}{ 170}
\datum{ 272}{ 178}
\datum{ 264}{ 184}
\datum{ 300}{ 168}
\datum{ 264}{ 188}
\datum{ 216}{ 212}
\datum{ 312}{ 166}
\datum{ 192}{ 226}
\datum{ 288}{ 182}
\datum{ 240}{ 206}
\datum{ 312}{ 172}
\datum{ 304}{ 176}
\datum{ 252}{ 202}
\datum{ -60}{ 358}
\datum{ 324}{ 168}
\datum{ 300}{ 180}
\datum{ 288}{ 186}
\datum{ 276}{ 192}
\datum{ 288}{ 188}
\datum{ 252}{ 206}
\datum{ 312}{ 178}
\datum{ 288}{ 190}
\datum{ 96}{ 286}
\datum{ 120}{ 276}
\datum{ 120}{ 278}
\datum{ 168}{ 256}
\datum{ 336}{ 178}
\datum{ 324}{ 184}
\datum{ 312}{ 190}
\datum{ 240}{ 226}
\datum{ 312}{ 192}
\datum{ 300}{ 198}
\datum{ 216}{ 240}
\datum{ 288}{ 206}
\datum{ 312}{ 196}
\datum{ 216}{ 244}
\datum{ -156}{ 430}
\datum{ 336}{ 188}
\datum{ 288}{ 214}
\datum{ 0}{ 358}
\datum{ 348}{ 186}
\datum{ 264}{ 228}
\datum{ 336}{ 194}
\datum{ 84}{ 322}
\datum{ 336}{ 198}
\datum{ 288}{ 222}
\datum{ 340}{ 198}
\datum{ 360}{ 190}
\datum{ 336}{ 202}
\datum{ 288}{ 226}
\datum{ 348}{ 198}
\datum{ 276}{ 234}
\datum{ 336}{ 206}
\datum{ 300}{ 226}
\datum{ 180}{ 286}
\datum{ 372}{ 194}
\datum{ 312}{ 224}
\datum{ 300}{ 230}
\datum{ 372}{ 202}
\datum{ 368}{ 204}
\datum{ 60}{ 358}
\datum{ 360}{ 212}
\datum{ 324}{ 232}
\datum{ 264}{ 262}
\datum{ 384}{ 204}
\datum{ 336}{ 230}
\datum{ 216}{ 292}
\datum{ 376}{ 214}
\datum{ 360}{ 228}
\datum{ 384}{ 218}
\datum{ 372}{ 226}
\datum{ 348}{ 238}
\datum{ 288}{ 270}
\datum{ 408}{ 212}
\datum{ 396}{ 222}
\datum{ 324}{ 262}
\datum{ 420}{ 218}
\datum{ 408}{ 224}
\datum{ 384}{ 242}
\datum{ 408}{ 232}
\datum{ 420}{ 230}
\datum{ 384}{ 250}
\datum{ 408}{ 240}
\datum{ 372}{ 258}
\datum{ -60}{ 474}
\datum{ 0}{ 446}
\datum{ 372}{ 262}
\datum{ 432}{ 238}
\datum{ 180}{ 366}
\datum{ 432}{ 242}
\datum{ 396}{ 262}
\datum{ 228}{ 346}
\datum{ 456}{ 234}
\datum{ 276}{ 330}
\datum{ 408}{ 268}
\datum{ 312}{ 318}
\datum{ 456}{ 248}
\datum{ 432}{ 266}
\datum{ 456}{ 256}
\datum{ 480}{ 246}
\datum{ 456}{ 264}
\datum{ 456}{ 272}
\datum{ 492}{ 256}
\datum{ 480}{ 262}
\datum{ 0}{ 502}
\datum{ 60}{ 474}
\datum{ 156}{ 430}
\datum{ 240}{ 394}
\datum{ 420}{ 306}
\datum{ 288}{ 374}
\datum{ 468}{ 286}
\datum{ 480}{ 286}
\datum{ 336}{ 358}
\datum{ 372}{ 346}
\datum{ 396}{ 340}
\datum{ 528}{ 278}
\datum{ 540}{ 274}
\datum{ 528}{ 286}
\datum{ 432}{ 334}
\datum{ 516}{ 302}
\datum{ 480}{ 334}
\datum{ 552}{ 306}
\datum{ 528}{ 334}
\datum{ 564}{ 322}
\datum{ 540}{ 334}
\datum{ 564}{ 330}
\datum{ 564}{ 340}
\datum{ 612}{ 330}
\datum{ 588}{ 346}
\datum{ 624}{ 330}
\datum{ 624}{ 358}
\datum{ 660}{ 366}
\datum{ 672}{ 374}
\datum{ 732}{ 386}
\datum{ 720}{ 394}
\datum{ 804}{ 430}
\datum{ 900}{ 474}
\datum{ 960}{ 502}

\begin{center}
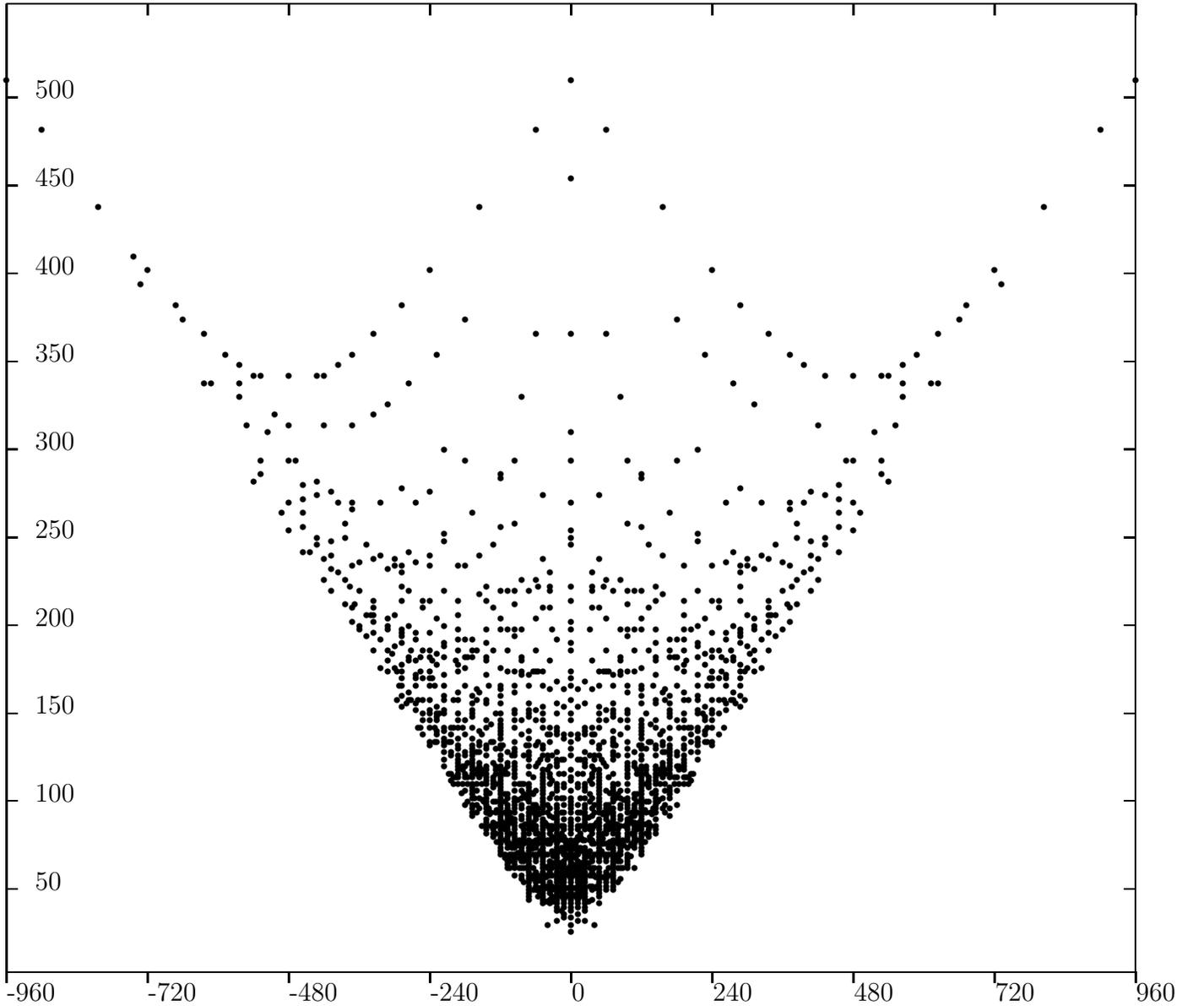
\parbox{6.4truein}{\noindent {\bf Fig. 4}~~{\it A plot of Euler
numbers against
 ${\bar n}_g+n_g$ for the 1900 odd spectra of all the LG potentials
and phase
orbifolds constructed.}}
\end{center}
\end{center}

It is obvious from Fig. 4 that the upper boundary of the distribution of
spectra is the same for the orbifolds as for the complete intersection
manifolds. Very likely this boundary is in fact a
property of the total moduli space of all
three--dimensional Calabi--Yau manifolds.
It would be interesting
to see whether all Calabi--Yau Hodge numbers fall into the limits defined
by the Figures 3 \& 4. As expected the lower part of the plot has shifted
since
new theories with smaller numbers for the total number of fields have
appeared.
It is to be expected that, as the construction becomes more complete, the
structure
of the lower part will change again. In fact there are well known manifolds
that lie below the models presented here; these however involve permutation
groups.


\vfill \eject
\part{ Construction of Landau--Ginzburg Theories and Weighted CICYs.}

\noindent
Even though there is some overlap between the sets of string vacua
described by Landau--Ginzburg vacua on the one hand and Calabi--Yau
manifolds on the other it is not, at present, clear whether the former
is contained in the latter. It is appropriate
justified to separate the discussion of these two classes somewhat.
In Sections 2 and 3 the emphasis will be on the explicit construction
of a set of CY manifolds embedded in weighted $\IP_4$ whereas
the remaining sections of Part I will be concerned
 with the complete class of LG vacua.

\vskip .1truein
\noindent
\section{Complete Intersection Manifolds in Weighted $\IP_4$.}

\noindent
This section contains some elements of the theory of hypersurfaces defined
by polynomials in weighted projective spaces. An extensive discussion
of these
spaces can be found in \cite{wproj}.

A weighted $\IP_4$ with weights $(k_1,k_2,k_3,k_4,k_5)$, which will be
denoted by
$\IP_{(k_1,k_2,k_3,k_4,k_5)}$, is most easily described in terms of 5
complex `homogeneous coordinates' $(z_1,z_2,z_3,z_4,z_5)$, not all zero,
which are subject to the identification
\beq
(z_1,...,z_5) \simeq (\l^{k_1}z_1,...,\l^{k_5}z_5)
\lleq{pro5}
for all nonzero $\l \in \IC$. Thus a weighted projective space is a
generalization
of ordinary projective space and $\IP_4 = \IP_{(1,1,1,1,1)}$ in
this notation.
In the following, when referring to a generic weighted $\IP_4$, we shall
frequently consider the weights to be understood and write $\IP_4$ for
$\IP_{(k_1,k_2,...,k_5)}$.

The first point to note concerning these spaces
is the fact that weighted projective spaces have
orbifold singularities owing to the identification (\ref{pro5}), except for
the case that the weights are all unity. This is most easily seen by setting
$z_j = \left ( \zeta_j \right)^{k_j}$ so that (\ref{pro5}) becomes
$(\zeta_1,...,\zeta_5) \simeq \l (\zeta_1,...,\zeta_5)$. However in virtue
of the definition of $\zeta_i$ we must also identify
$\zeta_j \simeq e^{2\pi i/k_j} \zeta_j$. So we see that
\beq
\IP_{(k_1,k_2,k_3,k_4,k_5)} = \frac{\IP_4}
 {\ZZ_{k_1}\times \cdots \times \ZZ_{k_5}}
\eeq
These identifications lead to singular sets. A simple example is
$\IP_{(1,1,1,2,5)}$ which has weights that are mutually prime. If we
take in turn $\l=-1$ and $\l=\a$ with $\a$ a fifth root of unity in
(\ref{pro5}) then we see that
\begin{eqnarray}
 (z_1,z_2,z_3,z_4,z_5) &\simeq & (-z_1,-z_2,-z_3,z_4,-z_5) \nonumber \\
 &\simeq &(\a z_1,\a z_2,\a z_3,\a^2 z_4,z_5).\llea{ex}
Consider now a neighborhood of the point $(0,0,0,1,0)$. We can take
coordinates on the neighborhood by setting $\l = z_4^{-1/2}$ and
writing $u_j = z_j/z_4^{1/2}$ for $j=1,2,3$ and $u_5 = z_5/z_4^{5/2}$.
We can therefore think of points in the neighborhood as corresponding
to $(u_1,u_2,u_3,1,u_5)$. However from the first of identifications
(\ref{ex})
it follows that
\beq
(u_1,u_2,u_3,1,u_5) \simeq (-u_1,-u_2,-u_3,1,-u_5)
\eeq
so that there is a $\ZZ_2$ identification on the space and hence on a
neighborhood of the point and the action fixes $(0,0,0,1,0)$. In the same
way there is a $\ZZ_5$ action which fixes $(0,0,0,0,1)$. In this case
the singular set consists of points owing to the fact that the weights
are mutually prime. Consider now $\IP_{(1,1,1,2,4)}$ then the
identification
is
\beq
(z_1,z_2,z_3,z_4,z_5) \simeq (\l z_1,\l z_2,\l z_3,\l^2 z_4,\l^4 z_5)
\eeq
If we set $\l = z_5^{-1/4}$ and
choose coordinates $w_j = z_j/z_5^{1/4}$, $w_4 = z_4/z_5^{1/2}$ then due
to the freedom to set $\l = -1$ in (5) we have
\beq
(w_1,w_2,w_3,w_4,1) \simeq (-w_1,-w_2,-w_3,w_4,1)
\eeq
and we see that we have a $\ZZ_2$ action which fixes the curve
$(0,0,0,z_4,z_5)$.
There is in addition a $\ZZ_4$ action with fixed point $(0,0,0,0,1)$ that
lies within this curve. In general there is a fixed point for each
weight that
is greater than unity, a fixed curve for every pair of weights $k_i,k_j$
whose greatest common factor, which we denote by $(k_i,k_j)$, is
greater than
unity, a fixed surface for each triple with $(k_i,k_j,k_l) > 1$ and so on.

We wish to study \cy\ hypersurfaces defined by polynomials in the
homogeneous coordinates. We require the polynomials to be transverse,
that is
$p=0$ and $dp=0$ have no common solution. Given the weights of the ambient
space the requirement of a vanishing first Chern class fixes the degree of
the polynomial as in the case of CICYs. To derive the explicit
condition we digress briefly on the Chern classes of the submanifold
${\cal M}$ defined by the equation $p=0$. We denote by ${\cal T}_{\IP_4}$
and
${\cal T}_{\cal M}$ the tangent spaces to the $\IP_4$ and ${\cal M}$ and by
${\cal N}$ the normal bundle of ${\cal M}$ in $\IP_4$. Proceeding in a
standard manner we have
\beq
{\cal T}_{\IP_4} = {\cal T}_{\cal M} \oplus {\cal N}
\eeq
which allows us to calculate the Chern polynomial of ${\cal M}$ once
those for
$\IP_4$ and ${\cal N}$ are known. The tangent space of $\IP_4$ may be
thought of
as the set of vectors
\beq
{\cal V} = {\cal V}^j \frac{\partial}{\partial z^j}
\eeq
which act on bona fide functions of the homogeneous coordinates, i.e.
functions
of degree zero. From Euler's theorem for homogeneous functions of
degree $m$
we have that
\beq
\sum_j k_j z^j \frac{\partial}{\partial z^j} f = m f.\lleq{euler}
So when acting on functions of degree zero we see that we may regard the
${\cal V}^j$ as independent apart from the identification
${\cal V}^j \simeq {\cal V}^j + k_jz^j$. This reduces the dimension of the
space of ${\cal V}^j$ to 4 as is necessary. It follows that
\beq
{\cal T}_{\IP_4} = {\cal O}(k_1) \oplus \cdots \oplus {\cal O}(k_5)/{\cal O}
\eeq
where ${\cal O}(k)$ is a line bundle with $c_1=kJ$ and $J$ is the K\"ahler
class. It is also the line
bundle whose fibre coordinates transform like a polynomial of degree $k$.
${\cal O}$ is the trivial bundle. Since ${\cal O}(k)$ is one dimensional
we have $c({\cal O}(k))= 1 +kJ$ and hence
\beq
c({\cal T}_{\IP_4}) = \prod_{j=1}^5 \left (1 + k_j J\right ).
\eeq
The defining polynomial $p$ can be regarded as a fibre coordinate on
${\cal N}$ so if $p$ is of degree $d$ we have ${\cal N} = {\cal O}(d)$ and
$c({\cal N}) = 1 + dJ$. Hence
\beq
c({\cal T}_{\cal M}) = \frac{ \prod_{j=1}^5 \left (1 + k_jJ \right )}
 { \left (1 + dJ \right )}
\lleq{ctot}
It follows that $c_1=0$ is the condition
\beq
d = \sum_j k_j~ .\lleq{c1}
We record here also an expression for $c_3$, obtained by expanding
(\ref{ctot})
to third order, which is useful for the
computation of the Euler number of these spaces
\beq
c_3 = - \frac{1}{3} \left (d^3 - \sum_{j=1}^5 k_j^3 \right )J^3~.
\eeq
There are, roughly speaking, two sorts of singularities that can arise
for a hypersurface defined by the vanishing of a polynomial $p$. The first
is that the locus $p=0$ intersects the $\ZZ_n$--singularities of the ambient
space. This does not pose a difficulty however as these can in general be
resolved.

The second is that the weights of a given $\IP_4$ may prevent all
polynomials
of the required degree from being transverse. This is in contradistinction
to CICYs where every configuration admits a smooth
representative (in fact almost all representatives are smooth \cite{gh1}).
To see this let
\beq
d_j:= d-k_j~,~j=1,\dots,5
\eeq
and expand $p$ in powers of $z_1$, say,
\beq
p=\sum_{r=0}^{a_1} C_r(z_m)\,z_1^r~,~~~m\neq 1~.\lleq{pexp}
It follows from Euler's Theorem (\ref{euler}) that $p=0$ and $dp=0$ if and
only if there
is a simultaneous solution of the equations
\beq
\frac{\del p}{\del z_j}=0~,~j=1,\dots,5~.
\lleq{transv5}
Applying this to (\ref{pexp}) we have
\begin{eqnarray}
\frac{\del p}{\del z_1}&=& \sum_{r=1}^{a_1} rC_r(z_m)\,z_1^{r-1} \nonumber \\
\frac{\del p}{\del z_m}&=& \sum_{r=0}^{a_1} \frac{\del C_r}{\del z_m}\,z_1^r~
\llea{partexp}
Now the degrees of the $C_r$ are fixed by (\ref{pexp})
\begin{eqnarray}
\deg\,(C_r) &=& d-rk_1=d_1-(r-1)k_1~,~{\rm for\ } r\ge 1~,\nonumber \\
\deg\left(\frac{\del C_r}{\del z^m}\right) &=&d_m-rk_1~.
\end{eqnarray}
and a polynomial of negative degree is understood to vanish identically.
Unless
at least one of the coefficients in (\ref{partexp}) has degree
precisely zero
then equations (\ref{transv5}) will be satisfied for $z_1=1$ and $z_m=0$
and $p$
 will not be transverse.
This leads to the following neccessary condition on the weights if
 $p$ is to be
transverse:

\noindent
{\sl For each $i$ there must exist a $j$ such that $k_i|d_j$
 ($k_i$ divides $d_j$).}

This condition is quite restrictive. For example it is immediate that
the only
manifolds of the form $\IP_{(1,1,1,1,k)}[k+4]$ that it allows are those
for the
four values $k=1,2,3,4$, and the assiduous reader may care to check
that apart
from these there only eleven other allowed cases of the form
$\IP_{(1,1,1,k,l)}[k+l+3]$, where $k+l+3$ indicates the degree of the
polynomial.

This criterion however is not a sufficient condition. A
counterexample being furnished by the configuration
$\IP_{(1,2,2,2,2)}[9]$ whose most general polynomial is of the form
\beq
p(z_1,z_2,z_3,z_4,z_5) = z_1 \tilde{p}(z_1,z_2,z_3,z_4,z_5).
\eeq
Transversality
requires that the equations $p=0$ and $dp=0$ have no common solution.
Consider
then a neighborhood $U_5$ say, on which we can take $z_5=1$. Taking the
differential gives
$dp =\nobreak \tilde{p}dz_1 + z_1 d\tilde{p}$ which has a zero for $z_1=0$
and $\tilde{p}=0$. These equations
give two constraints in 4 variables and therefore always have a solution.
Since the polynomial constraint is identically satisfied it follows that
this configuration cannot admit a transverse realization.
Thus there are further criteria which must be satisfied in order for a
polynomial to be transverse. In refs. \cite{cls} a list of
transverse polynomials was constructed. A little thought shows that
polynomials of
Fermat type for which $k_i|d_i$ for each $i$ {\sl are} transverse. These
are of the form
\beq
z_1^a + z_2^b + z_3^c + z_4^d + z_5^e~~~~~~~~~~~~~~~~~~~~~~~~~
{\thicklines \begin{picture}(150,30)
 \put(0,0){\circle*{5}}
 \put(0,7){\circle{12}}
 \put(26,0){\circle*{5}}
 \put(26,7){\circle{12}}
 \put(52,0){\circle*{5}}
 \put(52,7){\circle{12}}
 \put(78,0){\circle*{5}}
 \put(78,7){\circle{12}}
 \put(104,0){\circle*{5}}
 \put(104,7){\circle{12}}
 \end{picture}}
\eeq

\noindent
to which we have appended a diagrammatic shorthand.

There are other types which are also always transverse such as
\begin{eqnarray}
z_1^az_2 + z_2^b + z_3^c + z_4^d + z_5^e
& &~~~~~~~~~~~~{\thicklines \begin{picture}(150,30)
 \put(0,0){\circle*{5}}
 \put(0,0){\line(1,0){26}}
 \put(26,0){\circle*{5}}
 \put(26,7){\circle{12}}
 \put(52,0){\circle*{5}}
 \put(52,7){\circle{12}}
 \put(78,0){\circle*{5}}
 \put(78,7){\circle{12}}
 \put(104,0){\circle*{5}}
 \put(104,7){\circle{12}}
 \end{picture}}
\\
z_1^az_2 + z_2^bz_3 + z_3^cz_1 + z_4^d + z_5^e
& &~~~~~~~~~~~~{\thicklines {\begin{picture}(150,30)
 \put(0,0){\circle*{5}}
 \put(0,0){\line(1,0){26}}
 \put(26,0){\circle*{5}}
 \put(26,0){\line(1,0){26}}
 \put(52,0){\circle*{5}}
 \put(26,0){\oval(52,26)[t]}
 \put(78,0){\circle*{5}}
 \put(78,7){\circle{12}}
 \put(104,0){\circle*{5}}
 \put(104,7){\circle{12}}
 \end{picture}}}
\end{eqnarray}
Parts of these expressions corresponding to connected subdiagrams can also
be combined together to produce yet other transverse polynomials such as
\beq
{\thicklines \begin{picture}(150,30)
 \put(0,0){\circle*{5}}
 \put(0,0){\line(1,0){26}}
 \put(26,0){\circle*{5}}
 \put(26,7){\circle{12}}
 \put(52,0){\circle*{5}}
 \put(52,0){\line(1,0){26}}
 \put(78,0){\circle*{5}}
 \put(78,0){\oval(52,26)[t]}
 \put(78,0){\line(1,0){26}}
 \put(104,0){\circle*{5}}
 \end{picture}}
{\rm or}~~~~~~~~~~~~~
{\thicklines \begin{picture}(150,30)
 \put(0,0){\circle*{5}}
 \put(0,0){\line(1,0){26}}
 \put(26,0){\circle*{5}}
 \put(26,7){\circle{12}}
 \put(52,0){\circle*{5}}
 \put(52,0){\line(1,0){26}}
 \put(78,0){\circle*{5}}
 \put(78,7){\circle{12}}
 \put(104,0){\circle*{5}}
 \put(104,7){\circle{12}}
 \end{picture}}
\eeq

What is needed then is a classification of nondegenerate weighted
homogeneous
polynomials in five variables. Unfortunately, there exists as yet no such
classification and indeed its formulation seems to be a hard problem.

There {\it does} exist a classification of smooth polynomials in three
variables \cite{agzv} and what has been done in \cite{cls} is to extend
this to a partial classification of polynomials in five variables.
These constructions do not describe all possible
polynomials but they do respresent a minimal extension of Arnold's
classification to the case of five variables.
Table 1 contains the polynomial types implemented in \cite{cls}
to construct nondegenerate polynomials in five variables. By combining
the nineteen types listed below in the way described above one obtains
thirty different five dimensional catastrophes.

{\scriptsize
{\begin{center}
\begin{tabular}{|| l | l | l ||}
\hline
$\#$ & Polynomial Type & Diagram \tabroom \\
\hline
1 &$z_1^a$
&~~~~~{\thicklines \begin{picture}(150,20)
 \put(0,0){\circle*{5}}
 \put(0,7){\circle{12}}
 \end{picture}
 } \\ [.5mm]
\hline
2 &$z_1^az_2 + z_2^b$
&~~~~~{\thicklines \begin{picture}(150,20)
 \put(0,0){\circle*{5}}
 \put(0,0){\line(1,0){26}}
 \put(26,0){\circle*{5}}
 \put(26,7){\circle{12}}
 \end{picture}
 } \\ [.5mm]
\hline
3 &$z_1^az_2 + z_2^bz_1$
&~~~~~{\thicklines \begin{picture}(150,20)
 \put(0,0){\circle*{5}}
 \put(0,0){\line(1,0){26}}
 \put(26,0){\circle*{5}}
 \put(13,0){\oval(26,26)[t]}
 \end{picture}
 } \\ [.5mm]
\hline
4 &$z_1^az_2 + z_2^bz_3 + z_3^c$
&~~~~~{\thicklines {\begin{picture}(150,20)
 \put(0,0){\circle*{5}}
 \put(0,0){\line(1,0){26}}
 \put(26,0){\circle*{5}}
 \put(26,0){\line(1,0){26}}
 \put(52,0){\circle*{5}}
 \put(52,7){\circle{12}}
 \end{picture}}} \\ [.5mm]
\hline
5 &$z_1^az_2 + z_2^b + z_3^cz_2 + z_1^pz_3^q$
&~~~~~{\thicklines {\begin{picture}(150,20)
 \put(0,0){\circle*{5}}
 \put(0,0){\line(1,0){26}}
 \put(26,0){\circle*{5}}
 \put(26,7){\circle{12}}
 \put(26,0){\line(1,0){26}}
 \put(52,0){\circle*{6}}
 \end{picture}}} \\ [.5mm]
\hline
6 &$z_1^az_2 + z_2^bz_3 + z_3^cz_2 + z_1^pz_3^q$
&~~~~~{\thicklines {\begin{picture}(150,20)
 \put(0,0){\circle*{5}}
 \put(0,0){\line(1,0){26}}
 \put(26,0){\circle*{5}}
 \put(26,0){\line(1,0){26}}
 \put(52,0){\circle*{5}}
 \put(39,0){\oval(26,26)[t]}
 \end{picture}}} \\ [.5mm]
\hline
7 &$z_1^az_2 + z_2^bz_1 + z_3^cz_1$
&~~~~~{\thicklines {\begin{picture}(150,20)
 \put(0,0){\circle*{5}}
 \put(0,0){\line(1,0){26}}
 \put(26,0){\circle*{5}}
 \put(26,0){\line(1,0){26}}
 \put(52,0){\circle*{5}}
 \put(26,0){\oval(52,26)[t]}
 \end{picture}}} \\ [.5mm]
\hline
8 &$z_1^az_2 + z_2^bz_3 + z_3^cz_4 + z_4^d$
&~~~~~{\thicklines {\begin{picture}(150,20)
 \put(0,0){\circle*{5}}
 \put(0,0){\line(1,0){26}}
 \put(26,0){\circle*{5}}
 \put(26,0){\line(1,0){26}}
 \put(52,0){\circle*{5}}
 \put(52,0){\line(1,0){26}}
 \put(78,0){\circle*{5}}
 \put(78,7){\circle{12}}
 \end{picture}}} \\ [.5mm]
\hline
9 &$z_1^az_2 + z_2^bz_3 + z_3^c + z_4^dz_3 + z_2^pz_4^q$
&~~~~~{\thicklines {\begin{picture}(150,20)
 \put(0,0){\circle*{5}}
 \put(0,0){\line(1,0){26}}
 \put(26,0){\circle*{5}}
 \put(26,0){\line(1,0){26}}
 \put(52,0){\circle*{5}}
 \put(52,7){\circle{12}}
 \put(52,0){\line(1,0){26}}
 \put(78,0){\circle*{5}}
 \end{picture}}} \\ [.5mm]
\hline
10 &$z_1^az_2 + z_2^bz_3 + z_3^cz_4 + z_4^dz_3 + z_2^pz_4^q$
&~~~~~{\thicklines {\begin{picture}(150,20)
 \put(0,0){\circle*{5}}
 \put(0,0){\line(1,0){26}}
 \put(26,0){\circle*{5}}
 \put(26,0){\line(1,0){26}}
 \put(52,0){\circle*{5}}
 \put(52,0){\line(1,0){26}}
 \put(78,0){\circle*{5}}
 \put(65,0){\oval(26,26)[t]}
 \end{picture}}} \\ [.5mm]
\hline
11 &$z_1^az_2 + z_2^bz_3 + z_3^cz_4 + z_4^dz_2 + z_1^pz_4^q$
&~~~~~{\thicklines {\begin{picture}(150,20)
 \put(0,0){\circle*{5}}
 \put(0,0){\line(1,0){26}}
 \put(26,0){\circle*{5}}
 \put(26,0){\line(1,0){26}}
 \put(52,0){\circle*{5}}
 \put(52,0){\oval(52,26)[t]}
 \put(52,0){\line(1,0){26}}
 \put(78,0){\circle*{5}}
 \end{picture}}} \\ [.5mm]
\hline
12 &$z_1^az_2 + z_2^bz_3 + z_3^cz_4 + z_4^dz_1$
&~~~~~{\thicklines {\begin{picture}(150,20)
 \put(0,0){\circle*{5}}
 \put(0,0){\line(1,0){26}}
 \put(26,0){\circle*{5}}
 \put(26,0){\line(1,0){26}}
 \put(39,0){\oval(78,26)[t]}
 \put(52,0){\circle*{5}}
 \put(52,0){\line(1,0){26}}
 \put(78,0){\circle*{5}}
 \end{picture}}} \\ [.5mm]
\hline
13 &$z_1^az_2 + z_2^bz_3 + z_3^cz_4 + z_4^dz_5 + z_5^e$
&~~~~~{\thicklines {\begin{picture}(150,20)
 \put(0,0){\circle*{5}}
 \put(0,0){\line(1,0){26}}
 \put(26,0){\circle*{5}}
 \put(26,0){\line(1,0){26}}
 \put(52,0){\circle*{5}}
 \put(52,0){\line(1,0){26}}
 \put(78,0){\circle*{5}}
 \put(78,0){\line(1,0){25}}
 \put(104,0){\circle*{5}}
 \put(104,7){\circle{12}}
 \end{picture}}} \\ [.5mm]
\hline
14 &$z_1^az_2 + z_2^bz_3 + z_3^cz_4 + z_4^d + z_5^ez_4 + z_3^pz_5^q$
&~~~~~{\thicklines {\begin{picture}(150,20)
 \put(0,0){\circle*{5}}
 \put(0,0){\line(1,0){26}}
 \put(26,0){\circle*{5}}
 \put(26,0){\line(1,0){26}}
 \put(52,0){\circle*{5}}
 \put(52,0){\line(1,0){26}}
 \put(78,0){\circle*{5}}
 \put(78,7){\circle{12}}
 \put(78,0){\line(1,0){25}}
 \put(104,0){\circle*{5}}
 \end{picture}}} \\ [.5mm]
\hline
15 &$z_1^az_2 + z_2^bz_3 + z_3^c + z_4^dz_3 + z_5^ez_4 + z_2^pz_4^q$
&~~~~~{\thicklines {\begin{picture}(150,20)
 \put(0,0){\circle*{5}}
 \put(0,0){\line(1,0){26}}
 \put(26,0){\circle*{5}}
 \put(26,0){\line(1,0){26}}
 \put(52,0){\circle*{5}}
 \put(52,7){\circle{12}}
 \put(52,0){\line(1,0){26}}
 \put(78,0){\circle*{5}}
 \put(78,0){\line(1,0){25}}
 \put(104,0){\circle*{5}}
 \end{picture}}} \\ [.5mm]
\hline
16 &$z_1^az_2 + z_2^bz_3 + z_3^cz_4 + z_4^dz_5 + z_5^ez_4 + z_3^pz_5^q$
&~~~~~{\thicklines {\begin{picture}(150,20)
 \put(0,0){\circle*{5}}
 \put(0,0){\line(1,0){26}}
 \put(26,0){\circle*{5}}
 \put(26,0){\line(1,0){26}}
 \put(52,0){\circle*{5}}
 \put(52,0){\line(1,0){26}}
 \put(78,0){\circle*{5}}
 \put(78,0){\line(1,0){25}}
 \put(91,0){\oval(26,26)[t]}
 \put(104,0){\circle*{5}}
 \end{picture}}} \\ [.5mm]
\hline
17 &$z_1^az_2 + z_2^bz_3 + z_3^cz_4 + z_4^dz_5 + z_5^ez_3 + z_2^pz_5^q$
&~~~~~{\thicklines {\begin{picture}(150,20)
 \put(0,0){\circle*{5}}
 \put(0,0){\line(1,0){26}}
 \put(26,0){\circle*{5}}
 \put(26,0){\line(1,0){26}}
 \put(52,0){\circle*{5}}
 \put(52,0){\line(1,0){26}}
 \put(78,0){\circle*{5}}
 \put(78,0){\oval(52,26)[t]}
 \put(78,0){\line(1,0){25}}
 \put(104,0){\circle*{5}}
 \end{picture}}} \\ [.5mm]
\hline
18 &$z_1^az_2 + z_2^bz_3 + z_3^cz_4 + z_4^dz_5 + z_5^ez_2 + z_1^pz_5^q$
&~~~~~{\thicklines {\begin{picture}(150,20)
 \put(0,0){\circle*{5}}
 \put(0,0){\line(1,0){26}}
 \put(26,0){\circle*{5}}
 \put(26,0){\line(1,0){26}}
 \put(52,0){\circle*{5}}
 \put(52,0){\line(1,0){26}}
 \put(65,0){\oval(78,26)[t]}
 \put(78,0){\circle*{5}}
 \put(78,0){\line(1,0){25}}
 \put(104,0){\circle*{5}}
 \end{picture}}} \\ [.5mm]
\hline
19 &$z_1^az_2 + z_2^bz_3 + z_3^cz_4 + z_4^dz_5 + z_5^ez_1$
&~~~~~{\thicklines {\begin{picture}(150,20)
 \put(0,0){\circle*{5}}
 \put(0,0){\line(1,0){26}}
 \put(26,0){\circle*{5}}
 \put(26,0){\line(1,0){26}}
 \put(52,0){\circle*{5}}
 \put(52,0){\oval(104,26)[t]}
 \put(52,0){\line(1,0){26}}
 \put(78,0){\circle*{5}}
 \put(78,0){\line(1,0){25}}
 \put(104,0){\circle*{5}}
 \end{picture}}} \\ [.5mm]
\hline
\end{tabular}
\end{center}
}}

\begin{center}
\parbox{6.4truein}
{\noindent {\bf Table 1.}~~
 {\it The polynomial types that have been implemented}.}
\end{center}

The last problem that confronts the construction of all \cy\ manifolds
in weighted projective space is the question whether the list is finite.
Again we compare the new situation with the one in the case
of CICYs. There the condition of vanishing first Chern class leads to just
one configuration that can be found in the case
of one projective space and one polynomial, namely the quintic
$\IP_4[5]$. For a weighted $\IP_4$ however this condition leads to
the equation (\ref{c1}).
It seems at first that this equation has an infinite number of solutions.
In fact a little thought shows that for each of the thirty polynomial types
there are restrictions on the range of possible weights. For the polynomials
of Fermat type this is already known because of their relation
\cite{kms}\cite{m}\cite{gvw} to
certain types of conformal field theories which have all been constructed
\cite{ls12}\cite{ls3}\cite{lr}\cite{fkss1}. Consider a Fermat polynomial
\beq
p = z_1^{a_1} + z_2^{a_2} + z_3^{a_3} + z_4^{a_4} + z_5^{a_5}
\eeq
with weights $(k_1,....,k_5)$ and degree
\beq
d= k_1a_1 = k_2a_2 = k_3a_3 = k_4a_4 = k_4a_5~.
\eeq
The condition of vanishing first Chern class becomes
\beq
1 = \sum_{i=1}^5 \frac{1}{a_i} \lleq{cycon1}
It is possible to iteratively bound the $a_i$, which we take to be
ordered such that $a_i\leq a_{i+1}$, in virtue of the fact that
$\sum_i \frac{1}{a_i}$ is a decreasing function of all its arguments.
The smallest possible value for $a_1$ is 2.
The next step consists in finding the smallest possible
value for $a_2$. Using the lower bound on $a_1$ condition (\ref{cycon1})
becomes
\beq
\frac{1}{2} = \sum_{i=2}^5 \frac{1}{a_i},
\eeq
from which it follows that the smallest possible value for
$a_2$ is 3. Proceeding in this manner we end up with
the condition
\beq
\frac{1}{42} - \frac{1}{43} = \frac{1}{a_5}
\eeq
from which we find $a_5 = 1806$ which turns out to be the highest power
that arises in our polynomials.

As a second example consider polynomials of type 2,
\beq
p = z_1^{a_1} + z_2^{a_2} + z_3^{a_3} + z_4^{a_4} + z_5^{a_5}z_4.
\eeq
In this case we have
\beq
d= k_1a_1 = k_2a_2 = k_3a_3 = k_4a_4 = k_4 + k_5a_5.
\eeq
and the condition of vanishing first Chern class is now
\beq
1 = \sum_{i=1}^5 \frac{1}{a_i} - \frac{1}{a_4a_5}
\lleq{c1con2}

Taking again $a_1=2$ we proceed as above. Using the lower bound on $a_1$
condition (\ref{c1con2}) becomes
\beq
\frac{1}{2} = \sum_{i=2}^5 \frac{1}{a_i} - \frac{1}{a_4a_5}.
\eeq
{}From this equation it follows that the smallest possible value for
$a_2$ is 3. At the next stage we find $a_3=7$ and then $a_4=43$. Finally
we find the condition
\beq
\frac{1}{42\cdot 43} = \frac{1}{a_5} - \frac{1}{43\cdot a_5}
\eeq
whence $a_5 = 42^2 = 1764$. In a similar way we find constraints
on all other types of polynomials.

\vskip .1truein
\noindent
\section{Computation of the Spectrum}

Having constructed these 6,500 odd spaces one wants, of course, to know
 about the spectrum, especially about the number of light
generations. There are several methods available to compute
these numbers. First there is of course the geometrical analysis that can
be used to compute the independent Hodge numbers of a Calabi--Yau manifold.
Another method that is more useful for the class of spaces at hand are
techniques for computing the spectrum in the framework of Landau--Ginzburg
theories. In those cases for which an exactly solvable theory corresponding
to the model is known, techniques from conformal field theory are available
as well. Unfortunately for most of the theories constructed in the previous
section no exactly solvable model is known and hence the tools from
conformal field theory, even though most powerful, are not available here.

The very first step in a systematic analysis is, of course,
the determination
of the number of light generations, i.e. the computation of
the Euler number.
 This can be done by computing the integral
of the third Chern class using the fact that
$\int J^3 = 1/\prod k_j$ and taking into
account the contributions from the singularities \cite{yau}
\beq
\chi = -\frac{\frac{1}{3}\left (d^3 - \sum k_j^3 \right)d}{\prod k_j}
 - \sum_i \frac{\chi (S_i)}{n_i} + \sum_i n_i \chi (S_i)
\eeq
where $\chi (S_i)$ is the Euler number of the singular set $S_i$ and
$n_i$ is the order of the cyclic symmetry group $\ZZ_{n_i}$.

This can be illustrated with an example. Consider the manifold
\beq
\IP_{(4,4,5,5,7)}[25] ~~~~~~~~~
{\thicklines \begin{picture}(150,20)
 \put(0,0){\circle*{5}}
 \put(0,0){\line(1,0){26}}
 \put(26,0){\circle*{5}}
 \put(26,7){\circle{12}}
 \put(52,0){\circle*{5}}
 \put(52,0){\line(1,0){26}}
 \put(78,0){\circle*{5}}
 \put(78,0){\line(1,0){26}}
 \put(104,0){\circle*{5}}
 \put(104,7){\circle{12}}

 \end{picture}}.
\lleq{examp}
In this example we have three types of singular sets; first there is a
$\ZZ_4$--curve $C$ with Euler number $\chi (C) = 2$. Next there are five
$\ZZ_5$--points and finally the $\ZZ_7$ leads to one additional fixed point.
Put together with $\chi_s = -44\frac{5}{14}$ for the Euler number of the
singular space this gives $\chi = -6$ for this space.

It is clear from this example that the geometrical technique is rather
awkward to apply systematically as it entails a detailed analysis of the
singular sets $S_i$ for each manifold. These not only depend on the
divisibility
property of the weights but also on the type of the individual catastrophe
involved and therefore need to be determined on a case by case basis. This
is not easily automated.

It is much easier to use a result of Vafa \cite{v} on the Euler number of
$c=9$, $N=2$ Landau--Ginzburg models. His formula specialized to the
case at hand yields
\beq
\chi = \frac{1}{d} \sum_{l,r = 0}^{d-1}~~
 (-1)^{r+l+d}\prod_{lq_i,rq_i \in \ZZ} \frac{d-k_i}{k_i},
\eeq
where $q_i = k_i/d$.
Computing the Euler number for all 10,839 odd spaces leads to the results
of Figure 3. As already mentioned the Euler number $-960\leq \chi \leq 960$.
Among these
spaces are many that lead to 2, 3 and 4 light generations.

The next question then is how the Euler number actually splits up
into generations and antigenerations. Again it is possible to use both,
manifold techniques and Landau--Ginzburg type methods. As we have already
easy methods to compute the Euler number $\chi = 2(h^{(1,1)} - h^{(2,1)})$
we only need to compute either the number of generations of the number of
antigenerations to have the complete Hodge diamond.

We consider first the geometrical techniques and compute the number of
antigenerations. In a projective space there would be nothing to compute since
in this case the dimensions of this cohomology group is always one,
its only contribution coming from the K\"ahler form. In the case of weighted
projective spaces the K\"ahler form of course is not the only contribution
because the blow--ups of the singular sets introduce new (1,1)--forms.
The singular sets consist of points and/or curves. The techniques for blowing
up points have been discussed in \cite{ry} and the contributions
coming from blowing up curves have been discussed in \cite{s3}.
These are in fact the
only types of singularity that can arise if $p$ is transverse. In other words
that singular subsets of $M$ cannot have dimension 2 or 3.
First we show that the embedding $\IP_4$ cannot have singular sets of
dimension
greater than or equal to 3. Recall that the $\IP_4$ has a singular point for
each
$k_i>1$, a singular curve for each pair with $(k_i,k_j)>1$, a singular subset
of dimension 2 for each triple with $(k_i,k_j,k_l)>1$ \etc. In the definition
of $\IP_4$ we have assumed that the weights have no common factor so there
are no
singular sets of dimension 4. Consider the possibility of a singular set of
dimension 3. Such a set would correspond to weights such that
$$(k_2,k_3,k_4,k_5)=m>1~~{\rm but}~~m\not|k_1~.$$
Since $m$ does not divide $k_1$ but does divide the other $k$'s it cannot
divide
the degree $d=\sum_{j=1}^5k_j$. Every monomial of degree $d$ must
therefore contain at least one factor of $z_1$. Thus
$p(z)=z_1\tilde{p}(z)$ and
so is not transverse. On the other hand singular sets of dimension 2
can occur
and these will generically intersect $M$ in subsets of dimension 1.
It remains
to
show that a singular subset of dimension 2 cannot lie within $M$. To
this end
suppose there is a fixed point set of dimension 2 for the identification
(\ref{pro5})
and that this subset lies within $M$. We may choose coordinates $(x,y,z)$
such that, locally, the fixed point set is $(x,y,0)$. Suppose the
identification is represented by a matrix $A$. Since $A$ fixes $(x,y,0)$ it
has the form $$A=\pmatrix{1&0&a\cr 0&1&b\cr 0&0&c}~.$$
The three--form $dx\wedge dy\wedge dz$ transforms as
$$dx\wedge dy\wedge dz\longrightarrow\det A\,dx\wedge dy\wedge dz$$
we must have $\det\,A=1$ and hence $c=1$. Since $A$ is contained in a finite
group $A^n=1$ for some $n$, however
$$A^n=\pmatrix{1&0&na\cr 0&1&nb\cr 0&0&1}$$
so $a$ and $b$ must, in fact, vanish. Thus the only identification that
fixes a two--dimensional subset is the identity.

When resolving the singularities it needs to be checked that the blown up
manifolds are still Calabi--Yau manifolds. For the case of singular points
this
has been discussed in ref. \cite{ry} whereas the case of singular curves
was first analyzed in ref. \cite{s3}.

In order to resolve curves consider the action of a $\ZZ_n$ on a weighted
CICY leaving a curve invariant. In the three--dimensional Calabi--Yau manifold
the normal bundle of this curve has fibres $\IC_2$ and therefore this discrete
group induces an action on the fibres described by the matrix
\beq
\left (\matrix{~~\alpha^{mq}&0\cr 0&~~\alpha^m\cr} \right ) \lleq{normact}
where $\a$ is an n$^{th}$ root of unity, $0\leq m \leq n$, and $q,n$ have
no common divisor. This action has an isolated singularity which needs
to be resolved. The essential point is that the singularity of
$\IC_2/\ZZ_n$ can be described as the singular set of the surface
\begin{equation}
S :~~ z_3^n = z_1z_2^{n-q}
\end{equation}
in $\IC_3$ and therefore can be resolved by a construction that
is completely determined by the type $(n,q)$ of the action through the method
of continued fractions.
The expansion of $\frac{n}{q}$ in a continued fraction
\begin{eqnarray}
\frac{n}{q} &=& [b_1,...,b_s] \nonumber \\
 &=& b_1 - \frac{1}{b_2 - \frac{\Large 1}{ \ddots -
 \frac{\Large 1}{\Large b_s} }}
\end{eqnarray}

\noindent
determines the numbers $b_i$ which specify uniquely the plumbing process
which replaces the singularity. Furthermore, it also determines the additional
generators of the cohomology, since the number
of $\IP_1$'s necessary to resolve the singularity is precisely the number of
steps needed in the evaluation of $\frac{n}{q}= [b_1,..,b_s]$.
The reason for this is that the singularity is replaced by a bundle which
is constructed of $s+1$ patches with $s$ transition functions that are
specified by the $b_i$'s. Each of these glueing steps introduces a sphere
which in turn
supports a (1,1)--form. A standard shorthand notation for the geometry of
the blow--up is through what is called a Hirzebruch--Jung
tree \cite{hirz} which in the case of the blow--up of a $\ZZ_n$ action is
just an
SU($s+1$) Dynkin diagram with the negative values of the $b_i's$ attached to
the nodes. Each node in the diagram corresponds to a sphere and the diagram
shows which spheres intersect each other (in the case of the $\ZZ_n$--action
only the neighboring spheres intersect) whereas the $b_i$ determine the
intersection numbers.

In order to show that these blow--up procedures can be applied, we have to
show that the determinant (\ref{normact}) is always 1. This can be done by
checking the invariance of the holomorphic threeform $\Om$ under $\ZZ_n$.

Consider first manifolds of Fermat type. A singular curve in such a manifold
is signalled by three weights of the ambient space that are not coprime
\beq
(k_1,k_2,k_3) =n > 1
\eeq
($n$ does not divide $k_4,k_5$). The integer $n$ defines a $\ZZ_n$ discrete
group. The action of $\ZZ_n$ is given by
\beq
(z_1,z_2,z_3,z_4,z_5) \mapsto (z_1,z_2,z_3,\a^{k_4} z_4, \a^{k_5} z_5),
\eeq
where $\a$ is an $n$th root of unity. In order to show that the blown--up
manifold is still of Calabi--Yau type one needs to show that $\Om$
is invariant. In this case the action of $\ZZ_n$ on $\Om$ is
\beq
\Om \mapsto \a^{k_4+k_5} \Om.
\eeq
The condition for $\Om$ to be invariant is, therefore, that
\beq
\a^{k_4+k_5}=1,
\eeq
which is always true since for Fermat type polynomials
\beq
1=\a^d = \a^{\sum k_i} =\a^{k_4+k_5}.
\eeq

Now suppose the curve is embedded in a non--Fermat Calabi--Yau manifold.
Then
the possibility exists that there are only two weights for which
\beq
(k_1,k_2) =n > 1
\eeq
($n$ does not divide $k_3,k_4,k_5$), i.e. $k_1,k_2$ are not constrained by
the polynomial. In this case the coordinates parametrizing the curve occur
as
\beq
z_1^{l_1}z_p,~~~z_2^{l_2}z_q.
\eeq
The action of $\ZZ_n$ is given by
\beq
(z_1,z_2,z_3,z_4,z_5) \mapsto (z_1,z_2,\a^{k_2} z_3,\a^{k_4} z_4, \a^{k_5}
z_5),
\eeq
and the condition for $\Om$ to be invariant becomes
\beq
\a^{k_3+k_4+k_5}=\a^{k_p}=\a^{k_q}.
\eeq
As before
\beq
d=k_1l_1+k_p=k_2l_2+k_q=nr_1l_1+k+_p=nr_2l_2+k_q
\eeq
hence
\beq
\a^d = \a^{ k_3+k_4+k_5} =\a^{k_p}=\a^{k_q}
\eeq
which was to be shown.

Returning now to the discussion of the example $\IP_{(4,4,5,5,7)}[25]$, we
find from the results of ref. \cite{ry} that a $\ZZ_5$--point blow--up
of this
manifold contributes two (1,1)--forms, whereas a $\ZZ_7$--point blow--up
leads
to three additional generators of the second cohomology. Since
the $\ZZ_5$
singular set $\IP_1[5]$ consists of five points and the $\ZZ_7$ singular
set
consists just of one point all point blow--ups together contribute a total
of
thirteen (1,1)--forms. To find out how many (1,1)--forms the blow--up of
the
$\ZZ_4$--curve contributes we need to check the induced action of this
discrete group action on the normal bundle of the curve \cite{s3}.

In our example (\ref{examp}) this induced action
\beq
\IC_2 \ni (z_1,z_2) \longmapsto (\alpha z_1, \alpha^3 z_2)
\eeq
is of type $(n,q)=(4,3)$ and therefore the singularity of $\IC_2/\ZZ_4$
is equivalent to the singular set of the surface
\beq
 z_3^4 = z_1z_2 .
\eeq
The resolution is specified by the Hirzebruch--Jung tree
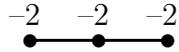
\begin{center}
{\thicklines {\begin{picture}(60,15)
 \put(0,6){--2}
 \put(8,0){\circle*{5}}
 \put(8,0){\line(1,0){26}}
 \put(26,6){--2}
 \put(34,0){\circle*{5}}
 \put(34,0){\line(1,0){26}}
 \put(52,6){--2}
 \put(60,0){\circle*{5}}
 \end{picture}}}
\end{center}
which in turn is determined completely by the continued fraction\beq
\frac{4}{3} = 2 - \frac{1}{2-\frac{1}{2}}.
\eeq
Therefore the blow--up of the curve contributes three more (1,1)--forms.
Taking into account the K\"ahler form of the ambient space then leads
to a total of 17 (1,1)--forms for this manifold.

As is evident however the manifold technique is awkward to implement in a
systematic
way and is better used as an independent check of the Landau--Ginzburg
type formulation of this problem by Vafa who constructs a Poincar\'e--type
polynomial for the $l^{\rm th}$ twisted sector
\beq
Tr_l ((t\bt)^{dJ_0})
= t^{d\left (Q_l+\frac{1}{6}c_T \right )}
 \bt^{d\left (-Q_l+\frac{1}{6}c_T \right )}
 \prod_{lq_i\in \ZZ}
 \left ( \frac{1- (t\bt)^{d-k_i}}{1-(t\bt)^{k_i}}\right)
\eeq
with
\begin{eqnarray}
 Q_l &=& \sum_{lq_i {\footnotesize{\not \in}} \ZZ}
 \left (lq_i - [lq_i] - \frac{1}{2} \right )~,\nonumber \\
 \frac{1}{6}c_T &=& \sum_{lq_i {\footnotesize{\not \in}} \ZZ}
 \left (\frac{1}{2} - q_i\right )
\end{eqnarray}
where $t$ and $\bt$ are formal variables, $d$ is the degree of the
Landau--Ginzburg
potential, the $q_i=k_i/d$ are the normalized weights of the fields and
$[lq_i]$ is the integer part of $lq_i$.
Expanding this polynomial in powers in $t$ and $\bt$ it is possible to read
off the contributions to the various cohomology groups from the different
sectors of the twisted LG--theory. The (2,1) forms for example are given by
the number of fields with charge (1,1), i.e. the coefficient of $(t\bt)^d$.
In general, the number of $(p,q)$ forms are given by the coefficients of
$t^{(3-p)d}\bt^{qd}$ in the Poincar\'e polynomials summed over all sectors
$l=0,1,\dots,d-1$.

\vskip .1truein
\noindent
\section{Landau--Ginzburg Vacua}

\noindent
In this section we briefly review the construction of all
 Landau--Ginzburg vacua based on superpotentials with an isolated
singularity at the origin.
Consider a string ground state based on a Landau--Ginzburg theory
which we
assume to be $N=2$ supersymmetric since we demand $N=1$ spacetime
supersymmetry.
Using a superspace formulation in terms of the coordinates
$(z,\bz,\th^+,\bth^+,\th^-,\bth^-)$ the action takes the form
\beq
\cA = \int d^2zd^4\th~K(\Phi_i,\bPhi_i) + \int d^2zd^2\th^- ~W(\Phi_i)
 + \int d^2zd^2\th^+ ~W(\bPhi_i)
\eeq
where $K$ is the K\"ahler potential and the superpotential $W$ is a
holomorphic function of the chiral superfields $\Phi_i$.
Since the ground states of the bosonic potential are the critical points
of the superpotential of the LG theory
 we demand their existence. The type of critical points we
need is determined by the fact that we wish to keep the fermions in
the theory massless; hence we assume that the critical points are
completely degenerate. Furthermore we require that all critical points
be isolated, since we wish to relate the finite dimensional ring of
monomials associated to such a singularity with the chiral ring of
physical states in the Landau--Ginzburg theory, in order to construct
the spectrum of the corresponding string vacuum.
Finally we demand that the theory is conformally invariant; from this
follows, relying on some assumptions regarding the renormalization
properties of the theory, that the Landau--Ginzburg potential is
quasihomogeneous. In other words, we require to be able to assign,
to each field
$\Phi_i$, a weight $q_i$ such that for any non--zero complex number
$\l \in \IC^{\star}$
\beq
W(\l^{q_1}\Phi_1,\dots,\l^{q_n}\Phi_n) =\l W(\Phi_1,\dots,\Phi_n).
\eeq
Thus we have formulated the class of potentials that we need to consider:
quasihomogeneous polynomials that have an isolated, completely
degenerate singularity (which we can always shift to the origin).

Associated to each of the superpotentials, $W(\Phi_i)$ is a
so--called catastrophe which is obtained by first truncating the
superfield $\Phi_i$ to its lowest bosonic component
$\phi_i(z,\bz)$, and then going to the
field theoretic limit of the string by assuming $\phi_i$ to be constant
$\phi_i=z_i$. Writing the weights as $q_i = k_i/d$, we will denote by
\beq
\IC_{(k_1,k_2,\dots,k_n)}[d]
\eeq
the set of all catastrophes described by the zero locus of polynomials
of degree $d$ in variables $z_i$ of weight $k_i$.

The affine varieties described by these polynomials are not compact
and hence it is necessary to implement a projection in order to
compactify these spaces. In Landau--Ginzburg language, this amounts to an
orbifolding of the theory with respect to a discrete group $\ZZ_d$ the
order
of which is the degree of the LG potential \cite{v}. The spectrum of the
orbifold theory will contain twisted states which, together with
the monomial ring of the potential, describe the complete spectrum of the
corresponding Calabi--Yau manifold. We will denote the orbifold
of a Landau--Ginzburg theory by
\beq
\IC^{\star}_{(k_1,k_2,\dots,k_n)}[d]
\eeq
and call it a configuration.

In manifold speak the projection should lead to a three--dimensional
K\"ahler manifold, with vanishing first Chern class. For a general
Landau--Ginzburg theory no unambiguous universal prescription for doing
so has been found and, as we will describe in Section 4, none can exist.
One way to compactify amounts to simply imposing projective
equivalence
\beq
(z_1,....,z_n) \equiv (\l^{k_1} z_1,.....,\l^{k_n} z_n)
\lleq{projn}
which embeds the hypersurface described by the zero locus of the
polynomial into a weighted projective space $\IP_{(k_1,k_2,\dots,k_n)}$
with weights $k_i$. The set of hypersurfaces of degree $d$ embedded
in weighted projective space will be denoted by
\beq
\IP_{(k_1,k_2,\dots,k_n)}[d].
\eeq
For a potential with five scaling variables this
construction is completely sufficient in order to pass from the
Landau--Ginzburg theory to a string vacuum \cite{m}\cite{gvw} provided
$d=\sum_{i=1}^5 k_i$, which is the condition that these hypersurfaces
have vanishing first Chern class. For more than five variables, however,
this type of compactification does not lead to a string vacuum.

Even though the precise relation between LG theories and CY manifolds
is not known for the most general case certain facts are known.
Since LG theories with five variables describe a CY manifold embedded in
a 4 complex dimensional weighted projective space one might expect
e.g. that LG potentials with 6, 7, etc., variables describe manifolds
embedded in 5, 6, etc., dimensional weighted projective spaces. This is
not correct.

In fact none of the models with more than five variables is related
to manifolds embedded in one weighted projective space. Instead they
describe Calabi--Yau manifolds embedded in products of weighted
projective space. A simple example is
furnished by the LG potential in six variables
\beq
W=\Phi_1 \Psi_1^2+\Phi_2 \Psi_2^2+\sum_{i=1}^3 \Phi_i^{12}+
\Phi_4^3
\eeq
which corresponds to the exactly solvable model described by the
tensor product of $N=2$ minimal theories at the levels
\beq
(22^2 \cdot 10\cdot 1)_{D^2\cdot A^2},
\eeq
where the subscripts indicate the affine invariants chosen for the
individual factors. This theory belongs to the LG configuration
\beq
\IC^{\star}_{(2,11,2,11,2,8)}[24]^{(3,243)}_{-480}
\lleq{lgform}
and is equivalent to the weighted complete intersection Calabi--Yau (CICY)
 manifold in the configuration
\beq
\matrix{\IP_{(1,1,1,4,6)}\cr \IP_{(1,1)}\hfill\cr}
 \left [\matrix{1&12\cr 2&0\cr}\right]^{(3,243)}_{-480}
\lleq{cyform}
described by the intersection of the zero locus of the two potentials
\bea
p_1 &=& x_1^2y_1+x_2^2y_2 \nonumber \\
p_2 &=& y_1^{12}+y_2^{12}+y_3^{12}+y_4^3+y_5^2.
\eea
Here we have added a trivial factor $\Phi_5^2$ to the potential and
taken the field theory limit via $\phi_i(z,\bz)=y_i$, where $\phi_i$
 is the lowest component of the chiral superfield $\Phi_i$. The first
column in the degree matrix (\ref{cyform}) indicates that the first
polynomial is of bidegree (2,1) in the coordinates $(x_i,y_j)$
of the product of the projective line $\IP_1$ and the weighted
projective space $\IP_{(1,1,1,4,6)}$ respectively, whereas the second
column shows that the second polynomial is independent of the
projective line and of degree 12 in the coordinates of the
weighted $\IP_4$. The superscripts in (\ref{lgform}) and (\ref{cyform})
describe the dimensions $(h^{(1,1)},h^{(2,1)})$ of the fields
corresponding to the cohomology groups $(H^{(1,1)},H^{(2,1)})$,
whereas the subscript is the Euler number of the configuration.

It should be noted however that Landau--Ginzburg potentials in six
variables do not describe the most general complete intersection in
products of weighted spaces, simply because not all of these manifolds
involve trivial factors, or put differently, quadratic monomials.
A simple example is the manifold that corresponds to the
Landau--Ginzburg theory
$(1\cdot 16^3)_{A\cdot E_7^3}$ with LG potential
\beq
W= \sum_{i=1}^3 \left(\Phi_i^3 + \Phi_i \Psi_i^3\right) + \Phi_4^3.
\eeq
This theory describes a
codimension--2 Calabi--Yau manifold embedded in
\beq
\matrix{\IP_2\cr \IP_3\cr} \left[\matrix{3 &0\cr
 1 &3\cr}\right].
\eeq
This space has 8 (1,1)--forms and 35 (2,1)--forms which correspond to
the possible complex deformations in the two polynomials
$p_1,p_2$ \cite{s1}.

Associated to this Calabi--Yau manifold in a product of ordinary
projective
spaces is an auxiliary algebraic manifold in a weighted six--dimensional
projective space
\beq
\IP_{(2,2,2,3,3,3,3)}[9].
\eeq
obtained via the na\"{\i}ve compactification (\ref{projn}) where the first
three
coordinates come from the fields $\Psi_i$ and the last four come from
the
$\Phi_i$. We want to compute the number of complex deformations of this
manifold, i.e. we want to compute the number of monomials of charge 1.

The most general monomial is of the form $\prod_i \Phi_i \prod \Psi_j$.
It is easy to show explicitly that there are precisely 35 monomials by
writing them down but the following remarks may suffice. There are
four different types of possible monomials, depending on whether they
contain the fields $\Phi_i$ not at all (I), linearly (II),
quadratically (III) or cubically (IV). First note
that monomials of type (I) do not contribute to the marginal operators.
Monomials of type (II) have to contain cubic monomials in the $\Psi$.
Since there are three fields $\Psi_i$ available, we obtain a total of
$ 40$ marginal operators of this type.
Because of the equation of motion nine of these are in fact in the ideal
and we are left with 31 complex deformations of this type. Monomials
quadratic in the $\Phi_i's$ again do not contribute, whereas those cubic
in the $\Phi_i$ fields contribute the remaining 4. Indeed, there are
20 cubic monomials in terms of the four $\Phi_i$ but using the equations
of motion one finds that 16 of those are in the ideal.

In other words, for the total of $60=40+20$ monomials of degree
9 (or charge 1) it is possible to fix the coefficients of 9 of the
40 by allowing linear field redefinitions of the first three coordinates
and also to fix the coefficients of 16 of the 20 via linear field
redefinitions of the last four coordinates. Hence even though the ambient
space is singular and the manifold hits these singularities in a $\IP_2$
and in a cubic surface $\IP_3[3]$ the resolution does not contribute any
complex deformations because these surfaces are simply connected.

It should be emphasized that this manifold is not the physical internal
part
of a string ground state, but that it plays an auxiliary role, which
allows to
discuss just one particular sector of the string vacuum, namely the
complex deformations.

 Before turning to the problem of constructing LG configurations
satisfying the constraints described above, we wish to make some remarks
regarding the validity of the
requirements formulated in the previous paragraph.

Even though the assumptions formulated in refs. \cite{m}\cite{vw} and
reviewed above seem rather reasonable, and previous work shows that the
set of such Landau--Ginzburg theories certainly is an interesting and
quite extensive class of models, it is clear that it is not the most
general class of (2,2) vacua. Although it provides a rather large set
of different models
\fnote{2}{The rather extravagant values that have been
 mentioned in the literature as the number of possible (2,2)
 vacua are based on extrapolations that do not take into account
 the problem of overcounting that is generic to all of these
 different constructions.},
 which contains many classes of previously constructed vacua
\fnote{3}{Such as vacua constructed tensor models based on the ADE minimal
 models \cite{m}\cite{gvw}\cite{ls12}\cite{ls3}\cite{fkss1}\cite{sy}
 and \break
 level--1 Kazama--Suzuki models \cite{lvwg}\cite{fiq2}\cite{s},
 as well as G/H LG theories \cite{ls5} related to Kazama--Suzuki
 models of higher levels.},
there are known vacua which cannot be described in this framework.

Perhaps the simplest example that does not fit into the classification
 above is that of the Calabi--Yau manifolds in
\beq
\IP_5[4~~2],
\eeq
described by the intersection of two hypersurfaces defined by a quartic
and a quadratic
polynomial in a five--dimensional
projective space $\IP_5$ because of the purely quadratic polynomial
that appears as one of the constraints defining the hypersurface.
 The requirement that the singularity
be completely degenerate seems, in fact, to exclude a
great many of the CICY manifolds,
the complete class of which was constructed in ref. \cite{cdls}.
An important set of manifolds in that class that does not fit into
the present framework
either is defined by
polynomials of bidegree (1,4) and (1,1)
\beq
\matrix{\IP_1\cr \IP_4\cr}\left[\matrix{1&1\cr 1&4\cr}\right].
\eeq
The superpotential $W=p_1+p_2$ is not quasihomogeneous, nor does it
have an isolated singularity
\fnote{4}{These manifolds are important in the context of possible
 phase transitions between Calabi--Yau string vacua \cite{cgh}
 via the process of splitting and contraction introduced in
 \cite{cdls}.}.
Thus it appears that there ought to be a generalization of the framework
described above, which allows a modified LG description of these and other
string vacua. This however we leave for future work.

\vskip .1truein
\noindent
\section{Transversality of Catastrophes}

\noindent
The most explicit way of constructing a Landau--Ginzburg vacuum is, of
course,
to exhibit a specific potential that satisfies all the conditions
imposed
by the requirement that it ought to describe a consistent ground state
 of the string.
 Even though much effort has
gone into the classification of singularities of the type described in
the
previous section,
such polynomials have not been classified yet. As already mentioned above
the mathematicians have classified
polynomials with at most three variables \cite{agzv}, which is two short
of
the lowest number of variables
that is needed in order to construct a vacuum that allows a formulation of
a four--dimensional low--energy effective theory
\fnote{5}{The complete list of K3 representations embedded in
$dim_{\IC} =3$ weighted projective space $\IP_{(k_1,k_2,k_3,k_4)}$ has
been obtained in \cite{reid}.}.

In Sections 1--3 a set of potentials in five variables was described
 which represents an obvious
generalization of the polynomials that appear in two--dimensional
catastrophes.
After imposing the conditions for these theories to describe
string vacua, only a finite number of the infinite number of LG theories
survive and all these solutions were constructed. It is clear that this
 classification of singularities is
not complete, even after restricting to five variables.
It is indeed easy to construct polynomials that are not contained in the
classification of \cite{cls}, a simple one being furnished by the
example \cite{orlik}
 \beq
\IP_{(15,10,6,1)}[45]\ni
\left\{z_1(z_1^2+z_2^3+z_3^5)+z_4^{45}+z_2^2z_3^4z_4=0\right\}.
\eeq
This polynomial is not of
any of the types analysed discussed above but it is nevertheless transverse.

Knowledge of the explicit form of the potential of a LG theory is very
useful information when it comes to the detailed analysis of such a
model.
It is however not necessary if only limited knowledge, such as the
computation of the spectrum of the theory, is required. In fact
the only ingredients necessary for the computation of the spectrum
of a LG vacuum \cite{v} are the anomalous dimensions of the scaling
fields
as well as the fact that in a configuration of weights
there exists a polynomial of appropriate degree with an
isolated singularity. However, it is much easier to check whether
there exists such a polynomial
in a configuration than to actually construct
such a potential. The reason is a theorem by Bertini \cite{algeom},
which asserts that
if a polynomial does have an isolated singularity on the base locus
then,
even though this potential may have worse singularities away from the
base locus, there exists a deformation of the original polynomial that
only
admits an isolated singularity anywhere. Hence we only have to find
criteria
that guarantee at worst an isolated singularity on the base locus.
It is precisely this problem that was addressed in the mathematics
literature \cite{f} at the same time as the explicit construction of LG
vacua was started in ref. \cite{cls}.
The main point of the argument in \cite{f} is the following.

Suppose we
wish to check whether a polynomial in $n$ variables $z_i$ with weights
$k_i$ has an isolated singularity, i.e. whether the condition
\beq
dp=\sum_i \frac{\del p}{\del z_i} dz_i = 0
\lleq{transv}
can be solved at the origin $z_1=\cdots = z_n=0$. According to
Bertini's theorem,
the singularities of a general element in $\IC_{(k_1,...,k_n)}[d]$
will lie
on the base locus, i.e., the intersection of the hypersurface and
all the
components of the base locus, described by coordinate planes of
dimension
$k=1, ..., n$. Let $\cP_k$ such a $k$--plane, which we may assume to be
described by setting to zero the coordinates $z_{k+1}=\cdots = z_n$.
Expand
the polynomials around the non--vanishing coordinates $z_1,...,z_k$
\beq
p(z_1,...,z_n) =
q_0(z_1,...,z_k)~ +~ \sum_{j=k+1}^n q_j(z_1,...,z_k)z_j + h.o.t.
\eeq
Clearly, if $q_0\neq 0$ then $\cP_k$ is not part of the base locus
and hence
the hypersurface is transverse. If on the other hand $q_j=0$, then
$\cP_k$ is part of the base locus and singular points
can occur on the intersection of the hypersurfaces defined by
$\cH_j=\{q_j=0\}$.
If, however, we can arrange this intersection to be empty, then the
potential is smooth on the base locus.

Thus we have found that the conditions for transversality in any
number of variables is the existence for any index set
$\cI_k=\{1,...,k\}$ of
\begin{itemize}
\begin{enumerate}
\item{either a monomial $z_1^{a_1}\cdots z_k^{a_k}$ of degree $d$}
\item{or of at least $k$ monomials $z_1^{a_1}\cdots z_k^{a_k}z_{e_i}$
with distinct $e_i$.}
\end{enumerate}
\end{itemize}

Assume on the other hand that neither of these conditions
holds for
all index sets and let $\cI_k$ be the subset for which they fail. Then
the potential has the form
\beq
p(z_1,...,z_n) = \sum_{j=k+1}^n q_j(z_1,...,z_k)z_j ~+~ \cdots
\eeq
with at most $k-1$ non--vanishing $q_j$. In this case the intersection
of the hypersurfaces $\cH_j$ will be positive and hence the polynomial
$p$ will not be transverse.

As an example for the considerable ease with which one can check whether
a given configuration allows for the existence of a
potential with an isolated singularity, consider the polynomial of Orlik
and Randall
\beq
p=z_1^3+z_1z_2^3+z_1z_3^5+z_4^{45}+z_2^2z_3^4z_4.
\eeq
Condition (\ref{transv}) is equivalent to the system of equations
\beq
\begin{array}{r l r l}
 0 &=~ 3z_1^2+z_2^3+z_3^5 , & 0 &=~ 3z_1z_2^2+2z_2z_3^4z_4 \\
 0 &=~ 5z_1z_3^4 + 4z_2^2z_3^3z_4, & 0 &=~ z_2^2z_3^4+45z_4^{44} .
\end{array}
\eeq
which, on the base locus, collapses to the trivial pair of
equations $z_2z_3=0=z_2^3+z_5^5$. Hence this configuration allows for
such a polynomial. To check the system away from the base locus
clearly is much more complicated.

By adding a fifth variable $z_5$ of weight 13 it is possible to define
a Calabi--Yau deformation class
$\IP_{(1,6,10,13,15)}[45]_{-72}^{(17,53)}$, a configuration not
considered in \cite{cls}.

\vskip .1truein
\noindent
\section{Finiteness Considerations}

 \noindent
The problem of finiteness has two parts: first one has to put a
constraint
on the number of scaling fields that can appear in the LG theory
and then one has to determine limits on the exponents with which the
variables occur in the superpotential. Both of these constraints follow
from the fact that the central charge of a Landau--Ginzburg theory with
fields of charge $q_i$
\beq
c=3\sum \left(1-2 q_i\right)=:\sum c_i
\lleq{cc}
has to be $c=9$ in order to describe a string vacuum.

It should be clear that without any additional input the number of
Landau--Ginzburg vacuum configurations that can be exhibited is infinite.
This is to be
expected simply because it is known from the construction of CICYs
\cite{cls} that it is often possible to rewrite a manifold in an infinite
number of ways and we ought to encounter similar things in the LG
framework. A trivial way to do this is to simply add mass terms that
do not contribute to the central charge. Even though trivial such
mass terms
are important and necessary for LG theories, not only in orbifold
constructions \cite{kss} but also in order to relate them to CY
manifolds. Consider e.g. the codimension--four Calabi--Yau manifold
\beq
\matrix{\IP_1 \cr \IP_2\cr \IP_2 \cr \IP_2\cr}
\left[\matrix{2 &0 &0 &0\cr
 1 &2 &0 &0\cr
 0 &1 &2 &0\cr
 0 &0 &1 &2\cr}\right]
\lleq{boggle}
with the defining polynomials
\beq
\begin{array}{r l r l}
p_1 &=~ \sum_{i=1}^2 u_i^2v_i, & p_2 &=~ \sum_{i=1}^3 v_i^2w_i \\
p_3 &=~ \sum_{i=1}^3 w_i^2x_i, & p_4 &=~ \sum_{i=1}^3 x_i^2
\end{array}
\lleq{bogglepollies}
the superpotential $W=\sum p_i$ of which lives in
\beq
\IC^{\star}_{(5,5,6,6,6,4,4,4,8,8,8)}[16]_{-32}^{10}
\eeq
and has an isolated singularity at the origin. All eleven variables
are coupled and hence
this example appears to involve three fields with zero central charge
in a nontrivial way.

It turns out, however, that the manifold (\ref{boggle}) is equivalent
to a manifold with nine variables. One way to see this is by making
 use of some topological identities introduced in \cite{cdls}. First
consider the well--known isomorphism
\beq
\IP_2[2] \equiv \IP_1,
\eeq
which allows us to rewrite the manifold above as
\beq
\matrix{\IP_1 \cr \IP_2\cr \IP_2 \cr \IP_1\cr}
\left[\matrix{2 &0 &0\cr
 1 &2 &0\cr
 0 &1 &2\cr
 0 &0 &2\cr}\right].
\eeq
Using the surface identity \cite{cdls}
\beq
\matrix{\IP_1 \cr \IP_2\cr} \left[\matrix{ 2\cr 1\cr}\right]=
\matrix{\IP_1 \cr \IP_1\cr}
\eeq
applied via the rule
\beq
\matrix{\IP_1 \cr \IP_2\cr X\cr} \left[\matrix{ 2 &0\cr
 1 &a\cr
 0 &M\cr}\right]=
\matrix{\IP_1 \cr \IP_1\cr X\cr} \left[\matrix{ a\cr
 a\cr
 M\cr}\right]
\eeq
shows that this space is in turn equivalent to
\beq
\matrix{\IP_1 \cr \IP_1\cr \IP_1 \cr\IP_2\cr}
\left[\matrix{2 &0 \cr
 2 &0 \cr
 0 &2 \cr
 1 &2 \cr}\right],
\lleq{nonlg}
a manifold with only nine homogeneous coordinates.
It should be noted that the LG potential of (\ref{nonlg}), defined
by the sum of the two polynomials, certainly does not
have an isolated singularity. Furthermore it is not possible to
 even assign weights to the fields such that the
central charge comes out to be nine!
It is thus possible, by applying topological identities, to extend
the applicability of Landau--Ginzburg theories to types of Calabi--Yau
manifolds that were hitherto completely inaccessible by the standard
formulation.

Further insight into the problem of redundancy in the construction of
LG potentials can be gained by an LG theoretic analysis of this
example. From the weights of the scaling variables in the LG configuration
above it is clear that the spectrum of this LG configuration remains
the same if the last three coordinates are set to zero. In the
potential
\beq
W=\sum_{i=1}^2 \left(u_i^2v_i+v_i^2w_i+w_i^2x_i+x_i^2\right) +
 (v_3^2w_3+w_3^2x_3+x_3^2)
\lleq{bogglelg}
 described by the CICY polynomials (\ref{bogglepollies}), these variables
cannot be set to zero because they are coupled to other fields; hence it
seems impossible to reduce the number of fields. Consider however the
following change of variables
\beq
y_i=x_i+\frac{1}{2}w_i^2,~~i=1,2,3.
\eeq
 It follows from these transformations that the potential defined by
by (\ref{bogglelg}) is equivalent to
\beq
W=\sum_{i=1}^2\left(u_i^2v_i+v_i^2w_i-
 \frac{1}{4}w_i^4\right)+v_3^2w_3-\frac{1}{4}w_3^4.
\eeq
 Adding a trivial factor and splitting this potential into three
separate polynomials
\beq
p_1=u_1^2v_1+u_2^2v_2,~~~p_2=v_1^2w_1+v_2^2w_2+v_3^2w_3,~~~
p_3=-\frac{1}{4}\left(w_1^4+w_2^4+w_3^4\right)+x^2
\eeq
we see that the original model is equivalent to a weighted
 configuration
\beq
\matrix{\IP_1 \hfill \cr \IP_2 \hfill \cr \IP_{(1,1,1,2)}\cr}
\left[\matrix{2 &0 &0\cr
 1 &2 &0\cr
 0 &1 &4\cr}\right],
\eeq
which again describes manifolds with nine variables.
Thus the original configuration (\ref{boggle}) is in fact equivalent
to two different (weighted) CICY representations
\beq
\matrix{\IP_1 \cr \IP_1\cr \IP_1 \cr\IP_2\cr}
\left[\matrix{2 &0 \cr
 2 &0 \cr
 0 &2 \cr
 1 &2 \cr}\right]
{}~~\equiv ~~
\matrix{\IP_1 \cr \IP_2\cr \IP_2 \cr \IP_2\cr}
\left[\matrix{2 &0 &0 &0\cr
 1 &2 &0 &0\cr
 0 &1 &2 &0\cr
 0 &0 &1 &2\cr}\right]
{}~~\equiv ~~
\matrix{\IP_1 \hfill \cr \IP_2 \hfill \cr \IP_{(1,1,1,2)}\cr}
\left[\matrix{2 &0 &0\cr
 1 &2 &0\cr
 0 &1 &4\cr}\right].
\eeq

To summarize the last few paragraphs, we have shown two things: first
that by adding trivial factors and coupling them to the remaining fields
we can give a Landau--Ginzburg description of a
larger class of Calabi--Yau manifolds than previously thought
possible. Furthermore we can use topological identities to obtain
an LG formulation of CY manifolds, which do not admit a
canonical LG potential at all.
Incidentally we have also shown that it is possible to relate complete
intersection manifolds embedded in products of projective spaces to
 weighted complete intersection manifolds
embedded in products of weighted projective space.

Similarly the number of fields can grow without bound if we not only
allow
fields that do not contribute to the central charge but also fields
with
a negative contribution. Again such fields provide redundant
descriptions
of simpler LG theories; they are nevertheless important for the LG/CY
relation and
occur in the constructions of splitting and contraction introduced in
ref. \cite{cdls}. Even though these
constructions were discussed in \cite{cdls} only in the context of
Calabi--Yau
manifolds embedded in products of ordinary projective spaces they
readily generalize to the more general framework of weighted projective
spaces.

In special circumstances, the splitting or contraction process does not
change
the spectrum of the theory and hence it provides another tool to relate
 LG potentials with at most nine variables manifolds with more than
nine
homogeneous coordinates. Consider e.g. the manifold embedded in
\beq
\matrix{\IP_{(1,1)} \hfill \cr \IP_{(1,1,1,1,3)}\cr}
\left[\matrix{2&0\cr 1&6\cr}\right]
_{-252}^{(2,128)}
\eeq
which is described by the zero locus of the two polynomials
\bea
p_1&=&x_1^2y_1+x_2^2y_2 \nn \\
p_2&=&y_1^6+y_2^6+y_3^6+y_4^6+y_5^2
\eea
that can be described by a superpotential $W=p_1+p_2$ defining a
Landau--Ginzburg theory in
\beq
\IC^{\star}_{(5,5,2,2,2,2,6)}[12]^{(2,128)}_{-252}.
\eeq
This manifold can be rewritten via an ineffective split in an infinite
sequence of different representations as
\beq
\matrix{\IP_{(1,1)} \hfill \cr \IP_{(1,1,1,1,3)}\cr}
\left[\matrix{2&0\cr 1&6\cr}\right]
\longrightarrow
\matrix{\IP_{(1,1)}\hfill\cr \IP_{(1,1)}\hfill\cr \IP_{(1,1,1,1,3)}\cr}
\left[\matrix{2&0&0\cr 1&1&0\cr 0&1&6\cr}\right]
\longrightarrow
\matrix{\IP_{(1,1)}\hfill\cr \IP_{(1,1)}\hfill\cr\IP_{(1,1)}\hfill \cr
 \IP_{(1,1,1,1,3)}\cr}
\left[\matrix{2&0&0&0\cr 1&1&0&0\cr 0&1&1&0\cr 0&0&1&6}\right]
\longrightarrow
\cdots
\eeq
which are described by LG potentials in the alternating classes
\beq
\IC^{\star 7}_{(5,5,2,2,2,2,6)}[12] \longrightarrow
\IC^{\star 9}_{(1,1,10,10,2,2,2,2,6)}[12] \longrightarrow
\IC^{\star 11}_{(5,5,2,2,10,10,2,2,2,2,6)}[12] \longrightarrow \cdots
\eeq
i.e. the infinite sequence belongs to the configurations
\beq
\IC^{\star 7+4k}_{(5,5,2,2,10,10,2,2,10,10,...,2,2,10,10,2,2,6)}[12]
\eeq
where the part $(2,2,10,10)$ occurs $k$ times, and
\beq
\IC^{\star 5+4k}_{(1,1,10,10,2,2,10,10,2,2,...10,10,2,2,2,2,6)}[12].
\eeq
where $(10,10,2,2)$ occurs $k$ times.

The construction above easily generalizes to a number of examples
which all belong to a class of spaces discussed in ref. \cite{s3}.
Consider manifolds embedded in
\beq
\matrix{\IP_{(1,1)} \hfill\cr \IP_{(k_1,k_1,k_3,k_4,k_5)}\cr}
\left[\matrix{2&0\cr k_1&k\cr}\right],
\eeq
where $k=k_1+k_3+k_4+k_5$. These spaces can be split
into the infinite sequences
\beq
\matrix{\IP_{(1,1)} \hfill \cr \IP_{(k_1,k_1,k_3,k_4,k_5)}\cr}
\left[\matrix{2&0\cr k_1&k\cr}\right]
\longrightarrow
\matrix{\IP_{(1,1)} \hfill \cr \IP_{(1,1)}\hfill \cr
 \IP_{(k_1,k_1,k_3,k_4,k_5)}\cr}
\left[\matrix{2&0&0\cr 1&1&0\cr 0&k_1&k\cr}\right]
\longrightarrow
\matrix{\IP_{(1,1)} \hfill \cr \IP_{(1,1)}\hfill\cr\IP_{(1,1)}\hfill \cr
 \IP_{(k_1,k_1,k_3,k_4,k_5)}\cr}
\left[\matrix{2&0&0&0\cr 1&1&0&0\cr 0&1&1&0\cr 0&0&k_1&k}\right]
\longrightarrow
\cdots
\eeq
If the weights are such that $k/k_i$ is an integer,
then it is easy to write down the tensor model that corresponds to
it (but this is not essential).
In such models the levels $l_i$ of the tensor model
 $l_1^2\cdot l_3\cdot l_4\cdot l_5$
in terms of the weights are given by
\beq
l_1=l_2=\frac{2k}{k_1}-2,~~~l_i=\frac{k}{k_i}-2,~i=3,4,5
\eeq
and the corresponding LG potentials live in
\bea
\IC^{\star 7}_{(k-k_1,k-k_1,2k_1,2k_1,2k_3,2k_4,2k_5)}[2k]
\longrightarrow
\IC^{\star 9}_{(k_1,k_1,2(k-k_1),2(k-k_1),2k_1,2k_1,2k_3,2k_4,2k_5)}[2k]
\longrightarrow \cdots
\eea
i.e. they belong to the sequences
\beq
\IC^{\star 7+4p}_{((k-k_1),(k-k_1),2k_1,2k_1,2(k-k_1),2(k-k_1),....,
 2k_1,2k_1,2k_3,2k_4,2k_5)}[2k]
\eeq
where the part $(2k_1,2k_1,2(k-k_1),2(k-k_1))$ occurs $p$ times, and
\beq
\IC^{\star 9+4k}_{(k_1,k_1,2(k-k_1),2(k-k_1),
2k_1,2k_1,...,2k_3,2k_4,2k_5)}[2k].
\eeq
where $(2(k-k_1),2(k-k_1),2k_1,2k_1)$ occurs $p$ times.

All these models are constructed in such a way that they have central
charge
nine, but in contrast to the example discussed previously,
there now appear fields with negative central charge.
In the case at hand, however, these dangerous fields
only occur in a {\it coupled} subpart of the theory; the smallest
subsystem which involves these fields and which one can isolate is in
fact a theory with positive central charge. In the series of splits
just described e.g., the fields with negative central charge that occur
in the
first split always appear in the subsystem described by the
configurations
\beq
\IC^{\star}_{(k_1,2(k-k_1),2k_1)}[2k]
\eeq
with potentials of the form
\beq
x^2y +yz+z^{2k/k_1}.
\eeq
Thus the contribution to the central charge of this sector becomes
\beq
c=3\left[\left(1-\frac{k_1}{k}\right)+\left(1-\frac{2(k-k_1)}{k}\right)+
 \left(1-\frac{2k_1}{k}\right)\right],
\lleq{ccc}
which is always positive. This formula suggests that it ought
to be possible to dispense with the variables $y,z$ altogether, as
their total contribution to the central charge adds up to zero, and
that this theory is equivalent to that of a single monomial of degree
$k/k_1$.

More generally one may consider the Landau--Ginzburg theory defined by
the potential
\beq
x_1^ax_2+x_2x_3+\cdots + x_{n-1}x_n + x_n^b.
\lleq{triv}
 From the central charge
\beq
c= \left\{ \begin{array}{l l}
 6\left(1-\frac{1}{a}\right)\left(1-\frac{1}{b}\right),
 & n~{\rm even} \\
 3\left(1-\frac{2}{ab}\right), &n~{\rm odd}
 \end{array} \right\},
\eeq as well as from the dimension of the chiral ring
\beq
{\rm dim}~R_n = \prod \left(\frac{1}{q_i}-1\right)
 = \left\{ \begin{array}{l l}
 ab-b-1,&n~{\rm even} \\
 ab-1, &n~{\rm odd}
 \end{array} \right\},
\eeq
we expect this theory to be equivalent to
\beq
x_1^ax_2+x_2x_3+\cdots x_{n-1}x_n + x_n^b =
\left\{ \begin{array}{l l}
 x_1^ax_n+x_n^b, &n~{\rm even} \\
 x_1^{ab}, &n~{\rm odd}
 \end{array} \right\}.
\lleq{ident}

Such relations are supported by the identification of rather different
 LG configurations such as
\beq
\IC^{\star}_{(1,1,1,6,6,6,3,3,3,3,3)}[9] \equiv \IC^{\star}_{(1,1,1,3,3)}[9]
\eeq
as well as many other identifications which are rather nontrivial
in the context of the associated manifolds.

It follows from the above considerations that we have to assume,
 in order to avoid redundant
reconstructions of LG theories, that the central charge
of all scaling fields of the potential should be positive.
 In order to relate the potentials to manifolds, we may then add
one or several trivial factors or more complicated theories with zero
central charge.

Using the above results we derive in the following more detailed
finiteness conditions. Observe first that from eq. (\ref{cc})
written as
\begin{equation}
\sum_{i=1}^r q_i=\left( {r\over 2}-{c\over 6}\right):=\hat c\label{cbed}
\end{equation}
we obtain $r>c/3$.

Now let $p$ be a polynomial of degree $d$ in $r$ variables.
For the index set $\cI_1$ the conditions for transversality
imply the existence of $n_i\in \IN^+$, $i=1,\ldots,r$ and of a map
$j:\cI_r \rightarrow \cI_r$ such that for all $i$ there exists
$j(i)$ such that
\begin{equation}
q_i={{1-q_{j(i)}}\over n_i}.
\lleq{qs1}
Let us first see how many non-trivial fields can occur at most.
Fields which have charge $q_i\le 1/3$
contribute $c_i\ge 1$ to the conformal anomaly.
Now consider fields with larger charge. Since we assume $c_i>0$, they are
in the range $1/3<q_i<1/2$.
Among these fields the transversality condition (1.) cannot hold,
because two of them are not enough and three of them are too many
fields in order to form a monomial of charge one. Transversality
condition (2.) implies that each of them has to occur together with a
partner field $z_{j(i)}$. These pairs contribute to the conformal
anomaly according to
(\ref{qs1},\ref{cc}) $c_i+c_{j(i)}>2$, so we
can conclude that $r\le c$.

In order to construct all transversal
LG potentials for a given $c$,
we choose a specific $r$ in the range obtained above and consider
all possible maps $j$ of which there are $r^r$.
Without restricting on the generality, we may
{\sl then} assume the $n_i$ to be ordered
$n_1\le \cdots \le n_r$.
Starting with (\ref{cbed}) we obtain via eq. (\ref{qs1}) and the
positivity of the charges a bound $n_1<r/c$. Now we choose $n_1$ in the
allowed range and use (\ref{qs1}) in order to eliminate the $q_1$, if
necessary in favour of the $q_2,\ldots,q_r$. This yields
an equation of the general form $(p=2)$
\begin{equation}
\sum_{i=p}^r w^{(p)}_i q_i=\hat c^{(p)}.\label{cbed2}
\end{equation}
If $\hat c^{(p)}\neq 0$, eq. (\ref{cbed2}) either is in contradiction
with the positivity of the charges or one can derive a finite bound for
$n_{p}$. Assume $\hat c^{(p)}>0$ and let $\cI_+$ be the
indices of the positive $w_i^{(p)}$; then one has
$n_{p}<(\sum_{i\in \cI_+} w_i^{(p)})/\hat c^{(p)}$; likewise if
$\hat c^{(p)}<0$ we have
$n_{p}<(\sum_{i\in\cI_-} w_i^{(p)})/\hat c^{(p)}$.

Consider the case $\hat c^{(p)}=0$.
If the $w_i^{(p)}$ are indefinite
we get no bound from (\ref{cbed2}). However we will show that the
existence of certain monomials, which are required by the transversality
conditions, implies a bound for $n_p$.
Let $\cI_a$ denote the indices of the already bounded
$n_i$ and $\cI_b$ the others. The charge of the field $z_a$ with
$a\in \cI_a$
will depend on the unknown charge of a field $z_{b(a)}$ with
$b\in \cI_b$ if
there is a chain of indices
$a_0=a,a_1=j(a),\dots,a_l=j(\ldots j(a)\ldots)=:b(a)$ linked by the
map $j$. The charge of $z_a$ is given by
\beq
q_a={1\over n_a}-{1\over n_a n_{a_1}}+
\cdots -{(-1)^l\over n_a\ldots n_{l-1}}+
{(-1)^l q_{b(a)}\over n_a\ldots n_{l-1}}.
\lleq{qform}
Indefiniteness of the $w_i^{(p)}$ can only occur if there are fields
$z_a$, $a\in \cI_a$, with odd $l$, i.e. the last term in (\ref{qform})
$s_a q_{b(a)}:=(-1)^l/(n_a\ldots n_{a_{l-1}} q_{b(a)})$ is negative.
Call the index set of these fields $\cI_a^-$.
Assume first that the transversality condition (1.) holds.
This implies the existence of $m_i\in \IN^+$ ($m_i<2 n_i$)
such that $\sum_{i\in \cI^-_a}
m_i q_i=1$. For the unknown $q_i$, $i\in \cI_b$, we get an equation
of the form $\sum w_i q_i=\epsilon$, which yields a bound for the lowest
$n_i$, $i\in \cI_b$, since $w_i>0$.
The lowest possible value for $\epsilon>0$ can be readily calculated from
the denominators occurring in (\ref{qform}).
If transversality condition (2.) applies, we have $|\cI_a^-|$ equations
of the form $\sum_{i\in \cI_a^-} m^{(j)}_i q_i=1-q_{e_j}$ which can be
rewritten as $\sum_{i\in \cI_b} w_i^{(j)} q_i=\epsilon^{(j)}$.
Only if all $w_i^{(j)}$ happen to be indefinite and all
$\epsilon^{(j)}$ are zero we get {\sl no bound} from this condition.
 Assuming this to be true we have
\beq
\sum_{i\in \cI_b} m^{(j)} s_i q_i-s_{e_j} q_{b(e_j)}=0,
\lleq{g1}
where $s_{e_j}:=1$ and $b(e_j):=b$ if $e_j\in \cI_b$. Note that
$\sum_i m^{(j)}_i\ge 2$ in order to avoid quadratic mass terms.
Now we can rewrite (\ref{cbed2}) in the form
\beq
\sum s_i q_{b(i)}=0.
\lleq{g2}
If one uses now (\ref{g1}) and $\sum_i m^{(j)}_i\ge 2$
in order to eliminate the negative $s_i$, one finds
 $\sum_i w_i q_i\le 0$ with $w_i>0$, which is in contradiction with
the positivity of the charges, hence we get a bound in any case.

This procedure of restricting the bound for $n_{p}$, given
$n_i,\dots,n_{p-1}$,
was implemented in a computer program. It allows all
configurations to be found without testing unnecessarily many combinations of
the $n_i$.
The actual upper bounds for the $(n_i,\dots,n_r)$
in the four--variable case are $(7,17,83,1805)$, and we have found $2390$
configurations which allow for transversal polynomials.
In the five--variable case the bounds are $(4,6,14,62,923)$ and
$5165$ configurations exist.
By adding a trivial mass term $z_5^2$ in the four--variable case, the
configurations mentioned so far lead to three--dimensional
Calabi--Yau manifolds described by a one polynomial constraint
in a four--dimensional weighted projective space.

The same figures for the six--variable case and seven--variable case are
$(3,3,5,11,41,482)$, $2567$ and $(2,2,3,4,8,26,242)$, $669$ respectively,
leading to a total of $3236$ combinations.
Likewise for eight-- and nine--variable potentials the bounds
become $(2,2,2,2,2,3,3,5,14)$ with $47$ examples and $(2,2,2,2,2,2,2,2,2)$
with
$1$ example respectively. The lists with all models can be found in
\cite{ks}.

\vskip .1truein
\noindent
\section{Results and Comparisons}

\noindent
We have constructed 10,839 Landau--Ginzburg theories with
2997 different spectra, i.e. pairs of generations and antigenerations.
The massless spectrum is very rough information about a theory and
it is likely that the degeneracy is lifted to a large degree when
additional information, such as the number of singlets and/or the
Yukawa couplings, becomes available. We expect the situation to be
very similar to the class of CICYs \cite{cdls}, which only leads to
some 250
different spectra \cite{ghl}, but for which a detailed analysis of the
Yukawa couplings \cite{ch} shows that it contains several thousand
distinct theories.

It is clear however that there is in fact some redundancy in this
class of Landau--Ginzburg theories even beyond the one discussed in
the previous sections. In the list there appear, for instance,
 two theories with
spectrum $(h^{(1,1)},h^{(2,1)},\chi)=(2,122,-240)$ involving five variables
\beq
\IP_{(2,2,2,1,7)}[14] \ni \{z_1^7+z_2^7+z_3^7+z_4^{14}+z_5^2=0\}
\eeq
and
\beq
\IP_{(1,1,1,1,3)}[7] \ni \{y_1^7+y_2^7+y_3^7+y_4^7+y_4y_5^2=0\}.
\eeq
Using the fractional transformations introduced in ref. \cite{ls4}
it is
easy to show that these two models are equivalent, even though
this is not obvious by just looking at the potentials. To see this
consider first the orbifold
\beq
\IP_{(2,2,2,1,7)}[14] {\large /} \ZZ_2: [~0~0~0~1~1]
\eeq
of the first model where the action indicated by
$\ZZ_2:~[~0~0~0~1~1] $
means that the first three coordinates remain invariant, whereas the
latter
two variables are to be multiplied with the generator of the cyclic
$\ZZ_2$, $\a=-1$.
Since the action of this $\ZZ_2$ on the weighted projective space
is part
of the projective equivalence, the orbifold is of course isomorphic
to the
original model. On the other hand the fractional transformation
 \beq
z_i=y_i,~i=1,2,3;~~~ z_4=y_4^{1/2},~~~ z_5=y_4^{1/2}y_5
\eeq
associated with this symmetry \cite{ls4} defines a 1--1 coordinate
transformation on the orbifold, which transforms the first theory into
the second; these are therefore equivalent as well.

Similarly the equivalences
\bea
\IP_{(2,2,2,3,9)}[18]_{-216}^{(4,112)}&=&\IP_{(1,1,1,3,3)}[9] \nonumber
\\
\IP_{(2,6,6,7,21)}[42]_{-96}^{(17,65)}&=&\IP_{(1,3,3,7,7)}[21]\nonumber
\\
\IP_{(2,5,14,14,35)}[70]_{-64}^{(27,59)} &=&\IP_{(1,5,7,7,15)}[35]
\eea
can be shown, as well as a number of others.

It should be noted that even though we now have constructed
Landau--Ginzburg potentials with an arbitrary number of scaling fields,
the basic range of the spectrum has not changed as compared with the
results of \cite{cls} where it was found that the spectra of all
6000 odd
theories constructed there lead to Euler numbers which fall in the
range
\fnote{6}{This should be compared with the result for the complete
 intersection Calabi--Yau manifolds where
\break $-200 \leq \chi \leq 0$ \cite{cdls}. See Figure 1.}
\beq
-960 \leq \chi \leq 960.
\eeq
In fact not only
do all the LG spectra fall into this range, all known Calabi--Yau
spectra
and all the spectra from exactly solvable tensor models are contained
in this range as well! This suggests that perhaps the spectra of all
string vacua based on $c=9$
will be found within this range. To put it differently, we conjecture that
the Euler numbers of all Calabi--Yau manifolds are contained in the
range $-960 \leq \chi \leq 960$.

Similarly to the results in \cite{cls}, the Hodge pairs do not pair up
completely.
In fact the mirror symmetry of the space of Landau--Ginzburg vacua is just
about $77\%$.
It thus appears that orbifolding is an essential ingredient
in the construction of a mirror--symmetric slice of the configuration space
of the heterotic string. It is in fact easy to produce examples of
orbifolds whose spectrum does not appear in our list of LG vacua.
An example of a mirror pair is furnished by the orbifold
\beq
\IP_{(1,1,1,1,2)}[6]^{(1,103)}_{-204} {\LARGE /} \ZZ_6:
\left[\matrix{3&2&1&0&0\cr}\right]
\eeq
which has the spectrum $(11,23,-24)$ and the space
\beq
\IP_{(1,1,1,1,2)}[6]^{(1,103)}_{-204} {\LARGE /} \ZZ_6\times \ZZ_3 :
\left[\matrix{3&2&1&0&0\cr
 0&0&1&0&2\cr}\right]
\eeq
which has the mirror--flipped spectrum $(23,11,24)$.
The space of LG orbifolds that has been constructed so far \cite{kss}
is indeed much closer to being mirror--symmetric than the space of
LG theories itself. Even though the construction in \cite{kss} is
incomplete, about 94\% of the Hodge numbers pair up.

By now there exist many different prescriptions to construct string
vacua and we would like to put the LG framework into the context of
other left--right--symmetric constructions.
Among the more prominent ones other than Calabi--Yau manifolds, which
are obviously closely related to LG theories, there are constructions which
have traditionally been called, somewhat misleadingly,
orbifolds (what is meant are
orbifolds of tori) \cite{dhvw}, free--fermion constructions
\cite{freeferm}, lattice constructions \cite{lls}, and interacting
exactly solvable models \cite{g}\cite{kasu}.

None of these classes are known completely, even though much effort
has gone into the exploration of some of them. Because powerful
computational tools are available, toroidal orbifolds have been analysed
in some detail \cite{orbis}\cite{orbiyuk} and much attention has
focused on the explicit construction of exactly solvable theories in
the
context of tensor models via $N=2$ superconformal minimal theories
\cite{ls12}\cite{ls3}\cite{fkss1} and Kazama--Suzuki models
\cite{fiq2}\cite{lvwg}\cite{ls5}\cite{s}\cite{fksv}
\cite{aafn} as well
as their in depth analysis \cite{lvwg}\cite{s2}\cite{fks}\cite{exyuk}.
In the context of
(2,2) vacua, orbifolds lead only to some tens of distinct models, whereas
the known
classes of exactly solvable theories lead only to a few hundred models
with distinct spectra. Similar results have been obtained so far via the
covariant--lattice approach \cite{lsw} and hence it is obvious that these
constructions do
not exhaust the configuration space of heterotic string by far. The
class of Landau--Ginzburg string vacua thus appears as a rather
extensive source of (2,2)--symmetric models.

Finally we should remark that the (2,2) Landau--Ginzburg theories
we have constructed here can be used to build a probably much
larger class
of (2,0) models along the lines described in \cite{dg}, using an
appropriate adaptation of the work of \cite{cdgp}\cite{dave}
\cite{klth}\cite{anamaria} in order to determine the instantons
on which the existence of a certain split of a vector bundle
has to be checked.

\vfill \eject

\part{ Mirror Pairs via Fractional Transformations.}

Even though the emerging mirror symmetry of the construction discussed in
Part I is clearly very suggestive there is a priori no reason why models
with mirror flipped spectra should be related at all. After all,
Hodge numbers do not classify manifolds topologically and knowing
the spectrum of a physical theory is only a very first step in the analysis
of its properties. It turns out that the detailed analysis of the models
rests both mathematically and physically on the computation of the Yukawa
couplings. This, in general, is a difficult business. It is therefore
gratifying that it is possible to proceed via a different line of thought
to relate different models, namely via a direct map between different
theories. In particular cases this rather general map turns out to be the
mirror map.

\vskip .1truein
\noindent
\section{ From A--Models to D--Models via Orbit Construction}

\noindent
It is well known from the ADE--classification of partition functions of
($N=2$)minimal models that the ${\rm D}$--type models are equivalent to
${\rm A}$--type models modded out by a $\ZZ_2$ discrete symmetry.
As a consequence subsets of vacua of the Heterotic String involving the
$N=2$ minimal series are also related via orbifold constructions
\fnote {3}{In this article orbifold construction will be understood
 to include twisted modes in the conformal field theory or
 Landau--Ginzburg construction and blow--up modes in the
 geometric formulation.}.
In terms of the corresponding Landau-Ginzburg potentials of the exact
models
this can be seen as follows. The potential corresponding to a model
at level
$k$ with a diagonal affine invariant is \cite{kms}
\beq
{\rm A}_{k+1}~~~\sim~~~\Phi_1^{k+2} + \Phi_2^2 \lleq{diag}
where, with hindsight, a trivial factor has been added.
The LG potential corresponding to the exact model at the same level
\fnote{4}{The ${\rm D}$--models only exist for even level $k=2n$.}
but with the ${\rm D}$--invariant is described by \cite{m}\cite{vw}

\beq
{\rm D}_{\frac{k}{2}+2}~~\sim~~~{\tilde {\Phi}}_1^{(k+2)/2} +
 \phiti_1 \phiti_2^2. \lleq{daff}
By defining the transformation of the scaling variables
\beq
\Phi_1 = \phiti_1^{1/2}~~~ {\rm and}~~~\Phi_2 = \phiti_1^{1/2}\phiti_2
\lleq{fraccy}
one can map the diagonal theory of (\ref{diag}) into the nondiagonal
model of
(\ref{daff}).
Moreover this transformation has a constant Jacobian and therefore one
might naively expect that
they are in fact equivalent descriptions of a given theory since
their path integrals are equivalent. It is clear however that this is not
the case as the two theories have a different spectrum:
the local ring of the diagonal theory consists of states
\beq
\cR_{{\rm A}_{k+1}}=\{1,\Phi_1, \Phi_1^2,\cdots,\Phi_1^k\}
\eeq
i.e. the theory has $k+1$ states. The ${\rm D}$--theory however has only
$\frac{k}{2}+2$ states in its spectrum
\beq
\cR_{{\rm D}_{\frac{k}{2}+2}}=
\{1,\phiti_2,\phiti_1,\cdots,\phiti_1^{\frac{k}{2}}\}.
\eeq
Hence these two theories are not the same. The resolution of this puzzle
comes
from the fact that the transformation of (\ref{fraccy}) is not a coordinate
transformation since it is not a bijection but is 2--1 as it stands.
To make it 1--1 we should identify
\beq
\Phi_i \sim -\Phi_i,~~~i=1,2.
\eeq
Of the $k+1$ states of ${\rm A}_{k+1}$--models $\frac{k}{2}+1$ are
invariant with respect to the action of this $\ZZ_2$. By including the one
twisted state we find precisely the states of the non--diagonal
${\rm D}$--theory.

This construction can immediately be applied to string compactification
proper.
It may seem unlikely at first that the simple modding of a $\ZZ_2$ should
account for the different behaviour that the various tensor models
exhibit under the exchange
of a diagonal invariant by a D--invariant. A quick look at the results
of refs.
\cite{ls3}\cite{fkss1} shows that this exchange generically changes the
spectrum
 in some
arbitrary way without any obvious systematics. However in some special
cases
the spectrum is flipped, exchanging generations and anti--generations while
in other models the spectrum does not change at all. The reason for this
erratic behaviour can be traced to the fact that the other coordinates
involved in the construction determine the fixed point structure of the
discrete group in an essential way.

Consider the following example where the exchange of the affine
invariant does
not change the spectrum
\fnote{5}{The notation of ref. \cite{ls3} is used.}
\beq
\IP_{(1,7,2,2,2)}[14]~~\ni~~(12\cdot 5^3)_{A_{13}\otimes A_6^3}~~
\longrightarrow
{}~~(12\cdot 5^3)_{D_{8}\otimes A_6^3}~~\in~~\IP_{(1,3,1,1,1)}[7].
\eeq
where we have added one trivial factor
\fnote{6}{An explanation for the necessity of this trivial
 factor and when it is to be added can be found in
 \cite{ls3}.}.
The reason that the spectrum does not change after the replacement of the
diagonal affine invariant by the D--invariant is rather obscure from the
point
of view of the conformal field theory described by the tensor model
$(12\cdot 5^3)$ with different affine invariants but can be understood
easily
from the orbifold point of view. In the manifold picture it simply follows
from the fact that the $\ZZ_2$--action $[1,1,0,0,0]$, generated by
\beq
(\Phi_1,\Phi_2,\Phi_3,\Phi_4,\Phi_5)\longrightarrow
(\a \Phi_1,\a \Phi_2,\Phi_3,\Phi_4,\Phi_5),
\eeq
(where $\a^2=1, \a\neq1$) is not an action at all on the Calabi--Yau
manifold embedded in the ambient
weighted projective space but it is part of the projective equivalence
transformation because we have
\beq
\IP_{(1,7,2,2,2)} = \IC_5/\sim
\eeq
with $(\Phi_1,\Phi_2,\Phi_3,\Phi_4,\Phi_5)\sim
(\l \Phi_1,\l^7 \Phi_2,\l^2 \Phi_3,\l^2 \Phi_4,\l^2 \Phi_5)$,
where $\l\in \IC^*$. Therefore the $\ZZ_2$ does not transform the
projective
coordinates at all but is trivial. In the conformal field theory picture
this translates into the fact that the action is actually part of the
modding that has to be done to implement the GSO projection. This example
shows that not all our models are independent but that some models appear
in different representations.

A more complicated example is furnished by the pair of minimal models
\beq
\IP_{(1,5,1,1,2)}[10]^1_{-288}~~=~~(8^3\cdot 3)_{A_{9}^3\otimes A_4}
\longrightarrow
(8^3\cdot 3)_{D_{6}\otimes A_9^2\otimes
A_4}~~=~~\IP_{(2,4,1,1,2)}[10]^3_{-192}.
\lleq{minmod}
The superscript denotes the dimension of the
second cohomology group $h^{(1,1)}={\rm dim}~{\rm H}^{(1,1)}(\cM)$ whereas
the subscript denotes the Euler number. It is again not obvious from the
point of view of the affine invariants involved why the spectrum changes the
way it does but the result is clear from the orbifold construction. Defining
the $\ZZ_2$--action as $[1,1,0,0,0]$ it follows that in this case the map is
nontrivial. Its fixed point set consists of two curves
$C_1=\IP_{(1,1,2)}[10]$ with Euler number $\chi_{C_1}=-30$ and
$C_2=\IP_{(1,5,2)}[10]$ with $\chi_{C_2}=-2$. The Euler number of the
resolved orbit manifold is therefore
\beq
\chi=-\frac{288}{2}-\frac{(-30-2)}{2}+2(-30-2)=-192.
\eeq
The cohomology can be computed as well. Given the Euler number we only need
to compute either the second cohomology group or the number of (2,1)--forms.
Using the results of \cite{s3} it is clear that each of the curves
contributes
one additional generator to the second cohomology, i.e. for the resolved
manifold we have $h^{(1,1)}=3$ and therefore the result of (\ref{minmod}).

These simple identifications of a class of Landau-Ginzburg potentials
with orbifolds of other potentials via fractional transformations can be
considered as a generalization
of the strange duality known from the exceptional singularities
of modality one \cite{agzv}. One of the `strangely dual pairs' appearing
in the
classification of Landau--Ginzburg potentials in 3 variables with an
isolated singularity at the origin is described by the two polynomials
\begin{eqnarray}
K_{14}~&:&\IP_{(3,12,8)}[24]~\ni~ \{\Phi_1^8+\Phi_2^2+\Phi_3^3=0\}\\
Q_{10}~&:&\IP_{(6,9,8)}[24]~\ni~\{\phiti_1^4+\phiti_1\phiti_2^2+\phiti_3^3=0\}.
\end{eqnarray}
The dimension of the chiral ring of these singularities is 10 and 14,
respectively, as indicated by the subscript of $Q$ and $K$. Each of these
catastrophes is characterized by two triplets of numbers, the Dolgachev
numbers
and the Gabrielov numbers. For the above singularity $Q_{10}$ the
corresponding pair of triplets is $\cD(Q_{10})=(2,3,9)$ and
$\cG(Q_{10})=(3,3,4)$ for the Dolgachev and Gabrielov numbers
respectively whereas the singularity $K_{14}$ leads to the triplets
$(3,3,4)$ and $(2,3,9)$ For the above pair $\cD(Q_{10})=\cG(K_{14})$ and
$\cG(Q_{10})=\cD(K_{14})$. It is in this sense that the two polynomials are
called dual
\fnote{7}{The precise nature of these numbers are not of
 concern here. More details can be found in
 ref. \cite{agzv}}.

{}From our previous results it is clear that an alternative way to relate
these two singularities flows from the description of affine
${\rm D}$--invariants as orbifolds of diagonal invariants.
The above polynomials can be viewed as the Landau--Ginzburg potentials
of the tensor product $(1\cdot 6)$
\beq
Q_{10}\sim (1\cdot 6)_{A_2\otimes D_5}~~{\rm and}~~
K_{14}\sim (1\cdot 6)_{A_2\otimes A_7}
\eeq
where we have added a trivial factor for the $K_{14}$ singularity.
Therefore it follows that
\beq
Q_{10}=K_{14}/\ZZ_2.
\eeq
More explicitly consider the chiral ring of $K_{14}$
\beq
{\cal R}_K=\{1,\Phi_1,\Phi_1^2,...,\Phi_1^6,
 \Phi_3,\Phi_1\Phi_3,...,\Phi_1^6\Phi_3\}.
\eeq
The action of the $\ZZ_2$ is defined as $[1,0,1]$ on the fields
$(\Phi_1,\Phi_2,\Phi_3)$. The two sectors of the orbifold consist of the
eight invariant states of $\cR_K$ and of two twisted states giving the
total of ten states necessary for $Q$.

This strangely dual pair can again be used as building block of
string compactifications. In the models constructed
in \cite{ls3}\cite{fkss1}
this involves vacua which have $(1\cdot 6)_{A_1\otimes A_7}$ or
$(1\cdot 6)_{A_1\otimes D_5}$ as subfactors.

The ADE models make up only a very small part of the string
compactifications constructed in \cite{cls} and it is natural to ask
whether it is
possible to relate spaces in this set to one another in a similar way
and if so
whether it is possible to construct {\it all} of these models as
orbifolds of
some basic set of polynomial types, e.g. Fermat type spaces. This is the
question to which will be addressed in the following section.

\vskip .1truein
\noindent
\section{Fractional Transformations}

\noindent
In the notation of \cite{cls} the transition from the diagonal affine
invariant to the
non--diagonal D--invariant can be formulated as the transition from
a Fermat
type polynomial with diagram
\beq
{\thicklines \begin{picture}(120,20)
 \put(0,0){\circle*{5}}
 \put(0,7){\circle{12}}
 \put(26,0){\circle*{5}}
 \put(26,7){\circle{12}}
 \end{picture}
 }
\Phi_1^a + \Phi_2^b
\lleq{gendiag}
to particular polynomials of the type of a tadpole
\beq
{\thicklines \begin{picture}(120,20)
 \put(0,0){\circle*{5}}
 \put(0,0){\line(1,0){26}}
 \put(26,0){\circle*{5}}
 \put(26,7){\circle{12}}
 \end{picture}
 }
\phiti_1^c\phiti_2 + \phiti_2^d
\lleq{gend}
Equations (\ref{diag},\ref{daff}) clearly describe only a very small
subset of these polynomials and one can ask whether transformations
generalizing (\ref{fraccy}) are possible. Indeed they are.

Consider the transformation
\beq
\Phi_1=\phiti_1^{c/a} \phiti_2^{1/a},~~~~~~~~
\Phi_2 = \phiti_2^{d/b}
\lleq{genfraccy}
which transforms the polynomial in (\ref{gendiag}) into the one of eq.
(\ref{gend}).
The next step is to find the constraints on the weights of the fields that
make the Jacobian of this transformation constant. They are given by
\beq
c=a~~~{\rm and}~~~ d=b\left[1-\frac{1}{a}\right].
\eeq
As in the case of the previous transformation however this transformation
is not well defined yet and we have to find the discrete group which makes
the transformation of variables well defined. Suppose that we have a map
\beq
\Phi_1 \longrightarrow \a \Phi_1 ~~~{\rm and}~~~
\Phi_2 \longrightarrow \b \Phi_2
\eeq
where $\a, \b$ are roots of unity. The condition that the change of variables
in eq. (\ref{genfraccy}) is invariant determines $\a$ as the generator of
$\ZZ_a$ and
$\b=\a^{a-1}$. Similarly observation regarding the inverse transformation
leads to the group that one has to mod out by in the nondiagonal theory.
Only after modding out these cyclic groups does
transformation (\ref{genfraccy}) become well defined.

The isomorphism can be summarized concisely with the following diagram

\vfill
\eject
\bea
& &\IC_{\left(\frac{b}{g_{ab}},\frac{a}{g_{ab}}\right)}
 \left[\frac{ab}{g_{ab}}\right]
 \ni \left \{z_1^a+z_2^b=0\right \}
 ~{\Big /}~ \ZZ_b: \left[\matrix{(b-1)&1}\right] ~~
 \nn \\ [3ex]
&\sim & \IC_{\left(\frac{b^2}{h_{ab}},\frac{a(b-1)-b}{h_{ab}}\right)}
 \left[\frac{ab(b-1)}{h_{ab}}\right]
 \ni \left \{y_1^{a(b-1)/b}+y_1y_2^b=0\right\}
 ~{\Big /}~ \ZZ_{b-1}: \left[\matrix{1&(b-2)}\right].
\llea{iso}
\vskip .1truein

\noindent
Here $g_{ab}$ is the greatest common divisor of $a$ and $b$ and
$h_{ab}$ is the greatest common divisor of $b^2$ and $(ab-a-b)$.
The action of a cyclic group $\ZZ_b$ of order $b$ denoted by
$[m~~n]$ indicates that the symmetry acts like
$(z_1,z_2) \mapsto (\a^m z_1, \a^n z_2)$ where $\a$ is the $b^{th}$ root
of unity.

It is easy to generalize this analysis to general transformations
from Fermat--type polynomials to tadpole--type polynomials with $N$ points
attached. Consider the transformation
\beq
\sum_{i=1}^N \Phi_i^{l_i} \longrightarrow
\left( \sum_{j=1}^{N-1} \phiti_j^{n_j} \phiti_{j+1}\right) + \phiti_N^{n_N}
\eeq
defined by
\begin{eqnarray}
\Phi_i &=& \phiti_i^{n_i/l_i} \phiti_{i+1}^{1/l_i} ~~~~~~~~i=1,..,N-1;
 \nonumber \\
\Phi_N &=& \phiti_N^{n_N/l_n}.
\llea{multifrac}
The condition that the Jacobian of this transformation is constant
again leads to constraints on the exponents:
\begin{eqnarray}
n_1 &=&l_1;\\
n_i &=&l_i\left[1 - \frac{1}{l_{i-1}}\right]~,~i=2,...,N.
\end{eqnarray}
Transformation (\ref{multifrac}) then becomes
\begin{eqnarray}
\Phi_1 &=& \phiti_1 \phiti_2^{1/l_1}; \\
\Phi_i &=& \phiti_i^{{(l_{i-1}-1)}/l_{i-1}} \phiti_{i+1}^{1/l_i}
 ~~~~~~~~~~i=2,..,N-1; \\
\Phi_N &=& \phiti_N^{{(l_{N-1}-1)}/l_{N-1}}.
\llea{multiexp}
As before, this transformation is not well defined, but one can find
a discrete group under which it is invariant. Let
\beq
\Phi_i \longrightarrow \alpha_i \Phi_i;
\lleq{scaling}
it then follows that under this transformation
\beq
\phiti_N \longrightarrow \a_N^{l_{N-1}/{(l_{N-1}-1)}} \phiti_N,
\eeq
i.e. we find the constraint
\beq
\a_N^{l_{N-1}/{(l_{N-1}-1)}}=1
\eeq
for the generator $\a_N$. Plugging the solutions of this constraint
iteratively into eqs. (\ref{multiexp}) one finds
that all generators $\a_i$ have to satisfy $\a_i^{l_i} = 1~,i=1,..,N-1$
for the transformation of variables to be invariant under the
identification (\ref{scaling}). The group by which one has to mod out is
$\prod_{i=1}^{N-1} \ZZ_{l_i}$. Again one has to similarly analyze the
inverse map. It is obvious that this generalization just amounts to
repeated application of the original isomorphism (\ref{iso}).

Consider e.g. the manifold
\beq
\IP_{(1,3,3,3,5)} [15]^3_{-144}=
\{\Phi_1^{15}+\Phi_2^5+\Phi_3^5+\Phi_4^5+\Phi_5^3=0\}.
\eeq
We will show now that the orbifold of this model with respect to two
cyclic groups $\ZZ_5$, the first generated by $[0,4,1,0,0]$
the second one by $\ZZ_5:[0,0,4,1,0]$, is isomorphic to the manifold
\beq
\IP_{(16,60,45,39,80)}[240]^{75}_{144}~=
 \{\phiti_1^{15}+\phiti_2^4+\phiti_2\phiti_3^4
 +\phiti_3\phiti_4^5+\phiti_5^3=0\}.
\eeq
Consider the first $\ZZ_5$. The fixed points of this action are determined
by the requirement that
\beq
(\Phi_1,\alpha^4 \Phi_2,\alpha \Phi_3,\Phi_4,\Phi_5)
= c(\Phi_1,\Phi_2,\Phi_3,\Phi_4,\Phi_5)
\eeq
where $c\in \IC^*$. The fixed point set consists of one curve
$\IP_{(1,3,5)}[15]$
with Euler number $ \chi= -6$ and one further fixed point $\IP_{(3,5)}[15]$.
The Euler number of the resolved orbifold therefore is
$\chi= \frac{-144}{5} - \frac{1}{5}(-6+1) + 5(-6+1) - \frac{1}{5} +5 =-48$.
The corresponding weighted CY--manifold of this space is described by
\beq
\IP_{(4,15,9,12,20)}[60]~=~\{\phi_1^{15}+\phi_2^4+\phi_2\phi_3^5+
 \phi_4^5+\phi_5^3=0\}.
\eeq
and the coordinate transformation is defined as follows
\beq
(\Phi_1,\Phi_2,...,\Phi_5) =
(\phi_1,\phi_2^{4/5},\phi_2^{1/5}\phi_3,\phi_4,\phi_5).
\eeq

Modding out by the other $\ZZ_5$ leads to a fixed point set of three curves
\beq
\IP_{(4,15,20)}[60]_{15},~~~\IP_{(15,9,20)}[60]_{16},
 ~~~\IP_{(15,12,20)}[60]_{11}.
\eeq
These three curves intersect in the point $\IP_{(15,20)}[60]=\IP_{(3,4)}[12]$
and the Euler number of the resolved manifold is given by
\beq
\chi = -\frac{48}{5} - \frac{1}{5}\left(15 + 11 + 16 -2 \times 5\right) +
 5\left(15 +11 + 16 -2\times 5\right) = 144.
\eeq
The term $-2\times 5$ comes from the fact that in all three curves we have
blown up the $\ZZ_5$ fixed point when in fact we only should have done
so once.

We have thus established that the discrete groups of the
LG--CY theories play a role comparable to the discrete symmetries of the
fusion rules in conformal field theories even though for most of the models
at hand the corresponding exact theory is not known. We have seen that
in general there are constraints on the weights of the potentials that
can be identified with with orbifolds of other models. Nevertheless, in the
types of potentials discussed above there was enough freedom to have
nontrivial identifications. This is not true in general as we will show in
the next section.

\vskip .1truein
\noindent
\section{ Loop Potentials}

\noindent
Consider potentials of the type
\beq
{\thicklines \begin{picture}(80,20)
 \put(0,0){\circle*{5}}
 \put(0,0){\line(1,0){26}}
 \put(26,0){\circle*{5}}
 \put(13,0){\oval(26,26)[t]}
 \end{picture}
 }
\phiti_1^c\phiti_2 + \phiti_2^d\phiti_1
\eeq
and more general polynomials of this type with an arbitrary number of
fields. For these polynomials it is possible to find a 1--1 coordinate
transformation from Fermat--type polynomials, but the
condition that the Jacobian is constant leads to the conclusion that
only deformations of the original space can be obtained in this way.

The transformation of a Fermat type polynomial into a loop--type
polynomial
\beq
\sum_{i=1}^{N} \Phi_i^{l_i} \longrightarrow
 \left( \sum_{i=1}^{N-1} \phiti_i^{n_i} \phiti_{i+1} \right) +
 \phiti_N^{n_N}\phiti_1.
\eeq
leads to the transformation of variables
\begin{eqnarray}
\Phi_i &=& \phiti_i^{n_i/l_i} \phiti_{i+1}^{1/l_i}~~~~~~~~i=1,..,N-1;
\nonumber \\
\Phi_N &=& \phiti_N^{n_N/l_N} \phiti_1^{1/l_N}.
\llea{loopfrac}
The condition that the Jacobian be constant
leads to the constraints
\begin{eqnarray}
n_1 &=& l_1\left[1 -\frac{1}{l_N}\right];\\
n_i &=& l_i\left[1 -\frac{1}{l_{i-1}}\right],~~~~~~~~~~~~~i=2,..,N;
\end{eqnarray}
so that the coordinate transformation (\ref{loopfrac}) becomes
\begin{eqnarray}
\Phi_1 &=& \phiti_1^{{(l_N-1)}/l_N} \phiti_2^{1/l_1}\\
\Phi_i &=& \phiti_i^{{(l_{i-1}-1)}/l_{i-1}} \phiti_{i+1}^{1/l_i},
{}~~~~~~i=2,..,N.
\end{eqnarray}
{}From this it follows however that the charges of the loop fields are
precisely those of the fields of the Fermat type polynomial, i.e.
the loop--type model is a trivial deformation of the original one.


\vfill \eject

\part{ Landau--Ginzburg Orbifolds.}

Even though the class of Heterotic Vacua described in Part I is the largest
constructed so far it clearly does not encompass all spectra that are known.
Consider e.g. the orbifold of the Fermat quintic in $\IP_4[5]$ with respect
to the action
\beq
\ZZ_5: \left[\matrix{ 3 &1 &1 &0 &0}\right]
\eeq
which leads to a space with spectrum $(h^{(1,1)},h^{(2,1)},\chi)=(21,17,8)$.
No complete intersection model with such Hodge numbers appears among the
results of \cite{ks}.
An obvious question of course is whether one could use the fractional
transformation discussed in Part II to {\it construct} the complete
intersection representation of this orbifold. Unfortunately this yields
a singular space.
Thus it is unclear at this point whether a complete intersection of this
orbifold exist. There are many examples of this type, some of which will
be mentioned below.

Orbifolding is important for the construction of mirrors as well because
in many examples the weighted CICY representation of a mirror is not known
whereas it is easy to construct the mirror as an orbifold. A simple example
is furnished by the manifold
\beq
\matrix{\IP_2\cr \IP_3\cr} \left[\matrix{3&0\cr 1&3\cr}\right].
\eeq
the mirror of which can easily be constructed as the orbifold with respect
to the action
\beq
\ZZ_9\times \ZZ_3 ~:~ \left[\matrix{6~&1~&3~&2~&0~&0~&6\cr
 0~&0~&0~&0~&0~&2~&1\cr}\right]
\eeq
when viewed as a symmetry on the corresponding LG theory embedded in
$\IC_{(3,2,3,2,3,2,3)}[9]$.
It is unclear however what the CICY representation of this mirror is.

A further motivation for the construction of orbifolds has been mentioned
already in the introduction. Namely in order to get some idea of how
`robust' the mirror property of the configuration space is it is useful
to implement different types of constructions. Until the proper framework
for mirror symmetry and its explicit form has been found this appears to be
 one feasible avenue for collecting support for the existence
\fnote{8}{There are of course also other reasons to look for
 new models but this is another story.}.

This last part of the review contains a brief description of the work
done in \cite{kss}
in which some 40odd types of actions have been considered
on all Landau--Ginzburg potentials that can be build from Fermat monomials,
single Tadpole potentials and single Loop potentials as well as arbitrary
combinations thereof.

\vskip .1truein
\noindent
\section{Actions of Symmetries: General Considerations}

\noindent
It is useful to first discuss some general aspects that are important
for group actions on Landau--Ginzburg theories that have been
orbifolded with respect to the U(1)--symmetry in order to describe
string vacua with $N=1$ spacetime supersymmetry.

An obvious question when considering orbifolds is whether
there is any a priori insight into
what spectra are possible for the orbifolds of a given model with respect
to a particular set of symmetries.
This question is of particular interest if the goal is to produce
orbifolds with presribed spectra, say models with a small number
of fields where the difference between the number of generations and
antigenerations is three.

Even though it is possible to formulate constraints on the
orbifold spectrum for
particular types of actions, we know of no constraints that hold
in full generality, or even for arbitrary cyclic actions.
One very simple class of symmetries are those without fixed points.
For such actions there are no twisted sectors and hence
there exists a simple formula expressing the Euler number $\chi_{orb}$ of the
orbifold in terms of the Euler characteristic $\chi$ of the covering space
and the order $|G|$ of the group
\beq
\chi_{orb} = \frac{\chi}{|G|}.
\eeq
The vast majority of actions however do have fixed points and hence the
result above does not apply very often.

For orbifolds with respect to cyclic groups of prime order there
exists a generalization of this result. For such group actions it was shown
in the first reference in \cite{orbis} that
\beq
\bn^g_{orb} - n^g_{orb} =
\left( |G|+1\right)\left(\bn^g_{inv} - n^g_{inv}\right)
-\left(\bn^g - n^g\right),
\eeq
where $n^g_{orb}$, $n^g_{inv}$, $n^g$ are the numbers of generations
of the orbifold theory, the invariant sector and the original LG theory,
respectively.

Consider then the problem of constructing an orbifold with a prescribed Euler
number $\chi_{orb}$ from a given theory.
Only for fixed point free actions will the order of the group be completely
specified as $|G|=\chi/\chi_{orb}$. It is important to realize that in
general the order of the group by which a theory is orbifolded does
{\it not} determine its spectrum -- the precise form of the action of the
symmetry is important.

Nevertheless we can derive {\it some} constraints on the order of the
action that we are looking for. Even though we don't know a priori what
the invariant
sector of the orbifold will be we do know that its associated Euler number
must be an integer
\beq
\chi_{inv} = \frac{\chi + \chi_{orbi}}{|G|+1}~~\in \IN.
\eeq
This simple condition does lead to restrictions for the order of the group.
Suppose, e.g., that we wish to check whether the quintic threefold admits
a three--generation orbifold: For the deformation class of the quintic
\beq
\IC_{(1,1,1,1,1)}[5]: ~\chi=-200
\eeq
the order of the discrete group in question must satisfy the constraint
$-206/(|G|+1) \in \ZZ$, implying $|G|=102$.
Hence there exists no three--generation orbifold of the quintic with respect
to a discrete group with prime order.
A counterexample for nonprime orders is furnished by the following theory
\beq
\IC_{(2,2,2,3,3,3,3)}[9]:~ (\bn^g,n^g,\c)=(8, 35,-54),
\lleq{3gen}
which corresponds to a CY theory embedded in a product of two projective
spaces by two polynomials of bidegree $(0,3)$ and $(3,1)$ \cite{s1}\cite{s2}
i.e. the Calabi--Yau manifold of this model is embedded in
an ambient space consisting of a product of two projective spaces
\beq
\matrix{\IP_2\cr \IP_3\cr} \left[\matrix{3&0\cr 1&3\cr}\right].
\eeq
Suppose we are searching for three--generation orbifolds of this space
with $\chi_{orb}=\pm 6$ . If $\chi=-6$ the constraint is not very restrictive
and allows a number of possible groups $|G|\in \{2,3,5,11,19,29\}$.
Even though it is not known whether any of these groups lead to a
three--generation model it {\it is} known that at a particular
point in the configuration space of (\ref{3gen}) described by the
superpotential
\beq W=\sum_{i=1}^3 (\Phi_i^3+\Phi_i \Psi_i^3)+\Phi_4^3 \eeq
a symmetry of order nine exists that leads to a three--generation model
\cite{s1}.

Our interest however is not restricted to models with particular spectra
for reasons explicated in the introduction. Hence we wish to implement
general types of actions regardless of their fixed point structure and
order. A general analysis of symmetries for an arbitrary Landau--Ginzburg
potential is beyond the scope of this paper; instead we restrict our
attention to the types of potentials that we have constructed explicitly.
Before we discuss these types we should remark upon a number of aspects
concerning actions on string vacua defined by LG--theories.

It is important to note that depending on the weights (or charges) of the
original LG theory it can and does happen that actions that take rather
different forms when considered as actions
on the LG theory actually are isomorphic when viewed as action of the
string vacuum proper because of the U(1) projection. It is easiest to
explain this with an example.
Consider the superpotential
\beq
W=\Phi_1^{18}+\Phi_2^{18}+\Phi_3^3+\Phi_4^3+\Phi_4\Phi_5^3
\eeq
which belongs to the configuration
$\IC_{(1,1,6,6,4)}[18]_{-204}^9$ (here the superscript denotes the
number of antigenerations and the subscript denotes the Euler number of the
configuration). At this
particular point in moduli space we can, e.g., consider the orbifolds with
respect to the actions
\bea
\ZZ_3&:&\left[\matrix{0&0&1&0&2\cr}\right],~~~(13, 79, -132) \nn \\
\ZZ_3&:&\left[\matrix{1&1&1&0&0\cr}\right],~~~(13, 79, -132) \nn \\
\ZZ_3&:&\left[\matrix{1&0&1&0&1\cr}\right],~~~(14, 44, -60) ,\label{z3}
\eea
where the notation $\ZZ_a: [p_1~\ldots~p_n]$ indicates that the
fields $\Phi_i$ transform with phases $(2\pi i p_i/a)$ under the
generator of the $\ZZ_a$ symmetry.
It is clear from the last action in (\ref{z3}) that the order of a group
is, in general, not suffient to determine the resulting orbifold spectrum
but that the specific form of the way the symmetry acts is essential.

Since the first two actions lead to the same spectrum we are led to ask
whether the two resulting orbifolds are equivalent.
Theories with the
same number of light fields need, of course, not be equivalent and to
show whether they are is, in general a rather involved analysis,
entailing the transformation behaviour of the fields and the computation
of the Yukawa couplings.

In the case at hand it is, however, very easy to check this question.
The first two actions only differ by the $6^{\rm th}$ power of the
canonical $\ZZ_{18}$ which is given by $\ZZ_3:\,[1~1~0~0~1]$.
Since the orbifolding with respect to this group is always present
in the construction of a LG vacuum the fist two orbifolds in
$eq.$~(\ref{z3}) are trivially equivalent.

Another important point is the role of trivial factors in the LG theories.
Given a superpotential $W_0$ with the correct central charge to define a
Heterotic String vacuum we always have the freedom to add trivial factors
to it
\beq
W=W_0 + \sum_i \Phi_i^2,
\eeq
since neither the central charge nor the chiral ring are changed by this
operation.
 As we restrict our attention to symmetries with unit determinant,
we gain, however, the possibility to cancel a negative sign of the determinant
by giving some $\Phi_i$ a nontrivial transformation property under a
$\ZZ_{2n}$. Adding a trivial factor hence changes the symmetry properties
of the LG--potential with regards to this class of symmetries.
\fnote{9}{In LG theories the determinant restriction is necessary for modular
 invariance and can be avoided by introducing discrete torsion \cite{iv}.}
If we wish to relate the vacuum described by the potential to a
Calabi--Yau manifold, consideration of trivial factors becomes essential
\cite{ls3}.
Consider e.g. the LG--potential
\beq
W_0=\Phi_1^{12}+\Phi_2^{12}+\Phi_3^6+\Phi_4^6
\eeq
which has $c=9$ and charges
$(\frac{1}{12},\frac{1}{12},\frac{1}{6},\frac{1}{6})$ and hence is a member
of the configuration $\IC_{(1,1,2,2)}[12]$. Only after adding the necessary
trivial factor 
this theory can be orbifolded with an action defined by
$\ZZ_2:[~1~0~0~0~1~]$ acting on the Fermat polynomial in
$\IC_{(1,1,2,2,6)}[12]$; this action leads to the orbifold spectrum
$(4,94,-180)$ and is not equivalent to any symmetry that
acts only on the first four variables with determinant 1. Neglecting the
addition of the quadratic term to the LG potential $W_0$
would have meant missing the above spectrum as one of the possible
orbifold results.

Finally it should be noted that obviously we have to make {\it some}
choice about which points in moduli space we wish to consider. Different
members of a moduli space have, in general, drastically different
symmetry properties. An example is the well known quintic theory
which we already mentioned. The most symmetric
point in the 101 dimensional space of complex deformations of the quintic is
described by the Fermat polynomial
\beq W= \sum_i \Phi_i^5,
\eeq
which has a discrete symmetry group of order $5!\cdot 5^4$. Any
deformation
breaks most of these symmetries but in some cases new symmetries appear
which
turn out to lead to new spectra. An example is furnished by the quintic
described by a combination of two 1--Tadpole polynomials and one Fermat
monomial
\beq
W=\Phi_1^5+\Phi_2^5+\Phi_2\Phi_3^4+\Phi_4^5+\Phi_4\Phi_5^4
\eeq
which, when orbifolded with respect to the symmetry
$\ZZ_2: [0~0~1~0~1]$ leads to a model with spectrum $(3,59,-112)$. This
spectrum cannot be obtained via orbifolding the Fermat quintic by any of
its supersymmetry preserving symmetries.

\vskip .1truein
\noindent
\section{Phase Actions: Implementation and Results}

Consider then a potential $W$ with $n$ order parameters
normalized such that the degree $d$ takes the lowest value such that
all order
parameters have integer weight.
In the following we discuss potentials of the type
\beq
W= \sum_i \Phi_i^{a_i} + \sum_j \left(\Phi_j^{e_j} + \Phi_j \Psi_j^{f_j}\right)
 + \sum_k \left(\Phi_k^{e_k}\Psi_k + \Phi_k \Psi_k^{f_k} \right)
\eeq
which consist of Fermat parts, tadpole parts and loop parts.

\noindent
{\bf FERMAT POTENTIALS}:
Clearly the potential $W=\sum_{i=1}^n \Phi_i^{a_i}$ is
invariant under $\prod_i \ZZ_{a_i}$, i.e. the phases of the individual fields,
acting like
\beq
\Phi_i \longrightarrow e^{2\pi i \frac{m_i}{a_i}} \Phi_i.
\eeq
For some divisor $a$ of ${\rm lcm}(a_1,\dots, a_n)$ and
$\frac{m_i}{a_i}=\frac{p_i}{a}$
we denote such an action by
\beq
\ZZ_a:~\left[\matrix{p_1&p_2 &\cdots &p_n\cr}\right],~~~0\leq~ p_i \leq~ a-1.
\eeq
and require that $a$ divides $\sum p_i$ in order to have determinant 1.

We have implemented such symmetries in the form
\beq
\ZZ_a:~\left[\matrix{(a-\sum_l i_l)&i_1&\cdots&i_p
 &(a-\sum_m j_m)&j_1&\cdots&j_q&\cdots} \right] \eeq
with the obvious divisibility conditions. For small $p$ and $q$
these symmetries can act on a large number of spaces and therefore lead to
many different orbifolds, but as $p,q$ get larger
the number of resulting orbifolds decreases rapidly.
We have stopped implementation of more complicated actions when the number of
results for the different orbifold Hodge pairs was of the order of a few tens.
As already mentioned above, the precise form of the action is very important
when considering symmetries with fixed points since the order itself is
not sufficient to determine the orbifold spectrum.

More complicated symmetries can be constructed via multiple actions
by multiplying single actions of the type described above
\beq
\prod_c \ZZ_{a_c} :~~
\left[\matrix{(a_c-\sum_l i_{c,l})&i_{c,1}&\cdots&i_{c,p}
 &(a_c-\sum_m j_{c,m})&j_{c,1}&\cdots&j_{c,q}&\cdots} \right].
\eeq
We have considered (an incomplete set of) actions of this type with up to
six twists (i.e. six $\ZZ_a$ factors). Again the precise form of the action
is rather important.

\vskip .2truein

\noindent
{\bf TADPOLE AND LOOP POLYNOMIALS}:~~
The action of the generator of the maximal phase symmetry within a tadpole
or loop sector is
\beq \ZZ_{\cO}: \left[\matrix{-f&1}\right], \eeq
where $\cO = ef$ or $ef-1$, respectively. If we want unit determinant within
one sector, we must take our generator to the $n^{\rm th}$ power with some
$n$ fulfilling $n(f-1)/\cO\in\ZZ$. With $\om =gcd(f-1,\cO)$ the action
of the resulting subgroup can be chosen to be
\beq
\ZZ_{\om}:~~ \left[\matrix{(\om-1)&1}\right].
\eeq

Other types of actions that we have considered for superpotentials consisting
of Fermat parts and tadpole/loop parts involve phases acting both on the
tadpole/loop part as well as on a number of Fermat monomials.
As was the case with pure Fermat polynomials we have also implemented
multiple actions of the type considered above.

We have implemented some forty different actions of the types
described in the previous paragraphs. These symmetries lead to a large
number of orbifolds not all of which are distinct however for reasons
explained in the previous section.
Our computations have concentrated on the number of generations and
anti--generations of these models and we have found some 1900 distinct
Hodge pairs. This set of spectra shows a mirror symmetry that is even
higher than the one exhibited by the complete intersection vacua: whereas
about 80
spectra have mirror partners!

It is obvious from this plot that there is a large overlap between the
results of \cite{cls} and the orbifolds constructed here. This might indicate
that the relation established in \cite{ls4} between orbifolds of
Landau--Ginzburg
theories and other Landau--Ginzburg theories is a general phenomenon and
not restricted to the particular classes of actions which were analysed in
\cite{ls4}.

Models with a low number of fields are clearly
of particular interest. There are two aspects to this question,
as mentioned in the introduction -- low numbers for the {\it difference}
of
generations and anti--generations
(more precisely one wants the number 3 here)
and low values for the {\it total} number of generations and
anti--generations.
As far as the latter are concerned the following `low--points' are the
`highlights' among the results for phase symmetry orbifolds.

The lowest models have $\chi=0$, more precisely the spectra (9,9,0)
and (11,11,0). These spectra appear many times in different
orbifolds of Fermat type; an example for the first one being
\beq
\IC_{(1,\ldots,1)}[9]/\ZZ_3^2: \left[\matrix{1&1&1&0&0&0&0&0&0\cr
 0&0&0&1&1&1&0&0&0\cr}\right] \eeq
or, even simpler,
\beq
\IC_{(4,4,4,4,4,4,3,3)}[12]/\ZZ_3: [1~1~1~0~0~0~0~0].
\eeq
The second one can be constructed e.g. as
\beq
\IC_{(4,3,3,3,3,2)}[12]/\ZZ_4^2: \left[\matrix{0&1&1&2&0&0\cr
 0&0&2&1&1&0\cr}\right].
\eeq

Other examples with a total of 22 generations and anti--generations are
the following orbifolds of the Fermat quintic:
\beq
\ZZ_5: \left[\matrix{0&1&2&3&4}\right],~~\qqd (1,21,-40)
\eeq
and
\beq
\ZZ_5^2: \left[\matrix{3&1&1&0&0\cr
 0&3&1&1&0\cr}\right],~~\qqd (21,1,40)
\eeq

\noindent
Of particular interest, of course, are three--generation models.
In the list of 3112 models there are no new such models aside from the
known three--generation models \cite{ks}.

Via orbifolding a number of such models can be found, which all, however,
have
a fairly large number of generations and antigenerations. We list those
in Table 2.

\begin{small}
\begin{center}
\begin{tabular}{||l|l l l r||}
\hline
\hline
$\#$ &Configuration &Potential &Action
&Spectrum \tabroom \\
\hline
\hline
1 &$\IC_{(9,2,5,9,2)}[27]_{-66}^{16}$
&$\Phi_1^3+\Phi_2^{11}\Phi_3+\Phi_2\Phi_3^5+\Phi_4^3+\Phi_4\Phi_5^9$
&$\ZZ_3~:[1~0~0~0~2]$ &$(18, 21, -6)$ \tabroom \\ [2ex]
2 &$\IC_{(9,2,5,3,8)}[27]_{-54}^{10}$
&$\Phi_1^3+\Phi_2^{11}\Phi_3+\Phi_2\Phi_3^5+\Phi_4^9+\Phi_4\Phi_5^3$
&$\ZZ_2~:[0~1~1~0~2]$ &$(21, 18, 6)$ \\ [2ex]
3 &$\IC_{(17,6,9,3,16)}[51]_{-102}^{15}$
&$\Phi_1^3+\Phi_2^7\Phi_3+\Phi_2\Phi_3^5+\Phi_4^{17}+\Phi_4\Phi_5^3$
&$\ZZ_2~:[0~1~1~0~0]$ &$(31, 34, -6)$ \\ [2ex]
4 &$\IC_{(15,15,2,9,4)}[45]_{-30}^{23}$
& $\Phi_1^3+\Phi_2^3+\Phi_2\Phi_3^{15}+\Phi_4^5+\Phi_4\Phi_5^9$
&$\ZZ_3~:[1~0~2~0~0]$ &$(23, 20, 6)$\\ [2ex]
5 &$\IC_{(15,15,10,3,2)}[45]_{-54}^{22}$
&$\Phi_1^3+\Phi_2^3+\Phi_2\Phi_3^3+\Phi_4^{15}+\Phi_4\Phi_5^{21}$
&$\ZZ_3~:[1~0~2~0~0]$ &$(35, 32, 6)$ \\ [2ex]
\hline
\hline
\end{tabular}
\end{center}
\end{small}

{\bf Table 2.} {\it Three--generation orbifold models; models which are
equivalent up to the U(1) projection are not listed separately.}

\noindent
By using the relation established in \cite{ls4} between LG/CY--theories
via fractional transformations it can be shown that the orbifold $\#1$ in
Table 2,
\beq
\IC_{(2,5,9,2,9)}[27]_{-66}^{16}/\ZZ_3:[~0~0~0~2~1],
\eeq
for which the covering model is described by the polynomial
\beq
W= \Phi_1^{11}\Phi_2+\Phi_1\Phi_2^5+\Phi_3^3+\Phi_3\Phi_4^9+\Phi_5^3,
\eeq
is isomorphic to the orbifold
\beq
\IC_{(2,5,9,3,8)}[27]_{-54}^{10}/\ZZ_2:[~0~0~0~1~1]
\eeq
where the covering theory is described by the polynomial
\beq
W= \Phi_1^{11}\Phi_2+\Phi_1\Phi_2^5+\Phi_3^3+\Phi_3\Phi_4^6
 +\Phi_4\Phi_5^3.
\eeq
The latter is a theory involving a subtheory with couplings among three
scaling fields and hence goes beyond the types of potentials we have
implemented.
This example indicates that more complicated examples than the ones
investigated here are likely to yield more (perhaps more realistic)
three generation models.

The covering spaces of all the three generation models are described by
either tadpole or loop type polynomials, and with our actions none of the
Fermat type polynomials leads to a three generation model.
It should be noted that these orbifolds exist only at particular points in
moduli space.

\vskip .1truein
\noindent
\section{ Conclusion}

\noindent
It is clear that the structure of the configuration space of the Heterotic
String is not particularly well understood and that much remains to be done;
it is apparent from
the results described above that what has been achieved so far at best is
 little more than scratching the surface. It might be hoped that once
a completely mirror symmetric part of the moduli space has been constructed
a representative part of the complete space has been uncovered.
Pursueing work
along the lines described above is certainly promising in this regard;
very likely it is possible to generate a mirror symmetric subspace by
all orbifolds of the Landau--Ginzburg theories listed in \cite{ks}
or, more generally, to construct all weighted complete intersection
Calabi--Yau
manifolds and their orbifolds.

Having done that it still remains to show that all potential mirror
partners
in this symmmetric subspace of ground states are in fact related.
Even though many types of LG--potentials constructed in \cite{cls} admit,
via fractional transformations, an interpretation as orbifolds not every
mirror potential can be constructed in this way at present. It is therefore
 clear that a generalization of this type of mirror map is necessary.

Aside from the question of mirror symmetry the orbifold technique is
extremely useful to get insight into both, the detailed structure
of the vacua constructed via such a classification of LG--potentials and
the relation between these vacua.
Properties of the ground states that are obscure from the point of view
of the
superpotentials alone or from the point of view of the partition function
of the underlying conformal field theory (if known) become rather obvious
in the orbifold picture.

\vskip .2truein

\noindent
\section*{Acknowledgement}

\noindent
Most of the work described here has been the result of the joint efforts
of several collaborations. I'm grateful to all the people involved, in
particular
Philip Candelas, Albrecht Klemm, Max Kreuzer and Monika Lynker.

\vfill \eject

\end{document}